\documentclass[11pt,twoside]{article} 

\usepackage{times,fancyhdr}
\usepackage{cite}
\usepackage{graphicx}
\usepackage{amsmath, amssymb, amsfonts}

\def\d#1#2{{\partial #1\over \partial #2 } }
\def\dd#1#2{{\partial^2 #1\over \partial #2^2 } }
\def\onehalf{{\scriptstyle\frac{1}{2}}}

\let\ro=\varrho

\def\R{\mathbb R}

\newtheorem{remark}{Remark}[section]

\date{}

\title{Transformation methods for evaluating approximations to the optimal
exercise boundary for linear and nonlinear Black-Scholes equations}
\author{Daniel \v{S}ev\v{c}ovi\v{c}$^1$}
\begin{document} 
\normalsize
\pagestyle{fancy}
\fancyhead{} 
\fancyhead[EC]{Daniel \v{S}ev\v{c}ovi\v{c}}
\fancyhead[EL,OR]{\thepage}
\fancyhead[OC]{Transformation methods for linear and nonlinear Black-Scholes equations}
\fancyfoot{} 
\renewcommand\headrulewidth{0.5pt}
\addtolength{\headheight}{2pt} 

\maketitle 
\small
\begin{center}
$^1$ Department of Applied Mathematics and Statistics,
\\
Faculty of Mathematics, Physics \& Informatics, Comenius University,
\\
842 48 Bratislava, Slovak Republic, \quad {\tt sevcovic@fmph.uniba.sk}
\end{center}
\par
\medbreak
\textbf{Abstract.}
The purpose of this survey chapter is to present a transformation technique
that can be used in analysis and numerical computation of the early exercise
boundary for an American style of vanilla options that can be modelled by
class of generalized Black-Scholes equations. We analyze qualitatively and
quantitatively the early exercise boundary for a linear as well as a class
of nonlinear Black-Scholes equations with a volatility coefficient which can
be a nonlinear function of the second derivative of the option price itself. A
motivation for studying the nonlinear Black-Scholes equation with a
nonlinear volatility arises from option pricing models taking into account
e.g. nontrivial transaction costs, investor's preferences, feedback and
illiquid markets effects and risk from a volatile (unprotected) portfolio.
We present a method how to transform the free boundary problem for the
early exercise boundary position into a solution of a time depending
nonlinear nonlocal parabolic equation defined on a fixed domain. We
furthermore propose an iterative numerical scheme that can be used in order to find
an approximation of the free boundary. In the case of a linear Black-Scholes
equation we are able to derive a nonlinear integral equation
for the position of the free boundary. We present results of numerical
approximation of the early exercise boundary for various types of linear and
nonlinear Black-Scholes equations and we discuss dependence of the free
boundary on model parameters. Finally, we discuss an application of the transformation method for the pricing equation for American type of Asian options. 

\normalsize

{\small
\tableofcontents
}

\section{Introduction}

According to the classical theory due to Black, Scholes and Merton 
the price of an option  in an idealized  financial market  can be computed  from a solution to the well-known Black--Scholes linear parabolic equation derived by Black and Scholes in \cite{BS}, and, independently by Merton  (see also Kwok \cite{Kw}, Dewynne {\em et al.} \cite{DH}, Hull \cite{H}, Wilmott {\em et al.} \cite{WDH}).
Recall that a European call (put) option is the right but not obligation to purchase (sell) an underlying  asset at the expiration price $E$ at the expiration time $T$.  Assuming that the underlying asset $S$ follows a geometric Brownian motion 
\begin{equation}
d S = (\ro-q) S dt + \sigma S d W,
\label{afm-geomBrown}
\end{equation}
where $\ro$ is a drift, $q$ is the asset dividend yield rate, $\sigma$ is the volatility of the asset and $W$ is the standard Wiener process (cf. \cite{Kw}),
one can derive a governing  partial differential equation for the price of an option. We remind ourselves that the equation  for option's price $V(S,t)$ is the following parabolic PDE:
\begin{equation}
\frac{\partial V}{\partial t} +  (r-q) S\frac{\partial V}{\partial S} + \frac{\sigma^2}{2} S^2 \frac{\partial^2 V}{\partial S^2} - r V =0
\label{c-BS}
\end{equation}
where $\sigma$ is the volatility of the underlying asset price process, $r>0$ is the interest rate of a zero-coupon bond, $q\ge 0$  is the dividend yield rate. A solution $V=V(S,t)$ represents the price of an option if the price of an underlying asset is $S>0$ at  time $t\in[0,T]$.

The case when the diffusion coefficient $\sigma>0$  in (\ref{c-BS}) is constant represents a classical Black--Scholes equation originally derived by Black and Scholes in \cite{BS}. On the other hand, if we assume the volatility coefficient $\sigma>0$ to be a function  of the solution $V$ itself then  (\ref{c-BS}) with such a diffusion coefficient represents a nonlinear generalization of the Black--Scholes equation. It is a purpose of this chapter to focus our attention to the case when the diffusion coefficient
$\sigma^2$ may depend on the time $T-t$ to expiry, the asset price $S$ 
and the second derivative $\partial^2_S V$ of the option price  (hereafter referred to as $\Gamma$), i.e.
\begin{equation}
\sigma = \sigma(S^2\partial^2_S V, S, T-t)\,.
\label{c-sigma}
\end{equation}
A motivation for studying the nonlinear Black--Scholes equation (\ref{c-BS}) with a volatility $\sigma$ having a general form (\ref{c-sigma}) arises from option pricing  models taking into account nontrivial transaction costs, market feedbacks and/or risk from a volatile  (unprotected) portfolio. Recall that the linear Black--Scholes equation with constant $\sigma$ has  been derived under several restrictive assumptions like e.g. frictionless, liquid and complete markets, etc.  We also recall that the linear Black--Scholes equation provides a perfectly replicated hedging portfolio. In the last decades some of these assumptions have been relaxed in order to model, for instance, the  presence of transaction costs (see e.g. Leland \cite{Le}, Hoggard {\em et al.} \cite{HWW}, Avellaneda and  Paras \cite{AP}), feedback and illiquid market effects due to large traders choosing given stock-trading  strategies (Frey and Patie \cite{FP}, Frey and Stremme \cite{FS}, During {\em et al.}\cite{DFJ}, Sch\"onbucher and  Wilmott \cite{SW}), imperfect replication and investor's preferences (Barles and Soner  \cite{BaSo}), risk from unprotected portfolio (Kratka \cite{Kr}, Janda\v{c}ka and \v{S}ev\v{c}ovi\v{c} \cite{JS} or \cite{Se2}). One of the first nonlinear models is the so-called Leland model (cf. \cite{Le}) for pricing call  and put options under the presence of transaction costs. It has been generalized for more complex  option strategies by Hoggard, Whaley and Wilmott  in \cite{HWW}. In this model the volatility $\sigma$ is given by
\begin{equation}
\sigma^2 (S^2\partial^2_S V, S, \tau) = \hat\sigma^2 ( 1 + \hbox{Le}\, \hbox{sgn}(\partial^2_S V) )
\label{c-leland}
\end{equation}
where  $\hat\sigma>0$ is a constant historical volatility of the underlying asset price process and $\hbox{Le}>0$ is the so-called Leland constant given by  $\hbox{Le} = \sqrt{2/\pi} C/(\hat\sigma \sqrt{\Delta t})$ where $C>0$ is a constant round trip transaction cost per unit dollar of transaction in the assets market and $\Delta t>0$ is the time-lag between portfolio adjustments. 

Notice that dependence of volatility adjustment on the second derivative of the price is quite natural. Indeed, in the idealized Black-Scholes theory, the optimal hedge is equal to $\pm \partial_S V$ and therefore one may expect more frequent transaction in  regions with the high second derivative $\partial^2_S V$ (cf. \cite{BS}).

A popular nonlinear generalization of the Black--Scholes equation has been proposed by Avellaneda, Levy, and Paras  \cite{AP2} for description of incomplete markets and uncertain but bounded volatility. In their model we have
\begin{equation}
\sigma^2(S^2\partial^2_S V, S, \tau) 
= \left\{
\begin{array}{cc}
\hat\sigma^2_1 & \quad \hbox{if} \ \partial^2_S V < 0, \\ 
\hat\sigma^2_2 & \quad \hbox{if} \ \partial^2_S V > 0,
\end{array}\right.
\label{c-avel}
\end{equation}
where $\sigma_1$ and $\sigma_2$ represent a lower and upper a-priori bound on the otherwise unspecified asset price volatility.

If transaction costs are taken into account perfect replication of the
contingent claim is no longer possible and further restrictions are needed 
in the model. By assuming that investor's preferences are characterized by an 
exponential utility function   Barles and Soner (cf. \cite{BaSo}) derived a 
nonlinear Black--Scholes equation with the volatility $\sigma$ given by
\begin{equation}
\sigma^2(S^2\partial^2_S V, S, \tau) 
= \hat\sigma^2 \left(1+\Psi(a^2 e^{r \tau} S^2\partial^2_S V)\right)
\label{c-barles}
\end{equation}
where $\Psi$ is a solution to the 
ODE: $\Psi^\prime(x) = (\Psi(x)+1)/(2 \sqrt{x\Psi(x)} -x), \Psi(0)=0,$
and $a>0$ is a given constant representing risk aversion. 
Notice that $\Psi(x)=O(x^\frac{1}{3})$ for 
$x\to 0$ and $\Psi(x)=O(x)$ for $x\to\infty$. 

Another popular model has been derived for the case when the asset dynamics 
takes into account the presence of feedback and illiquid market effects. Frey and Stremme 
(cf. \cite{FS,FP}) introduced directly the asset price dynamics in the case
when a large trader chooses a given stock-trading strategy (see also \cite{SW}). 
The diffusion coefficient $\sigma$ is again nonconstant and it can be expressed as:
\begin{equation}
\sigma^2(S^2\partial^2_S V, S, \tau) = \hat\sigma^2 \left(1-\varrho \lambda(S) S\partial^2_S V\right)^{-2}
\label{c-frey}
\end{equation}
where $\hat\sigma^2, \varrho>0$ are constants and $\lambda(S)$ is a strictly convex function,  $\lambda(S)\ge 1$. Interestingly enough, explicit solutions to the Black--Scholes equation with varying volatility as in (\ref{c-frey}) have been derived by Bordag and Chankova\cite{BC} and Bordag and Frey \cite{BF}.

The last example of the Black--Scholes  equation with a nonlinearly depending volatility is the so-called Risk Adjusted Pricing Methodology model proposed by Kratka in \cite{Kr} and revisited by Janda\v{c}ka and \v{S}ev\v{c}ovi\v{c} in \cite{JS}.  In order to maintain (imperfect) replication of a portfolio by the delta hedge one has to make  frequent portfolio adjustments leading to a substantial increase in transaction costs.  On the other hand, rare portfolio adjustments may lead to an increase of the risk arising  from a volatile (unprotected) portfolio. In the RAPM model the aim is to  optimize  the time-lag $\Delta t$ between consecutive portfolio adjustments.  By choosing $\Delta t>0$ in such way that the sum of the rate of transaction costs and the rate of a risk from unprotected portfolio is minimal  one can find the optimal time lag $\Delta t>0$. In the RAPM model, it turns out that the volatility is again nonconstant and  it has the following form:
\begin{equation}
\sigma^2(S^2\partial^2_S V, S, \tau) = \hat\sigma^2 \left(1 + \mu (S\partial^2_S V)^{\frac{1}{3}} \right)\,. 
\label{c-RAPM}
\end{equation}
Here $\hat\sigma^2>0$ is a constant historical volatility of the asset price returns and $\mu=3(C^2 R/2\pi)^{\frac{1}{3}}$ where $C, R\ge 0$ are  nonnegative constant representing the transaction cost measure and the risk premium measure, resp.  (see \cite{JS} for details).

Notice that all the above mentioned nonlinear models are consistent with the original  Black--Scholes equation in the case the additional model parameters (e.g. Le, $a$,  $\varrho$, $\mu$) are vanishing. If plain call or put vanilla options are concerned then the function $V(S,t)$ is convex in $S$ variable and therefore each of  the above mentioned models  has a diffusion coefficient strictly larger than $\hat\sigma^2$ leading to a larger values of computed option  prices. They can be therefore identified with higher Ask option prices, i.e. offers to sell an option.  Furthermore, these models have been considered and analyzed mostly for European style of options, i.e.  options that can be exercised only at the maturity $t=T$. On the other hand, American  options  are much more common in  financial markets as they allow for exercising of an option anytime before the expiry $T$. In the case of an American call option  a solution to equation (\ref{c-BS}) is defined on a time dependent domain 
$0<S<S_f(t),\ 0<t<T$. 
It is subject to the boundary conditions
\begin{equation}
V(0,t) = 0\,,\ \ V(S_f(t), t)= S_f(t)-E\,, \ \ \partial_S V (S_f(t),t) = 1\,,
\label{c-bccall}
\end{equation}
and terminal pay-off condition at expiry $t=T$
\begin{equation}
V(S,T)=\max(S-E, 0)
\label{c-tccall}
\end{equation}
where $E>0$ is a strike price (cf. \cite{DH,Kw}). 
One of important problems in this field is the analysis of the early exercise boundary 
$S_f(t)$ and  the optimal stopping time (an inverse function to $S_f(t)$) 
for American call (or put) options on stocks paying a continuous dividend yield with a rate $q>0$ (or $q\ge 0$). However, an exact analytical expression for the free boundary profile is not even known for the case when the volatility $\sigma$ is constant. Many authors have investigated  various approximation models leading to approximate expressions for valuing American call and put options: analytic approximations  (Barone--Adesi and Whaley \cite{BW},  Kuske and Keller \cite{KK}, Dewynne {\em et al.} \cite{DH}, Geske {\em et al.} \cite{GJ,GR}, MacMillan \cite{Mac}, Mynemi \cite{M});  methods of reduction to a nonlinear integral equation  (Alobaidi \cite{A},  Kwok \cite{Kw}, Mallier {\em et al.} \cite{Ma,MA}, \v{S}ev\v{c}ovi\v{c} \cite{Se}, Stamicar {\em et al.} \cite{SSC}). In recent papers \cite{Zhu1,Zhu2}, Zhu derived a closed form of the free boundary position in terms of on infinite series. We also refer to a recent survey paper by Chadam \cite{Ch} focusing on free boundary problems in mathematical finance. 

We remind ourselves that in the case of a constant volatility there are, in principle, two ways how to solve numerically the free boundary problem for the value of an American call resp. put option and the position of the early exercise  boundary. The first class of algorithms is based on reformulation of the problem in terms of a variational inequality (see Kwok \cite{Kw} and references therein). The variational inequality can be then solved numerically by the so-called Projected Super Over Relaxation method (PSOR for short). An advantage of this method is that it gives us immediately the value of a solution. A disadvantage is that one has to solve large systems of linear equations  iteratively taking into account the obstacle for a solution, and, secondly, the free  boundary position should be deduced from the solution a posteriori. Moreover, the PSOR method is not directly applicable for solving the problem (\ref{c-BS})-(\ref{c-bccall}) when the diffusion coefficient $\sigma$ may depend on the second derivative of a solution itself. The second class of methods is based on derivation of a nonlinear integral equation for the position of the free boundary  without the need of knowing the option price itself (see e.g. Evans, Kuske-Keller \cite{EKK,KK}, Mallier and Alobaidi \cite{A,Ma,MA},  \v{S}ev\v{c}ovi\v{c} {\em et al.} \cite{Se,SSC}, Chadam {\em et al.} \cite{Ch,CC}. In this approach an advantage is that only a single equation for the free boundary has to be solved provided that $\sigma$ is constant; a disadvantage is that the method is based on integral transformation techniques and therefore the assumption $\sigma$ is constant is crucial. 

 Higher order finite difference approximations of the free boundary problems for call or put options are discussed in recent papers by Ankudinova and Ehrhardt \cite{AE1,AE2}, Ehrhardt and Mickens \cite{EM} and Zhao {\em et al.} \cite{ZCD}.
Other interesting analytical and numerical methods for evaluating the early exercise boundary have been recently studied by e.g. Milstein {\em et al.} \cite{MS} (Monte--Carlo methods), Widdicks {\em et al.} \cite{WD} (singular pertubation techniques and asymptotic expansions), Cho {\em et al.} \cite{KKK} (parameter estimation methods), Grandits and Schachinger {\em et al.} \cite{GS} (tracking of a discontinuity method), Imai {\em et al.} \cite{IIS} (numerical method for generalized Lelend model). 

In this survey we recall an iterative numerical algorithm for solving the free boundary problem for an American type of options in the case the volatility $\sigma$ may depend  on the option and asset values as well as on the time $T-t$ to expiry as well as for American type of Asian options. A key idea of this method  consists in transformation of the free boundary problem into a semilinear  parabolic equation  defined  on a fixed spatial domain coupled with a nonlocal algebraic constraint  equation for the free boundary position. It has been proposed and analyzes by the author in a series of papers \cite{SSC,Se,Se2,Se3}. Since the resulting parabolic equation contains a strong convective term we make use of the operator splitting method in order to overcome numerical difficulties. A full space-time discretization of the problem leads to a system  of semi-linear algebraic equations that can be solved by an iterative procedure at each time level. 

The rest of the chapter has the following organization: in the next section we recall the well known nonlinear generalization of Black--Scholes equations due to Frey and Stremme, Barles and Soner and Kratka, Janda\v{c}ka and \v{S}ev\v{c}ovi\v{c}, resp.  We also present qualitative and quantitative properties of the nonlinear models for pricing European style of options  with special focus on the Risk adjusted pricing methodology (RAPM) due to Kratka \cite{Kr}, Janda\v{c}ka and \v{S}ev\v{c}ovi\v{c} \cite{JS}. Section 3 is devoted to the free boundary problem for pricing American options by means of a linear Black--Scholes equation, i.e. $\sigma>0$ is constant. This is important in order to understand important steps of the fixed domain transformation method. A resulting system of transformed equations consists of a  nonlocal parabolic equation defined on a fixed domain with time depending coefficients and an algebraic constraint equation for the free boundary position. Since the volatility $\sigma$ is constant in this case, by using Sine and Cosine Fourier transformations the system of transformed equations can be further simplified and reduced to a single nonlinear integral equation for the free boundary position. We show how this integral equation can be utilized in order to obtain qualitative properties of the free boundary position (early exercise behavior, long time behavior) as well as quantitative properties (a fast and stable numerical scheme for computing the free boundary function). In Section 4 we discuss  a transformation method applied to a class of nonlinear Black--Scholes equations. We are able to derive a similar system of transformed parabolic--algebraic equation to the one from Section 3. However, as the volatility is no longer a constant and it may depend on the solution itself the resulting system of equations can not be reduced to a single integral equation for the free boundary and it has to be solved numerically. We propose a numerical method based on the finite difference approximation combined with an operator splitting technique for numerical approximation of the solution and computation of the free boundary position. Several numerical results for nonlinear Black--Scholes equations with volatility functions $\sigma$ defined as in (\ref{c-barles}) and (\ref{c-RAPM}) are presented. We also compare our methodology with well-known methods for evaluation of approximation to the free boundary position in the case the volatility  $\sigma$ is constant. We analyze dependence of the free boundary position with respect to various parameters entering expressions  (\ref{c-barles}) and (\ref{c-RAPM}). Finally, Section 5 is devoted to a recent application of the transformation method in the case of American style of Asian options in which the strike price is an arithmetical average of underlying asset prices. Although the volatility $\sigma$ is assumed to be constant, due to a specific character of the Black--Scholes equation for pricing Asian option the resulting transformed system of equation cannot be further reduced and has to be solved numerically by a slight modification of the  numerical method discussed in Section 4. We finish the last section by presentation of several illustrative examples describing the early exercise boundary for American style of Asian call options with floating strike. 

\section{Risk adjusted methodology model}

The aim of this section is to present one of nonlinear generalizations of the classical Black--Scholes equation with a volatility $\sigma$ of the form (\ref{c-sigma}) in a more detail. We focus on the so-called  Risk adjusted pricing methodology model due to Kratka \cite{Kr} and straightforward generalization by Janda\v{c}ka and \v{S}ev\v{c}ovi\v{c} \cite{JS} (see also \cite{Se2}). In this model both the risk arising from nontrivial transaction costs as well as the risk from unprotected volatile portfolio are taken into account. Their sum representing the total risk is subject of minimization.  The original model was proposed by \cite{Kr}. In \cite{JS} we modified Kratka's approach by considering a different measure for risk arising from unprotected portfolio in order to  construct a model which is scale invariant and mathematically well posed.  These two important features were missing in the original model of Kratka. The model is based  on the Black--Scholes parabolic PDE in which transaction costs are described by the Hoggard, Whalley  and Wilmott extension of the Leland model (cf. \cite{HWW,Kw,H}) whereas  the risk from a volatile portfolio is described by the average value of the variance of  the synthesized portfolio. Transaction costs as well as the volatile portfolio risk depend  on the time-lag between two consecutive transactions. We define the total risk premium as a sum of transaction costs and the risk cost from the unprotected volatile portfolio. By minimizing  the total risk premium functional we obtain  the optimal length of the hedge interval. 

Concerning the dynamics of an underlying asset we will 
assume that the asset price $S=S(t), t\ge 0,$ follows a geometric Brownian motion (\ref{afm-geomBrown})  with a drift $\rho$, standard deviation $\hat\sigma>0$ and it may pay continuous dividends, i.e. $dS = (\rho - q) S dt + \hat\sigma S dW$ where $dW$ denotes the differential of the standard Wiener process
and $q\ge0$ is a continuous dividend yield rate. 
This assumption is usually made when deriving the classical Black--Scholes
equation (see e.g. \cite{H,Kw}). 
Similarly as in the derivation of the classical Black--Scholes equation
we construct a synthesized portfolio $\Pi$ consisting of a one
option with a price $V$ and $\delta$ assets with a price $S$ per one asset: 
\begin{equation}
\Pi = V + \delta S\,.
\label{afm-syntport}
\end{equation}
We recall that the key idea in the Black--Scholes theory is to examine 
the differential $\Delta\Pi$ of equation (\ref{afm-syntport}). The right-hand side of  
(\ref{afm-syntport}) can be differentiated by using It\^o's formula whereas 
portfolio's  increment
$\Delta \Pi(t)=\Pi(t+\Delta t) - \Pi (t)$ of the left-hand side can
be expressed as follows:
\begin{equation}
\Delta\Pi = r \Pi \Delta t + \delta q S\Delta t
\label{afm-port-simple}
\end{equation}
where $r>0$ is a risk-free interest rate of a zero-coupon bond. In the real 
world, such a simplified  assumption is not satisfied and a new term measuring 
the total risk should be added to (\ref{afm-port-simple}). More precisely, the change
of the portfolio $\Pi$ is composed of two parts: the risk-free interest 
rate part $r \Pi \Delta t$ and the total risk premium:
$r_{R} S \Delta t$ where $r_R$ is a risk premium per unit asset price. 
We consider a short positioned call option. Therefore the writer of an option is exposed to this total risk. Hence we are going to price the higher Ask option price -- an offer to sell an option. It means that $\Delta\Pi = r \Pi \Delta t - r_{R} S \Delta t$.
The total risk premium $r_{R}$  consists of the transaction risk premium 
$r_{TC}$ and the portfolio volatility risk premium $r_{VP}$, i.e. 
$r_{R}=r_{TC}+r_{VP}$. Hence

\begin{equation}
\Delta\Pi = r \Pi \Delta t + \delta q S \Delta t 
- (r_{TC} + r_{VP}) S \Delta t\,.
\label{afm-port-full}
\end{equation}
Our next goal is to show how these risk premium measures $r_{TC},$ $r_{VP}$ 
depend on the time lag and other quantities, like e.g. $\hat\sigma, S, V,$ and derivatives of  $V.$ The problem can be decomposed in two parts: modeling the  transaction costs measure $r_{TC}$ and   volatile portfolio risk measure $r_{VP}$. 

We begin with modeling transaction costs. In practice, we have to adjust our portfolio by frequent buying and selling  of assets. In the presence of nontrivial transaction costs, continuous  portfolio adjustments may lead to infinite total transaction costs. A natural  way how to consider transaction costs within the frame of the Black--Scholes  theory is to follow the well known Leland approach extended by Hoggard, Whalley and Wilmott (cf. \cite{HWW,Kw}). We will recall an idea how to incorporate the effect of transaction costs into the governing  equation. More precisely, we will derive the coefficient of transaction costs  $r_{TC}$ occurring in (\ref{afm-port-full}). Let us denote by $C$ the round trip transaction cost per unit dollar of
transaction. Then 
\begin{equation}
C=(S_{ask}-S_{bid})/S 
\label{afm-Ccoeff}
\end{equation}
where $S_{ask}$ and $S_{bid}$  are the so-called Ask and Bid prices of the 
asset, i.e. the market price offers for selling and buying assets, resp.
Here $S=(S_{ask}+S_{bid})/2$ denotes the mid value of the underlying asset price. 

In order to derive the term $r_{TC}$ in (\ref{afm-port-full}) measuring
transaction costs we will assume, for a moment, that there is no 
risk from volatile portfolio, i.e. $r_{VP}=0$. Then
$\Delta V +\delta \Delta S = \Delta \Pi = r\Pi\Delta t +\delta q S\Delta t
+ r_{TC} S \Delta t$. Following Leland's approach (cf. \cite{HWW}),
using It\^o's formula and assuming $\delta$-hedging of a synthetised 
portfolio $\Pi$ one can derive that 
the coefficient $r_{TC}$ of transaction costs is given by the formula:
\begin{equation}
r_{TC}=\frac{ C \hat\sigma S}{\sqrt{2\pi}} \left| \partial^2_S V \right|
\frac{1}{\sqrt{\Delta t} }
\label{afm-rTC}
\end{equation}
(see \cite[Eq. (3)]{HWW}). It leads to the well known Leland generalization of the Black--Scholes equation (\ref{c-BS}) in which the diffusion coefficient is given by (\ref{c-leland}) (see \cite{Le,HWW} for details).

Next we focus our attention to the problem how to incorporate
a risk from a volatile portfolio into the model. In the case when a portfolio 
consisting of options and assets is highly volatile,  an investor usually
asks for a price compensation. Notice that exposure to risk is higher 
when the time-lag between portfolio adjustments is higher. We shall 
propose a measure of such a risk based on the volatility of a 
fluctuating portfolio. It can be measured by the variance
of relative increments of the replicating portfolio $\Pi=V+\delta S$, 
i.e. by the term $var((\Delta\Pi)/S)$. Hence it is reasonable to
define the measure $r_{VP}$ of the portfolio volatility risk  as follows:
\begin{equation}
r_{VP}=R\, \frac{var\left( \frac{\Delta \Pi}{S}\right) }{\Delta  t}\,.
\label{afm-r-coeff}
\end{equation}
In other words, $r_{VP}$ is proportional to the variance of a relative
change of a portfolio per time interval $\Delta t$. A constant $R$
represents the so-called {\it risk premium coefficient}. It can be interpreted as the marginal 
value of investor's exposure to a risk. If we apply It\^o's formula to the
differential $\Delta\Pi = \Delta V+\delta \Delta S$ we obtain 
$\Delta\Pi = 
\left( \partial_S V +\delta \right) \hat\sigma S \Delta W 
+ \onehalf \hat\sigma^2 S^2 \Gamma (\Delta W)^2
+ {\mathcal G}$ 
where $\Gamma=\partial_S^2 V$ and ${\mathcal G}=(\partial_S V +\delta)\rho S \Delta t 
+\partial_t V \Delta t$ is a deterministic term, i.e $E({\mathcal G}) = {\mathcal G}$
in the lowest order $\Delta t$ - term approximation. Thus 
\[
\Delta\Pi - E(\Delta\Pi) 
= \left( \partial_S V +\delta \right) \hat\sigma S \phi \sqrt{\Delta t}
+ \frac12 \hat\sigma^2 S^2 (\phi^2 -1)\Gamma  \Delta t
\]
where $\phi$ is a random variable with the standard normal distribution such
that $\Delta W = \phi \sqrt{\Delta t}$. Hence the variance of  
$\Delta\Pi$ can be computed as follows:
\[
var(\Delta \Pi)=E\left( [\Delta \Pi-E(\Delta \Pi)]^2 \right) 
= E\left( [ ( \partial_S V +\delta ) \hat\sigma S \phi \sqrt{\Delta t}
+\onehalf\hat\sigma^2 S^2\Gamma 
\left( \phi ^2-1 \right) \Delta t ] ^2 \right)\,.
\]
Similarly, as in the derivation of the transaction costs measure $r_{TC}$ we 
assume the $\delta$-hedging of portfolio adjustments, i.e. we choose 
$\delta=-\partial_S V$. Since $E((\phi^2-1)^2) = 2$ we obtain an expression 
for the risk premium $r_{VP}$ in the form:
\begin{equation}
r_{VP}=\frac12 R\hat\sigma^4 S^2\Gamma ^2\Delta t\,.
\label{afm-rVP}
\end{equation}
Notice that in our approach the increase in the time-lag $\Delta t$ 
between consecutive 
transactions leads to a linear increase of the risk from a volatile portfolio
where the coefficient of proportionality depends the asset price 
$S$, option's Gamma, $\Gamma=\partial^2_S V$, as well as the constant
historical volatility $\hat\sigma$ and the risk premium coefficient $R$. 

\subsection{Risk adjusted Black--Scholes equation}

\begin{figure}
\begin{center}
\includegraphics[width=0.45\textwidth]{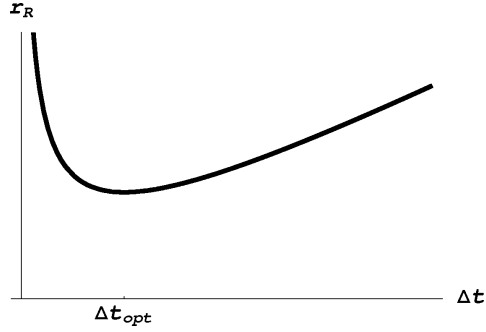}
\end{center}
\caption{\small 
The total risk premium $r_R = r_{TC}+r_{VP}$ as a function of the
time-lag $\Delta t$ between two consecutive portfolio adjustments.}
\label{optimalnetau}
\end{figure}

The total risk premium $r_R =  r_{TC} + r_{VP}$ consists of two parts: 
transaction costs premium $r_{TC}$ and the risk from a volatile portfolio 
$r_{VP}$ premium  defined as in (\ref{afm-rTC}) and  (\ref{afm-rVP}), resp. We
assume that an investor is risk aversive and he/she wants to minimize the 
value of the total risk premium 
$r_R$. For this purpose one has to choose the optimal time-lag $\Delta
t$ between two consecutive portfolio adjustments. As both $r_{TC}$ as well as
$r_{VP}$ depend on the time-lag $\Delta t$ so does the total risk premium
$r_R$. In order to find the optimal value of $\Delta t$ we have to minimize
the following function:
\[
\Delta t \mapsto r_R = r_{TC} + r_{VP}  = 
\frac{C|\Gamma| \hat\sigma S }{\sqrt{2\pi}} \frac{1}{\sqrt{\Delta  t}}
+
\frac{1}{2} R\hat\sigma^4 S^2\Gamma ^2 \Delta t\,.
\]
The unique minimum of the function $\Delta t \mapsto r_R(\Delta t)$ is attained at the time-lag
$\Delta t_{opt} = K^2/(\hat\sigma^2|S\Gamma|^\frac23)$ where
$K = (C/(R\sqrt{2\pi })^\frac13$.
For the minimal value of the function $\Delta t\mapsto r_R(\Delta t)$ we
have
\begin{equation}
r_R(\Delta t_{opt}) =  \frac32 \left(\frac{C^2 R}{2\pi}\right)^\frac13
\hat\sigma^2 |S\Gamma|^\frac43\,.
\label{afm-rRmin}
\end{equation}
Taking into account both transaction costs as well as risk from a volatile 
portfolio effects we have shown that the equation for the change $\Delta\Pi$ 
of a portfolio $\Pi$ read as:
\[
\Delta V +\delta \Delta S =\Delta \Pi \Delta t 
= r \Pi \Delta t +\delta q S \Delta t - r_R S\Delta t
\]
where $r_R$ represents the total risk premium, $r_R = r_{TC} + r_{VP}$. Applying It\^o's lemma to a smooth  function $V=V(S,t)$ and assuming the $\delta$-hedging strategy for 
the portfolio adjustments we finally obtain the following generalization of 
the Black--Scholes equation for valuing options:
\[
\frac{\partial V}{\partial t}
+ \frac{\hat\sigma^2}{2} S^2 \frac{\partial^2 V}{\partial S^2}
+(r-q) S \frac{\partial V}{\partial S}  -r V - r_{R} S = 0\,.
\]
By taking the optimal value of the total risk coefficient $r_R$ derived as in 
(\ref{afm-rRmin}) the option price $V$ is a solution to the following 
nonlinear parabolic equation:
\par({\it Risk adjusted Black--Scholes equation})
\begin{equation}
\frac{\partial V}{\partial t} + \frac{\hat\sigma^2}{2} S^2 
\left(1 +  \mu (S\partial^2_S V)^\frac13\right) \frac{\partial^2 V}{\partial S^2} 
+ (r-q)S \frac{\partial V}{\partial S} -r V =0\,,
\label{afm-generalizedBS}
\end{equation}
where  $\mu = 3\left(\frac{C^2 R}{2\pi}\right)^\frac13$ and $u^p$ with $u=S\partial^2_S V$ and $p=1/3$ stands for the signed power function, i.e. $u^p=|u|^{p-1} u$.  In the case there are neither transaction costs ($C=0$) nor the risk from a volatile portfolio ($R=0$) we have $\mu =0$. Then equation 
(\ref{afm-generalizedBS}) reduces to the original Black--Scholes linear  parabolic equation (\ref{c-BS}). We note that equation (\ref{afm-generalizedBS}) is a backward parabolic PDE if and only if the function  $\beta(H)=\frac{\hat\sigma^2}{2}(1+\mu H^\frac13) H$ is an increasing function in the variable $H:=S\Gamma=S\partial^2_S V$. It is clearly satisfied if $\mu\ge 0$ and $H\ge 0$. 

As it is usual in the classical Black-Scholes
theory for European style of options (cf. \cite{H,Kw}) we consider the change of independent variables:
$x:=\ln\left(\frac{S}{E}\right)\,,\ \  x\in \R\,,  \qquad \tau:=T-t\,, \ \ \tau\in(0,T)\,.$
As equation (\ref{afm-generalizedBS}) contains the term $S\Gamma=S\partial^2_S V$  it is convenient to introduce the following transformation: $H(x,\tau) := S \Gamma = S\partial^2_S V(S,t)$. 
Now we are in position to derive an equation for the function $H$. It turns out that the function $H(x,\tau)$ is a solution to a 
nonlinear parabolic equation subject to the initial and boundary conditions. More 
precisely, by taking the second derivative of equation (\ref{afm-generalizedBS}) 
with respect to $x$ we obtain, after some calculations, that $H=H(x,\tau)$ is 
a solution to the quasilinear parabolic equation 
\begin{equation}
\frac{\partial H}{\partial \tau}
= \frac{\partial^2}{\partial x^2} \beta(H) + \frac{\partial}{\partial x}\beta(H)
+ r \frac{\partial H}{\partial x}\,,
\label{jam-equationH}
\end{equation}
$\tau\in(0, T), x\in \R$ (see \cite{JS}).  Henceforth, we will refer to (\ref{jam-equationH}) as a {\it $\Gamma$ equation}. A solution $H$ to (\ref{jam-equationH}) is subjected to the initial condition at 
$\tau=0$: \begin{equation}
H(x,0) = \bar H(x)\,, \quad x\in\R\,,
\label{jam-equationHinit}
\end{equation}
where  $\bar H(x)$ is the Dirac $\delta$ function $\bar H(x)=\delta(x)$.  For the purpose of numerical approximation we approximate the initial Dirac delta function  by $\bar H(x)=N^\prime(d)/(\sigma\sqrt{\tau^*})$ where $\tau^*>0$ is sufficiently small, $N(d)$ is the cumulative distribution function of the normal distribution, and $d=\left(x+ (r-q-\sigma^2/2)\tau^* \right/\sigma\sqrt{\tau^*})$.  It corresponds to the value $H= S\partial^2_S V$ of a call (put) option valued by a linear Black--Scholes equation with a constant volatility $\sigma>0$ at the time $T-\tau^*$ close to expiry $T$ when the time parameter $0<\tau^*\ll 1$ is sufficiently small. In the case of call or put options the function $H$ is  subjected to boundary conditions at $x=\pm\infty$, 
\begin{equation}
H(-\infty, \tau) = H(\infty, \tau) = 0\,, \quad \tau\in(0,T)\,.
\label{jam-equationHbound}
\end{equation}
A numerical discretization scheme  of the $\Gamma$ equation (\ref{jam-equationH}) based on finite volume approximation has been discussed in \cite{JS} in more details. 

\subsection{Pricing of European style of options by the RAPM model}

\begin{figure}
\begin{center}
\includegraphics[width=0.45\textwidth]{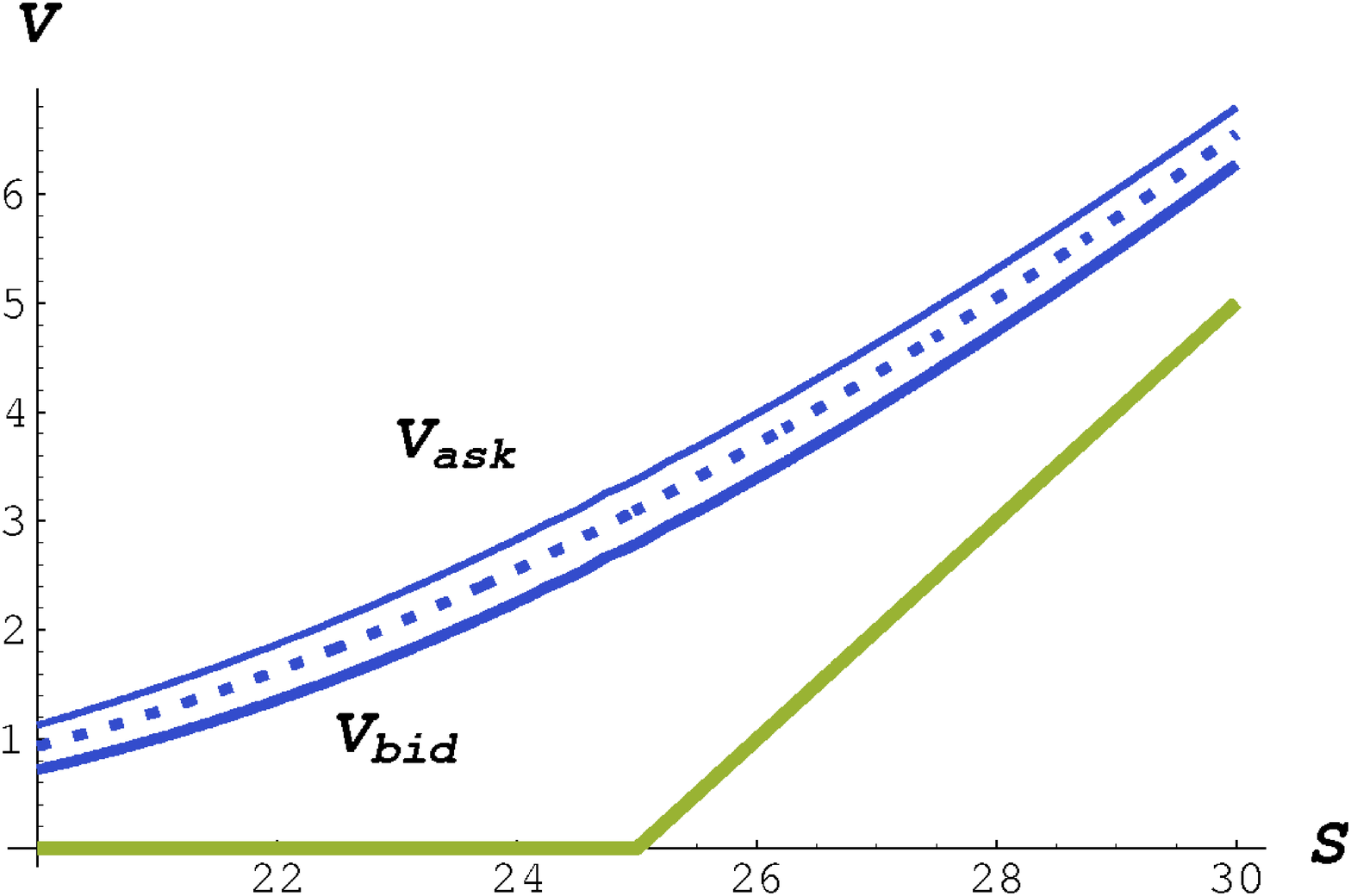}
\hskip 0.5truecm
\includegraphics[width=0.45\textwidth]{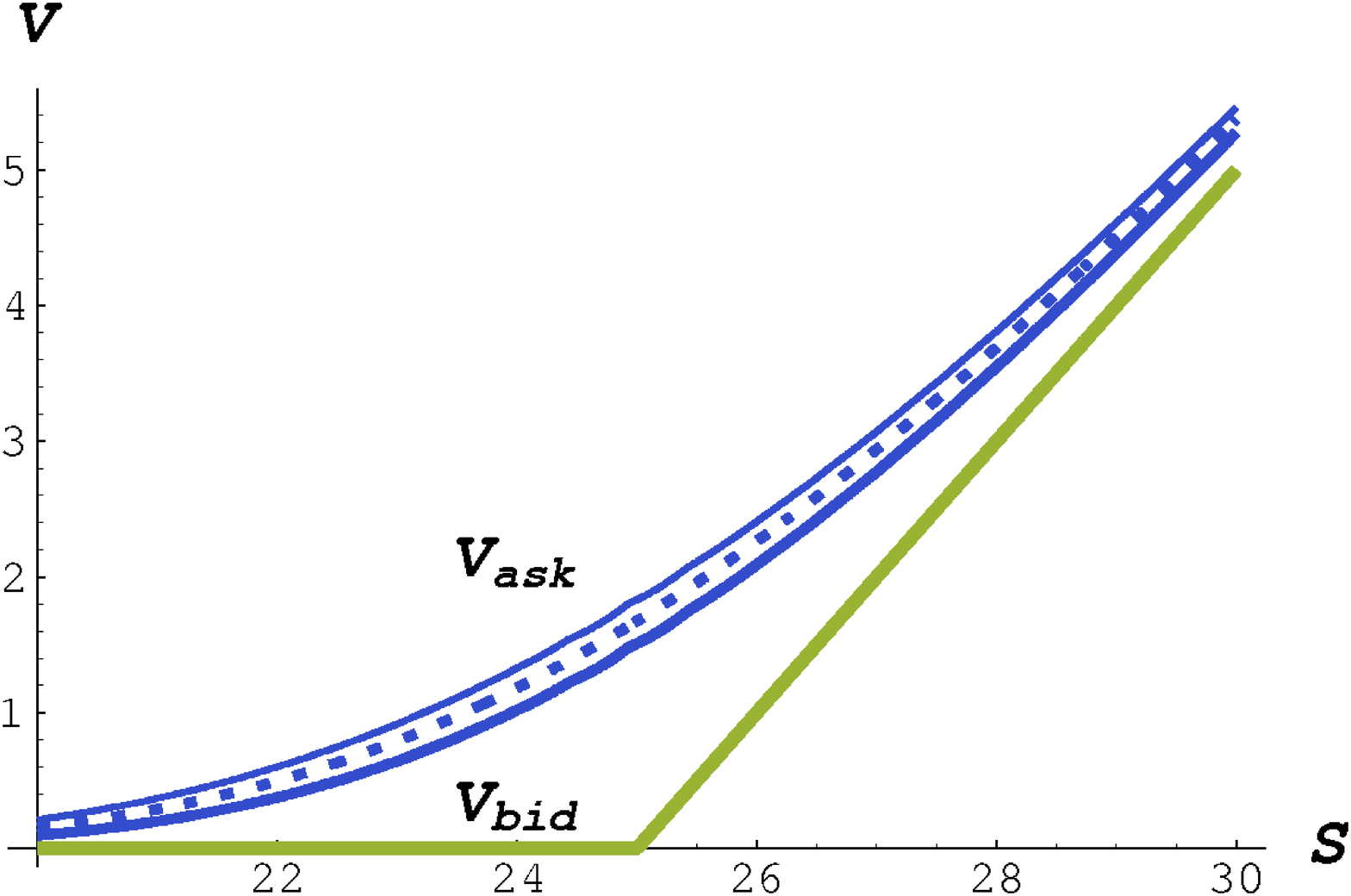}
\end{center}

\caption{\small
A comparison of Bid and Ask option prices computed by means of the RAPM model. 
The middle dotted line is the option price computed from the Black-Scholes
equation. We chose $\sigma=0.3, \mu=0.2, r=0.011, E=25$ and $T=1$ (left) and
$T=0.3$ (right).
}
\label{jam-fig:vbidask}
\end{figure}

Let us denote $V(S,t; C, \hat\sigma, R)$ the value of a solution to
(\ref{afm-generalizedBS}) with parameters $C, \hat\sigma, R$. Suppose that the coefficient of transaction costs $C$ is known from and is given by (\ref{afm-Ccoeff}). In real option market data we can observe different Bid and Ask prices for an option, $V_{bid}<V_{ask}$, resp.  Let us denote by $V_{mid}$ the mid  value, i.e. $V_{mid}= \onehalf(V_{bid}+V_{ask})$. By the RAPM model we are able to explain such a Bid-Ask spread in option prices.  The higher Ask price corresponds to a solution to the RAPM model with some nontrivial risk premium $R>0$ and $C>0$ whereas the mid value $V_{mid}$ corresponds to a solution $V(S,t)$ for  vanishing risk premium $R=0$, i.e. to a solution of the linear Black-Scholes equation (\ref{c-BS}). An illustrative example of Bid-Ask spreads captured by the RAPM model is shown in Fig.~\ref{jam-fig:vbidask}.

To calibrate the RAPM model we  seek for a couple $(\hat\sigma_{RAPM}, R)$ 
of implied RAPM volatility $\sigma_{RAPM}$ and risk premium $R$ such that $V_{ask}=V(S,t; C, \hat\sigma_{RAPM}, R)$ and $V_{mid}=V(S,t; C, \hat\sigma_{RAPM}, 0)$. Such a system of nonlinear equation for $\sigma_{RAPM}$ and $R$ can be solved by by means of the Newton-Kantorovich iterative method (cf. \cite{JS}). 

\begin{figure}
\begin{center}
\includegraphics[width=0.45\textwidth]{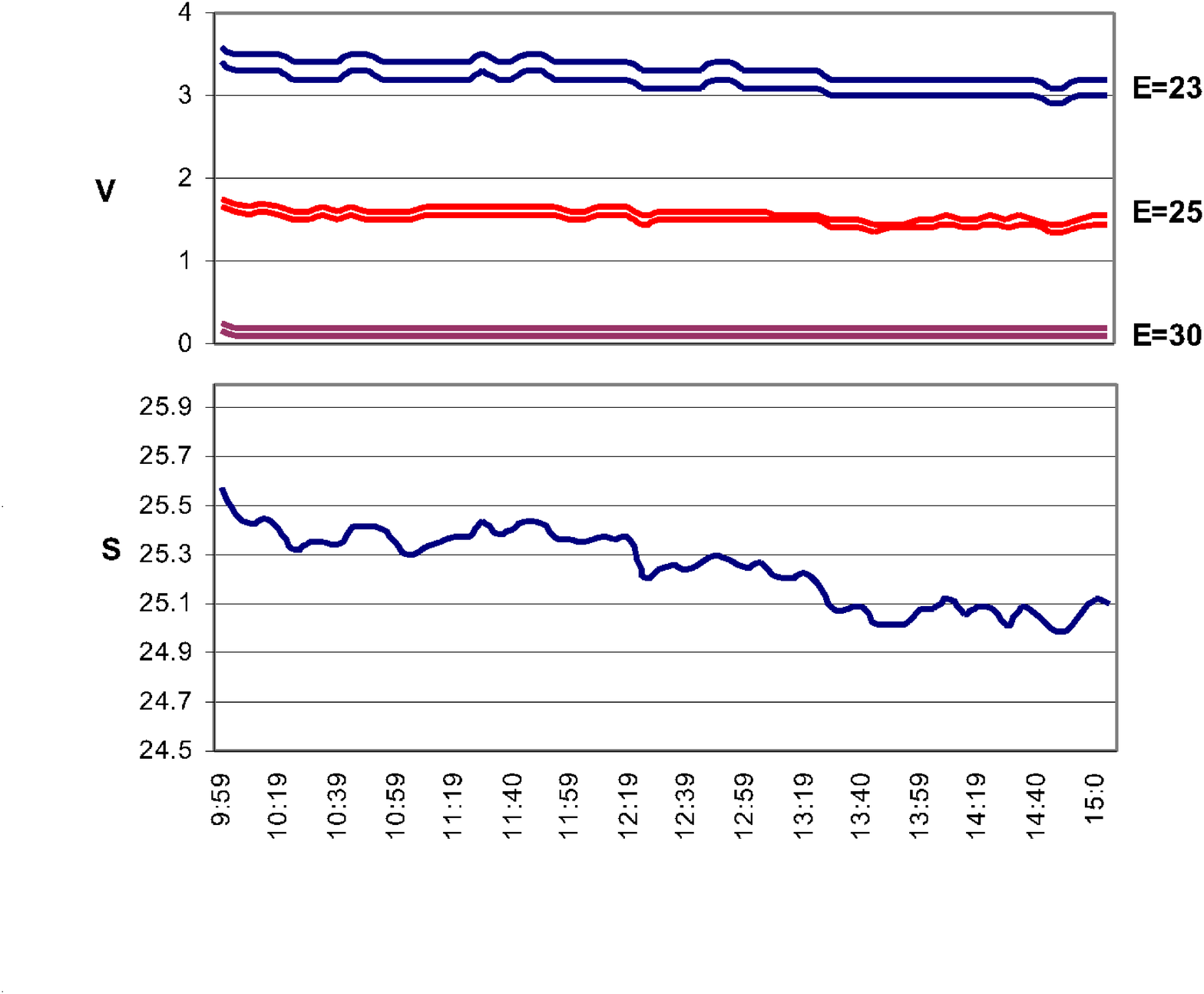}
\hglue 1truecm
\includegraphics[width=0.45\textwidth]{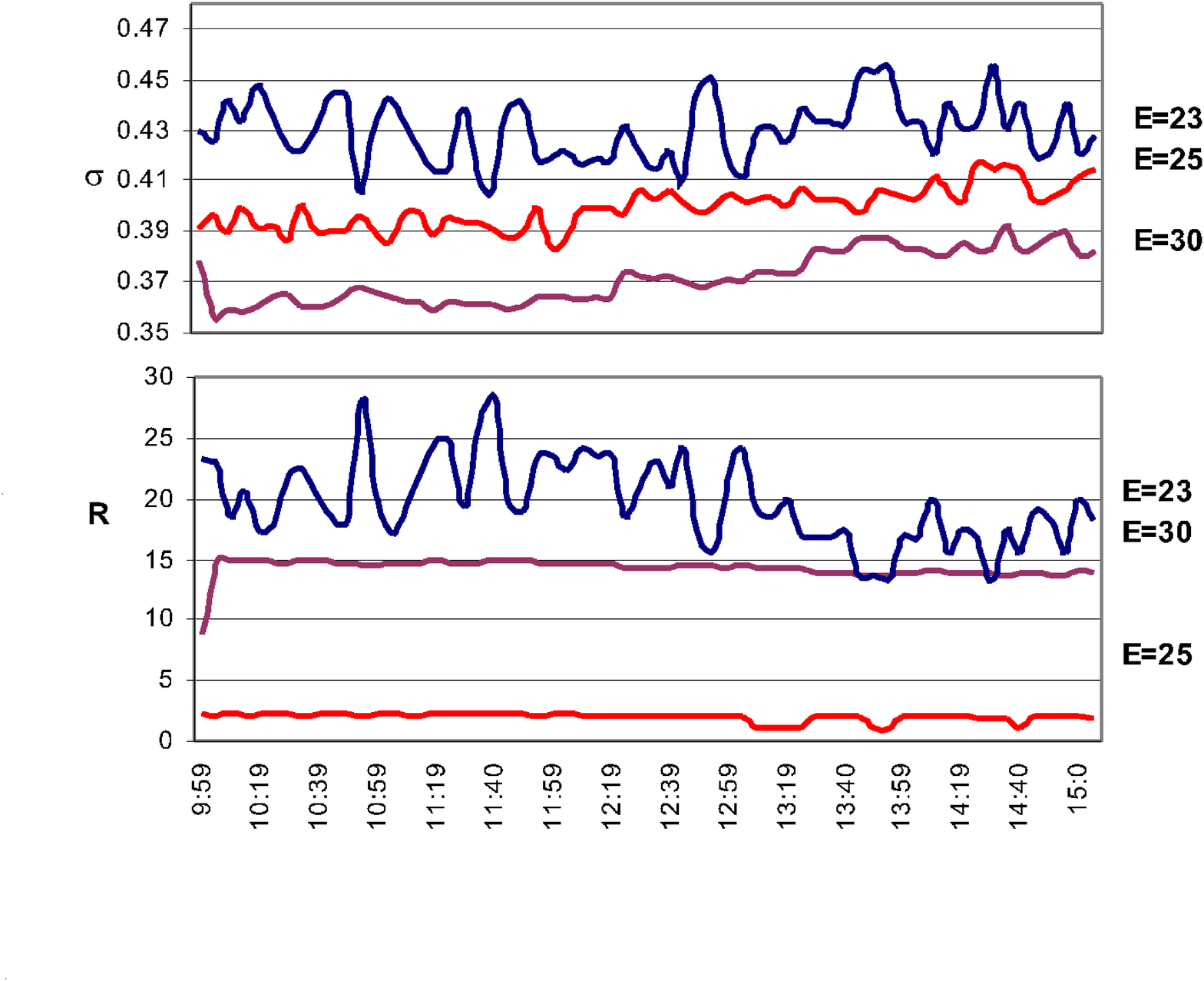}
\end{center}

\caption{\small
Intra-day behavior of Microsoft stocks (April 4, 2003) and shortly expiring
call options with expiry date April 19, 2003.
Computed  implied volatilities $\hat\sigma_{RAPM}$
and risk premium coefficients $R$.
}
\label{afm-fig:msft1}
\end{figure}

As an example we considered sample data sets for call options on  Microsoft stocks.
We considered a flat interest rate $r=0.02$, a constant transaction cost
coefficient $C=0.01$ estimated from (\ref{afm-Ccoeff}), and
we assumed that the underlying asset pays no dividends, i.e. $q=0$. 
In Fig.~\ref{afm-fig:msft1} we present results of calibration of implied couple
$(\hat\sigma_{RAPM}, R)$. Interestingly enough, two call options with higher strike prices $E=25, 30$ had almost constant implied risk premium $R$. On the other the risk premium  of an option with lowest $E=23$ was fluctuating and it had highest average of $R$.

\begin{figure}

\begin{center}
\includegraphics[width=0.45\textwidth]{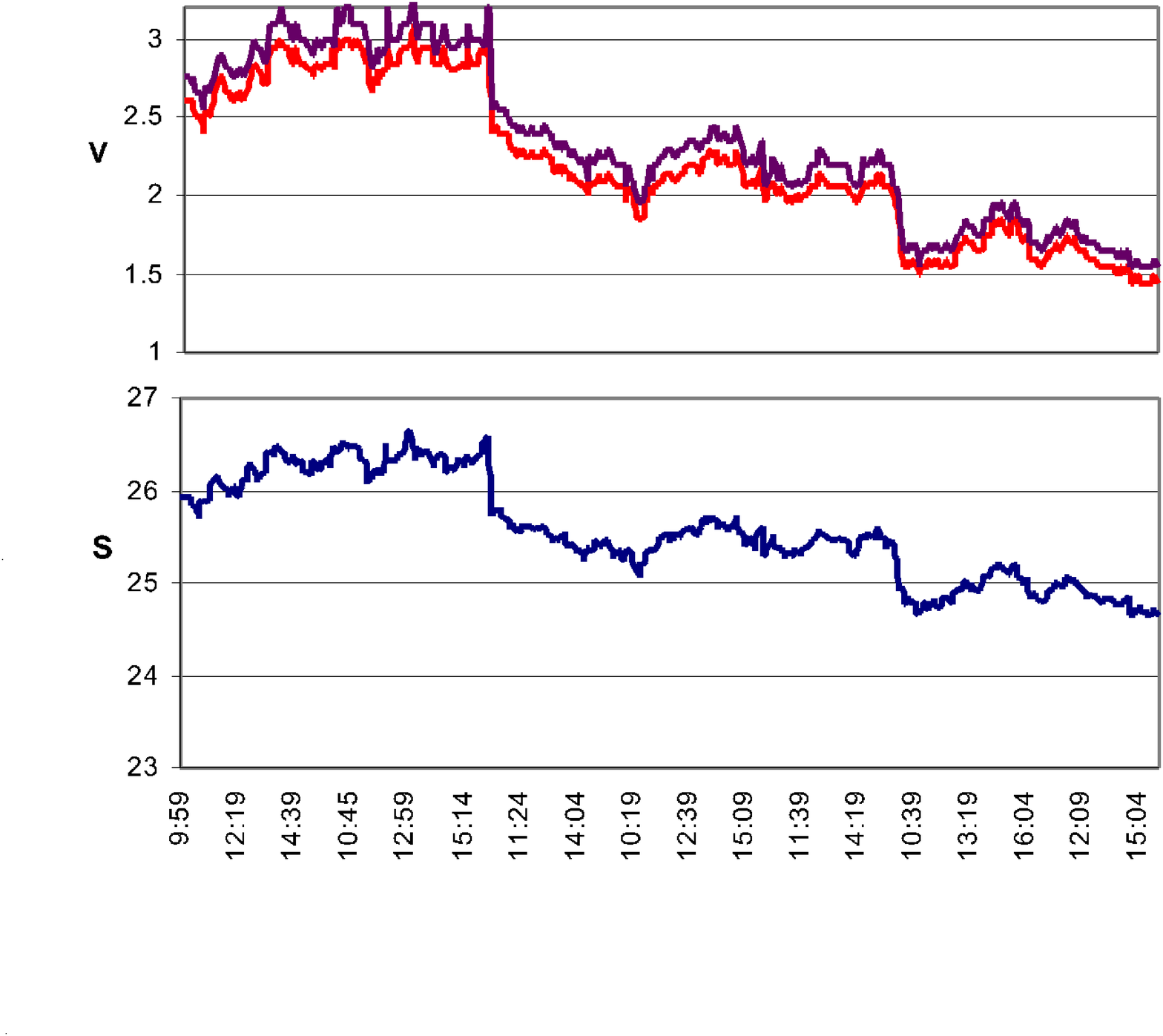}
\includegraphics[width=0.45\textwidth]{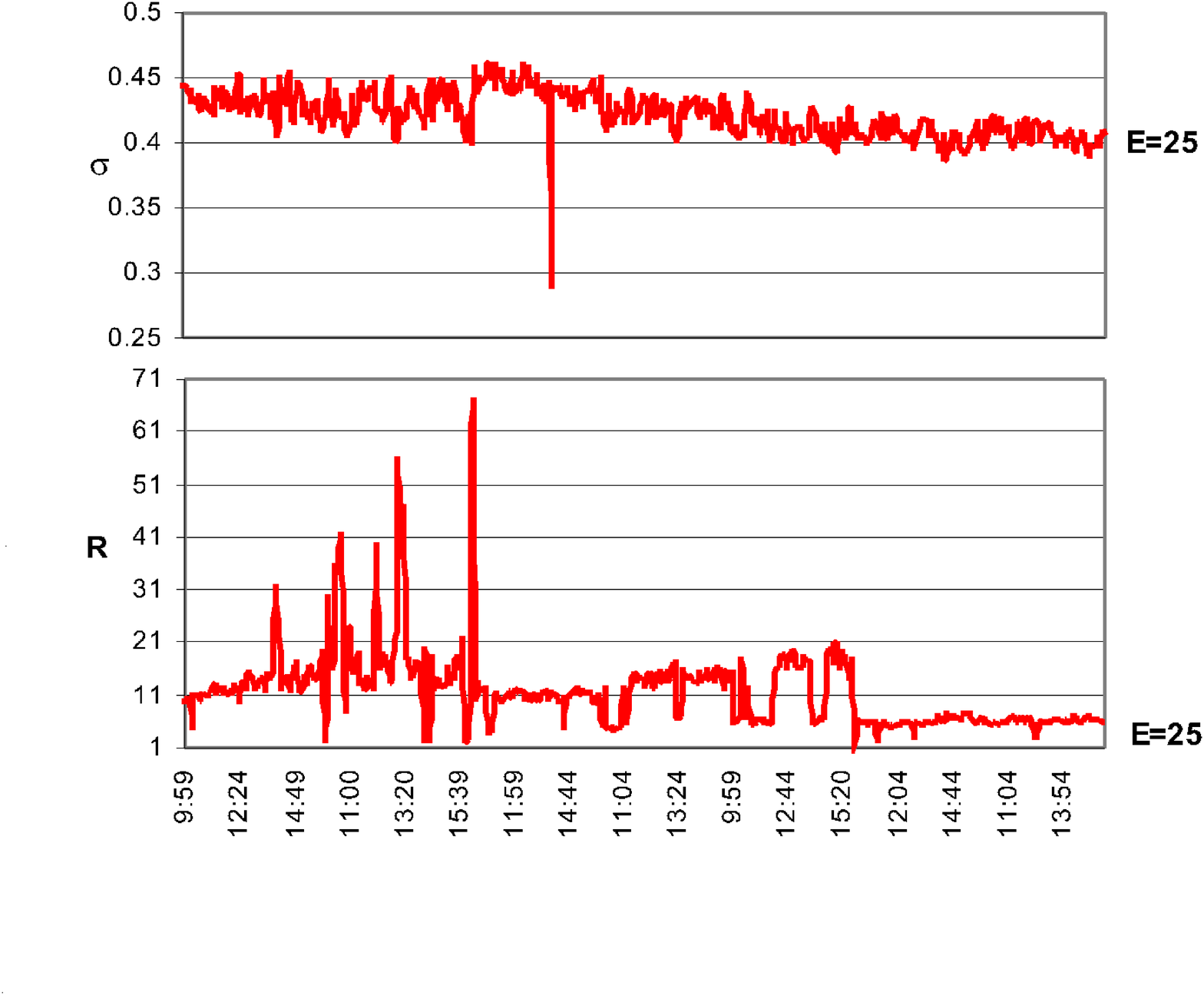}
\end{center}

\caption{\small
One week  behavior of Microsoft stocks (March 20 - 27, 2003) and
call options with expiration date April 19, 2003.
Computed  implied volatilities $\hat\sigma_{RAPM}$
and risk premiums $R$.
}
\label{afm-fig:msft2}
\end{figure}

Finally, in Fig.~\ref{afm-fig:msft2} we present one week behavior of
implied volatilities and risk premium
coefficients for the Microsoft call option on $E=25$  expiring at
$T=$ April 19, 2003. In the beginning of the investigated period the risk
premium coefficient $R$ was rather high and fluctuating. On the other hand,
it tends to a flat value of $R\approx 5$ at the end of the week. Interesting
feature can be observed at the end of the second day when both stock and option
prices went suddenly down. The time series analysis of the 
implied volatility $\hat\sigma_{RAPM}$ from first two days was unable to predict
such a behavior. On the other, high fluctuation in the implied risk premium $R$
during first two days can send a signal to an investor that sudden changes
can be expected in the near future.

\section{Transformation method for a linear Black--Scholes equation}

One of the interesting problems in this field is the analysis
of the early exercise boundary and the optimal stopping time for American
options on stocks paying a continuous dividend. It can be easily  reduced
to a problem of solving a certain free boundary problem for the Black--Scholes
equation (cf. Black \& Scholes~\cite{BS}). However, the exact analytical expression 
for the free boundary profile is not known yet. Many authors have investigated various approximate
models leading to approximate expressions for valuing American call and
put options (see e.g. \cite{GJ, GR, J, K1, KK, M, R} and recent papers by Zhu \cite{Zhu1,Zhu2}, Alobaidi \emph{et al.}\cite{A,Ma,MA}, Stamicar \emph{et al.} \cite{SSC} and the survey paper by Chadam \cite{Ch} and other references therein). For the purpose of studying the free boundary profile
near expiry, many different integral equations have been derived
\cite{BW, CJM, H, Mac}. 

Let us recall that the equation governing the time evolution of the price
$V(S,t)$ of the American call option is the following parabolic PDE:
\begin{eqnarray}
\label{euro-1.1}
&&\d{V}{t} +  (r-q) S\d{V}{S} + {\sigma^2\over 2} S^2 \dd{V}{S} - r V =0\,,
\qquad  0<S<S_f(t),\ 0<t<T\,,
\nonumber
\\
&&V(0,t)=0,\ V(S_f(t), t)= S_f(t)-E\,, \ \d{V}{S}(S_f(t),t)=1\,,
\\
&&V(S,T)=\max(S-E, 0)\,,
\nonumber
\end{eqnarray}
defined on a time-dependent domain $S\in(0,S_f(t) )$, where $t\in(0,T)$.
As usual  $S>0$ is the stock price, $E>0$ is the exercise price,
$r>0$ is the risk-free rate, $q>0$ is the continuous stock dividend rate
and \[
\sigma\equiv const >0
\] 
is a constant volatility of the underlying stock process.

The main purpose of this section is to  present an alternative integral equation which will provide
an accurate numerical method for calculating the early exercise boundary
near expiry. The derivation of the nonlinear integral equation is based on
the Fourier transform. A solution to this integral
equation is the free boundary profile. The novelty of this approach consists
in three steps:
\begin{enumerate}
\item The fixed domain transformation.
\item Derivation of a parabolic PDE for the so-called synthetic
portfolio with a nonlinear nonlocal constraint between the free boundary position and the solution of this parabolic equation itself.
\item Construction of a solution by means of Fourier sine and cosine integral
transforms.
\end{enumerate}

Throughout this section we restrict our attention to the case when $r>q>0$. It is well
known that, for  $r>q>0$, the free boundary  $\varrho(\tau)=S_f(T-\tau)$
starts at $\varrho(0)=r E/q$, whereas  $\varrho(0)= E$ for the case $r\le q$
(cf. Dewynne {\em et al}.~\cite{DH}, Kwok \cite{Kw}). Thus, the free boundary profile develops an initial jump
in the case  $r>q>0$. Notice that the case $0<r\le q$ can be also treated by
other methods based on integral equations. Kwok~\cite{Kw} derived
another integral equation which covers both cases $0<r\le q$, as well as
$r>q>0$ (see equation (\ref{euro-3.13}) and  Remark~\ref{euro-rem-3.1}). However, in the latter case equation
(\ref{euro-3.13}) becomes  singular as $t\to T^-$, leading to numerical instabilities
near expiry.

In the rest of this section to investigate the behavior of the free
boundary $S_f(t)$. We present a method developed by \v{S}ev\v{c}ovi\v{c} in \cite{Se} of reducing the free boundary problem for (\ref{euro-1.1}) to a nonlinear integral equation with a singular kernel. Notice that our method of reducing the free boundary problem to a nonlinear integral equation can be also successfully adopted for valuing the American put option paying no dividends (cf. Stamicar {\em et al}.~\cite{SSC}).

\subsection{Fixed domain transformation of the free boundary problem}
In this section we will perform a fixed domain transformation of the
free boundary problem (\ref{euro-1.1}) into a parabolic equation defined on
a fixed spatial domain. As it will be shown below, imposing of the
free boundary condition will result in a nonlinear time-dependent term
involved in the resulting equation. To transform equation (\ref{euro-1.1}) defined on a time dependent spatial domain $(0, S_f(t))$, we  introduce the following change of variables:
\begin{equation}
\tau= T-t, \quad x=\ln\left({\ro(\tau)\over S}\right)
\qquad \hbox{where}\ \ \ro(\tau)=S_f(T-\tau).
\label{euro-2.1}
\end{equation}
Clearly, $\tau\in(0,T)$ and $x\in(0,\infty)$ whenever $S\in(0,S_f(t))$.
Let us furthermore define the auxiliary function
$\Pi=\Pi(x,\tau)$ as follows:
\begin{equation}
\Pi(x,\tau)= V(S,t) - S \d{V}{S}(S,t).
\label{euro-2.2}
\end{equation}
Notice that the quantity $\Pi$ has an important financial meaning as it
is a synthetic portfolio consisting of one long option and $\Delta= \d{V}{S}$ underlying short positioned stocks. It follows from (\ref{euro-2.1}) that
\begin{eqnarray}
\label{euro-2.3}
&& \d{\Pi}{x}= S^2\dd{V}{S}, \ \ \  \dd{\Pi}{x} + 2 \d{\Pi}{x}=
-S^3 {\partial^3 V\over\partial S^3},
\nonumber
\\
&& \d{\Pi}{\tau}+ {\dot\ro \over\ro}\d{\Pi}{x} =
S {\partial^2V\over\partial S\partial t} - \d{V}{t},
\end{eqnarray}
where $\dot\ro=d\ro/d\tau$. Now assuming that $V=V(S,t)$ is a sufficiently smooth solution of (\ref{euro-1.1}), we may differentiate (\ref{euro-1.1}) with respect to $S$. Plugging expressions (\ref{euro-2.3}) into (\ref{euro-1.1}), we finally obtain that the function $\Pi=\Pi(x,\tau)$ is a solution of the parabolic equation
\begin{equation}
\d{\Pi}{\tau} + a(\tau)\d{\Pi}{x} - {\sigma^2\over 2} \dd{\Pi}{x} + r \Pi=0,
\label{euro-2.4}
\end{equation}
$x\in(0,\infty),\ \tau\in(0,T),$ with a time-dependent coefficient
\[
a(\tau)= {\dot\ro(\tau)\over\ro(\tau)} + r -q - {\sigma^2\over 2} .
\]
It follows from the boundary condition $V(S_f(t), t)= S_f(t) - E$ and
$V_S(S_f(t), t)= 1$ that
\begin{equation}
\Pi(0, \tau) = -E, \qquad \Pi(+\infty, \tau) =0\ .
\label{euro-2.5}
\end{equation}
The initial condition $\Pi(x,0)$ can be deduced from the pay-off
diagram for $V(S,T)$. We obtain
\begin{equation}
\Pi(x,0)=
\left\{
   \begin{array}{lll}
-E & \hbox{for} \ x<\ln\left({\ro(0)\over E}\right)\hfil
\\
0 & \hbox{otherwise.}\hfil
\end{array}
\right.
\label{euro-2.6}
\end{equation}
Notice that equation (\ref{euro-2.4}) is a parabolic PDE with a time-dependent
coefficient $a(\tau)$. In what follows, we will show
the function $a(\tau)$ depends upon a solution $\Pi$ itself.
This dependence is non-local in the spatial variable $x$. Moreover,
the initial position of the interface $\ro(0)$ enters the initial
condition $\Pi(x,0)$. Therefore we have to determine the relationship
between the solution $\Pi(x,\tau)$ and the free boundary function
$\ro(\tau)$ first. To this end, we make use of the boundary condition
imposed on $V$ at the interface $S=S_f(t)$. Since
$S_f(t)-E=V(S_f(t),t)$ we have
\[
{d\over dt} S_f(t)
= \d{V}{S}(S_f(t),t) {d\over dt} S_f(t) + \d{V}{t}(S_f(t), t)\,.
\]
As $\d{V}{S}(S_f(t),t)=1$ we obtain $\d{V}{t}(S,t)=0$ at $S=S_f(t)$.
Assuming the function  $\Pi_x$ has a trace at $x=0$, and
taking into account (\ref{euro-2.3}), we may conclude that, for any $t=T-\tau\in[0,T)$,
\[
S^2\dd{V}{S}(S,t) \to  \d{\Pi}{x}(0,\tau), \quad
S\d{V}{S}(S,t)\to \ro(\tau)\ \hbox{as} \ S\to S_f(t)^-\ .
\]
If $\d{V}{t}(S, t) \to \d{V}{t}(S_f(t), t)=0$ as $S\to S_f(t)^-$, then
it follows from the Black--Scholes equation (\ref{euro-1.1}) that
\[
(r-q)\ro(\tau) +{\sigma^2\over 2} \d{\Pi}{x}(0,\tau) -r(\ro(\tau)-E)
\]
\[
= \lim_{S\to S_f(t)^-}
    \left(
    \d{V}{t}(S,t) +  (r-q) S\d{V}{S}(S,t) + {\sigma^2\over 2} S^2
        \dd{V}{S}(S,t) - r V(S,t)
    \right) = 0.
\]
As a consequence we obtain a nonlocal algebraic constraint between the free boundary function $\ro(\tau)$ and the boundary trace $\partial_x\Pi (0,\tau)$ of the derivative of the solution $\Pi$ itself:
\begin{equation}
\ro(\tau)= {r E\over q} + {\sigma^2\over 2q}\d{\Pi}{x}(0,\tau)
\ \hbox{for} \  0<\tau\le T.
\label{euro-2.7}
\end{equation}
It remains to determine the initial position of the interface $\ro(0)$.
According to Dewynne {\em et al}.~\cite{DH} (see also Kwok \cite{Kw}), the initial position
$\ro(0)$ of the free boundary is $rE/q$ for the case $0<q<r$. Alternatively, we
can derive this condition from (\ref{euro-2.4})--(\ref{euro-2.6}) by assuming the smoothness of $\Pi$
in the $x$ variable up to the boundary $x=0$ uniformly for
$\tau\to 0^+$. More precisely, we assume that
\[
 \lim_{\tau\to0^+}\d{\Pi}{x}(0,\tau)
=\lim_{\tau\to0^+,x\to0^+}\d{\Pi}{x}(x,\tau)
=\lim_{x\to0^+}\d{\Pi}{x}(x,0) = 0
\]
because $\Pi(x,0)=-E$ for $x$ close to $0^+$. By (\ref{euro-2.7}) we obtain
\begin{equation}
\ro(0)= {rE\over q}.
\label{euro-2.8}
\end{equation}
In summary, we have shown that, under suitable regularity assumptions
imposed on a solution $\Pi$ to
(\ref{euro-2.4}), (\ref{euro-2.5}), (\ref{euro-2.6}),
the free boundary problem (\ref{euro-1.1})
can be transformed into the initial boundary value problem for parabolic PDE
\begin{eqnarray}
\label{euro-2.9}
&&\d{\Pi}{\tau} + a(\tau)\d{\Pi}{x} - {\sigma^2\over 2} \dd{\Pi}{x} + r \Pi=0,
\nonumber
\\
&&\Pi(0, \tau) = -E, \qquad \Pi(+\infty, \tau) =0, \qquad x>0,
\tau\in(0,T),
\\
&&\Pi(x,0)= \left\{
\begin{array}{lll}
-E & \hbox{for} \ x<\ln\left({r/q}\right),\hfil
\\
 0 & \hbox{otherwise},\hfil
\end{array}
\right.
\nonumber
\end{eqnarray}
where
$a(\tau)= {\dot\ro(\tau)\over\ro(\tau)} + r -q - {\sigma^2\over 2}$ and
\begin{equation}
\ro(\tau)={r E\over q} + {\sigma^2\over 2q}\d{\Pi}{x}(0,\tau), \ \
\ro(0)={rE\over q}\ .
\label{euro-2.10}
\end{equation}
We emphasize that the problem (\ref{euro-2.9}) constitutes a nonlinear parabolic equation with a nonlocal constraint given by (\ref{euro-2.10}).

\medskip
\begin{remark}\label{euro-rem-2.1}
In our derivation of the free boundary
function $\ro(\tau)$  and its initial condition $\ro(0)$ we did not assume that the solution $V(S,t)$ is $C^2$ smooth up to the free boundary $S=S_f(t)$. Such an assumption  would lead to an obvious contradiction $\Gamma:=\partial^2 V/\partial S^2 = 0$ at $S=S_f(t)$. On the other hand, the jump in $\Gamma$ at the free boundary is the only driving force for the evolution of the free boundary function $\ro$ (see (\ref{euro-2.7})). Construction of a PDE for the synthetic portfolio function $\Pi$ is a crucial step in our approach because
the derivative $\partial_x\Pi$ admits a trace at the boundary $x=0$ and the
unknown free boundary function $\varrho$ can be determined via
(\ref{euro-2.10})
\end{remark}

\subsection{Reduction to a nonlinear integral equation}

The main purpose of this section is to show how the fully nonlinear nonlocal
problem (\ref{euro-2.9})--(\ref{euro-2.10}) can reduced to a single nonlinear integral equation for
$\ro(\tau)$ giving the explicit formula for the solution $\Pi(x,\tau)$ to
(\ref{euro-2.9}). The idea is to apply the Fourier sine and cosine integral
transforms (cf. Stein \& Weiss~\cite{SW1}).
Let us recall that for any Lebesgue integrable function $f\in L^1(\R^+)$ the sine and cosine transformations are defined as follows:
${\cal F}_S(f) (\omega) = \int_0^\infty f(x)\,\sin\omega x\, dx$,
${\cal F}_C(f) (\omega) = \int_0^\infty f(x)\,\cos\omega x\, dx$.
Their inverse transforms are given by
${\cal F}^{-1}_S(g) (x) = {2\over\pi}\int_0^\infty g(\omega)\,\sin\omega x\, d\omega,$
${\cal F}^{-1}_C(g) (x) = {2\over\pi}\int_0^\infty g(\omega)\,\cos\omega x\, d\omega$. 
Now we suppose that the function $\ro(\tau)$ and subsequently $a(\tau)$ are
already know. Let $\Pi=\Pi(x,\tau)$ be a solution of (\ref{euro-2.9}) corresponding to a given function $a(\tau)$. Let us denote
\begin{equation}
p(\omega,\tau) = {\cal F}_S( \Pi(.,\tau))(\omega),\qquad
q(\omega,\tau) = {\cal F}_C( \Pi(.,\tau))(\omega)
\label{euro-3.2}
\end{equation}
where $\omega\in \R^+,\tau\in(0,T)$. By applying the sine and cosine integral
transforms, taking into account their basic properties and (\ref{euro-2.10}),
we finally obtain a linear non-autonomous $\omega$-parameterized system of ODEs
\begin{eqnarray}
\label{euro-3.3}
&& {d\over d\tau} p(\omega,\tau)  - a(\tau)\omega q(\omega,\tau)
+ \alpha(\omega) p(\omega,\tau) = -E\omega {\sigma^2\over 2},
\nonumber
\\
&& {d\over d\tau} q(\omega,\tau)  + a(\tau)\omega p(\omega,\tau)
+ \alpha(\omega) q(\omega,\tau) = -E a(\tau) -q\ro(\tau) +r E,
\end{eqnarray}
for the sine and cosine transforms of $\Pi$ (see (\ref{euro-3.2})) where
\[
\alpha(\omega)={1\over 2}(\sigma^2\omega^2 + 2 r).
\]
The system of equations (\ref{euro-3.3}) is subject to initial conditions at $\tau=0$,
$p(\omega,0)={\cal F}_S( \Pi(.,0)(\omega)$, \
$q(\omega,0)={\cal F}_C( \Pi(.,0)(\omega)$. In the case of the call option,
we deduce from the initial condition for $\Pi$ (see (\ref{euro-2.9})) that
\begin{equation}
p(\omega,0)={E\over\omega} \left( \cos\left(\omega\ln{r\over q}\right) -1 \right),
\qquad q(\omega,0)=-{E\over\omega} \sin\left(\omega\ln{r\over q}\right).
\label{euro-3.4}
\end{equation}
Taking into account (\ref{euro-3.4}) and by using the variation of constants formula for solving linear
non-autonomous ODEs, we obtain an explicit formula for
$p(\omega,\tau) = -E \omega^{-1} +\tilde p(\omega,\tau)$ where
\[
\tilde p(\omega,\tau) =
{E\over \omega} e^{-\alpha(\omega)\tau} \cos(\omega(A(\tau,0)+\ln(r/q)))
\]
\begin{equation}
+\int_0^\tau {\rm e}^{-\alpha(\omega)(\tau-s)}
            \left[{r E\over \omega} \cos(\omega A(\tau,s))
                  + (r E - q \ro(s)) \sin(\omega A(\tau,s))
            \right]ds.
\label{euro-3.5}
\end{equation}
Here we have denoted by $A$ the function defined as
\begin{equation}
A(\tau,s) = \int_s^\tau a(\xi)\,d\xi = \ln{\ro(\tau)\over\ro(s)}
+\left(r-q-{\sigma^2\over 2}\right)(\tau-s)\ .
\label{euro-3.6}
\end{equation}
As ${\cal F}^{-1}_S(\omega^{-1}) =1$ we have
$\Pi(x,\tau)= {\cal F}^{-1}_S(p(\omega,\tau)) =
-E + {2\over\pi} \int_0^\infty \tilde p(\omega,\tau)\sin(\omega
x)\,d\omega$. From (\ref{euro-2.10}) we conclude that the free boundary function
$\varrho$ satisfies the following equation:
\begin{equation}
\ro(\tau) = {r E\over q} +{\sigma^2\over q \pi}\int_0^\infty \omega\tilde
p(\omega,\tau)\, d\omega .
\label{euro-3.7}
\end{equation}
Taking into account (\ref{euro-3.5}) and
(\ref{euro-3.7}) we end up
with the following nonlinear singular integral equation for the free boundary
function $\ro(\tau)$:
\begin{eqnarray}
\label{euro-3.8}
\ro(\tau) &=& {r E\over q} \biggl( 1+ {\sigma\over r\sqrt{2\pi\tau}}
\ \exp\left(-r\tau -{(A(\tau,0)+\ln(r/q))^2\over 2\sigma^2\tau}\right)
\\
&& + \ {1\over\sqrt{2\pi}}\int_0^\tau
\biggl[\sigma+{1\over\sigma}\biggl(1-{q\ro(s) \over r E} \biggr)
{A(\tau,s)\over\tau-s}
\biggr]
{\exp\left(-r(\tau-s) -{A(\tau,s)^2\over 2\sigma^2(\tau-s)}\right)\over\sqrt{\tau-s}}
\, ds \biggr),
\nonumber
\end{eqnarray}
where the function $A$ depends upon $\ro$ via equation (\ref{euro-3.6}). To
simplify this integral equation, we introduce a new auxiliary function
$H:[0,\sqrt{T}]\to \R$ as follows:
\begin{equation}
\ro(\tau)= {r E\over q} \left( 1 + \sigma\sqrt{2} H(\sqrt\tau)\right).
\label{euro-3.9}
\end{equation}
Using the change of variables $s=\xi^2\cos^2\theta$, one can rewrite the
integral equation (\ref{euro-3.7}) in terms of the function $H$ as follows:
\begin{equation}
H(\xi) = f_H(\xi)  + {1\over\sqrt{\pi} }
\int_0^{\pi\over2}
    \left[
    \xi\cos\theta - 2\hbox{cotg}\,\theta\  H(\xi\cos\theta) g_H(\xi,\theta)
    \right]
    {\rm e}^{- r\xi^2\sin^2\theta - g_H^2(\xi,\theta) } \, d\theta ,
\label{euro-3.10}
\end{equation}
where
\begin{equation}
g_H(\xi,\theta)= {1\over\sigma\sqrt{2}}
{1\over\xi\sin\theta}
        \ln
        \left(
        {
        1+\sigma\sqrt{2} H(\xi)
        \over
        1+\sigma\sqrt{2} H(\xi\cos\theta)
        }
        \right)
+\ {\Lambda\over\sqrt{2}} \xi\sin\theta
\label{euro-3.11}
\end{equation}
for $\xi\in[0,\sqrt T],\ \theta\in(0,\pi/2)$,
\[
\Lambda= {r-q\over \sigma}  -{\sigma\over 2}
\]
and
\begin{equation}
f_H(\xi) = {1\over 2 r\sqrt{\pi}\xi }
\ {\rm e}^{- r\xi^2 -
\left(g_H(\xi,{\pi\over 2}) +{1\over \xi} {1\over\sigma\sqrt{2}}\ln(r/q)
\right)^2 }.
\label{euro-3.12}
\end{equation}
Notice that equations (\ref{euro-3.8}) and (\ref{euro-3.10}) are integral equations with a singular
kernel (cf. Gripenberg {\em et al}.~\cite{GLS}).

\medskip
\begin{remark}\label{euro-rem-3.1}
Kwok~\cite{Kw} derived another integral  equation for
the early exercise boundary for the American call option on a stock
paying continuous dividend. According to Kwok~\cite[Section 4.2.3]{Kw},
$\ro(\tau)$ satisfies the integral equation
\begin{eqnarray}
\label{euro-3.13}
\ro(\tau) && = E +\ro(\tau)e^{-q\tau} N(d) -
E {\rm e}^{-r\tau} N(d-\sigma\sqrt{\tau})
\nonumber
\\
&&+\int_0^\tau q \ro(\tau)e^{-q \xi} N(d_\xi)
- r E {\rm e}^{-r \xi} N(d_\xi-\sigma\sqrt{\xi}) d\xi
\end{eqnarray}
where
\[
d = {1\over \sigma\sqrt{\tau}}\ln\left({\ro(\tau)\over E}\right)
+\Lambda \sqrt{2\tau}, \quad
d_\xi = {1\over \sigma\sqrt{\xi}}\ln\left({\ro(\tau)\over \ro(\tau-\xi)}\right)
+\Lambda \sqrt{2\xi}
\]
and $N(u)$ is the cumulative distribution function for the normal
distribution. The above integral equation covers both cases:
$r\le q$ as well as $r>q$. However, in the case $r>q$ this  equation becomes
singular as $\tau\to 0^+$. 
\end{remark}

\medskip

In the rest of this section we  derive a formula for pricing American
call options based on the solution $\ro$ to the integral equation
(\ref{euro-3.10}).
With regard to (\ref{euro-2.2}), we have
\[
\d{}{S} \left( S^{-1} V(S,t) \right) =
- S^{-2} \Pi\left( \ln\left(S^{-1} \ro(T-t)\right), \, T-t \right).
\]
Taking into account the boundary condition $V(S_f(t), t) = S_f(t)-E$ and
integrating the above equation from $S$ to $S_f(t)=\ro(T-t)$, we obtain
\begin{equation}
V(S,T-\tau)= {S\over\ro(\tau)} \left(
\ro(\tau) -E + \int_0^{\ln{\ro(\tau)\over S}} {\rm e}^x \Pi(x,\tau) \, dx
\right).
\label{euro-3.14}
\end{equation}
It is straightforward to verify that $V$ given by (\ref{euro-3.14} is indeed a solution to the free boundary problem (\ref{euro-1.1}). Inserting the expressions (\ref{euro-3.5}) and (\ref{euro-3.7}) (recall that $\Pi(x,\tau)=-E +{2\over\pi}\int_0^\infty\tilde p(\omega,\tau)\sin\omega x\, d\omega$)
into (\ref{euro-3.14}), we end up with the formula for pricing the American call option:
\begin{eqnarray}
\label{euro-3.17}
V(S,T-\tau) &&=  S- E +
{S\over\ro(\tau)} E \ I_2(A(\tau,0)+\ln(r/q), \ln(\ro(\tau)/S), \tau)
\nonumber
\\
&&+ {S\over\ro(\tau)} \int_0^\tau
\biggl[
r E\  I_2(A(\tau,s), \ln(\ro(\tau)/S), \tau-s)
\\
&&+(r E - q\ro(s)) \ I_1(A(\tau,s), \ln(\ro(\tau)/S), \tau-s)
\biggr]\, ds
\nonumber
\end{eqnarray}
for any $S\in(0,S_f(t))$ and $t\in[0,T]$,  where
$A(\tau,s) = \ln{\ro(\tau)\over\ro(s)} +(r-q-{\sigma^2\over 2})(\tau-s)$.
Here
\begin{eqnarray}
\label{euro-3.16}
&& I_1(A,L) = { e^{-(r-\sigma^2/2)\tau} \over 2}
\left[
e^A M\left( {-A -\sigma^2\tau\over \sigma\sqrt{2\tau}}, {L\over
\sigma\sqrt{2\tau} }  \right)
-
e^{-A} M\left( {A -\sigma^2\tau\over \sigma\sqrt{2\tau}}, {L\over
\sigma\sqrt{2\tau} }  \right)
\right],
\nonumber
\\
&& I_2(A,L) = { e^{-r\tau}  e^L \over 2}
M\left( {A -L\over \sigma\sqrt{2\tau}}, {2 L\over
\sigma\sqrt{2\tau} }  \right)
\\
&&\qquad -  {e^{-(r-\sigma^2/2)\tau} \over 2}
\left[
e^A M\left( {-A -\sigma^2\tau\over \sigma\sqrt{2\tau}}, {L\over
\sigma\sqrt{2\tau} }  \right)
+
e^{-A} M\left( {A -\sigma^2\tau\over \sigma\sqrt{2\tau}}, {L\over
\sigma\sqrt{2\tau} }  \right)
\right]
\nonumber
\end{eqnarray}
and
\[
M(x,y) = \hbox{erf}(x+y)-\hbox{erf}(x) = {2\over\sqrt\pi} \int_x^{x+y}
e^{-\xi^2}\,d\xi .
\]
We will refer to (\ref{euro-3.17}) as the semi-explicit formula for pricing the
American call option. We use the term `semi-explicit' because
(\ref{euro-3.17}) contains
the free boundary function $\ro(\tau)=S_f(T-\tau)$ which has to be determined
first by solving the nonlinear integral equation (\ref{euro-3.10}).

\subsection{Numerical experiments}
In this section we focus on  numerical experiments. We will compute the free boundary profile
\begin{equation}
S_f(t)=\ro(T-t) = {r E\over q}\left(1+\sigma\sqrt{2}H(\sqrt{T-t})\right)
\label{euro-4.1}
\end{equation}
(see (\ref{euro-3.4})) by solving the nonlinear integral equation (\ref{euro-3.10}). We also present a comparison of the results obtained by our methods to those obtained by other known methods for solving the American call option problem.

The computation of a solution of the nonlinear integral equation is based on an iterative method. We will construct a sequence of approximate solutions to (\ref{euro-3.10}). Let $H^0$ be an initial approximation of a solution to (\ref{euro-3.10}). If we suppose $H^0(\xi)=h_1 \xi$, then, plugging this ansatz into (\ref{euro-3.10}) yields the well-known first order approximation of a solution $H(\xi)$ in the form
\[
H^0(\xi) = 0.451381\, \xi
\]
i.e. $\ro^0(\tau) = {r E\over q}\left(1+\,0.638349\,\sigma\sqrt{\tau}\right)$
(cf. \v{S}ev\v{c}ovi\v{c} \cite{Se}). This asymptotics is in agreement with that of  Dewynne {\em et al}.~\cite{DH}. For $n=0,1, ... $ we will define the $n+1$ approximation $H^{n+1}$ as follows:
\[
H^{n+1}(\xi) =
\]
\begin{equation}
f_{H^n}(\xi)  + {1\over\sqrt{\pi} }
\int_0^{\pi\over2}
    \left[
    \xi\cos\theta - 2\hbox{cotg}\,\theta\  H^n(\xi\cos\theta) g_{H^n}(\xi,\theta)
    \right]
    {\rm e}^{- r\xi^2\sin^2\theta - g_{H^n}^2(\xi,\theta) } \, d\theta
\label{euro-4.2}
\end{equation}
for $\xi\in[0,\sqrt{T}]$. With regard to (\ref{euro-3.11}) and
(\ref{euro-3.12}), we have
\[
g_{H^n}(\xi,\theta)= {1\over\sigma\sqrt{2}}
{1\over\xi\sin\theta}
        \ln
        \left(
        {
        1+\sigma\sqrt{2} H^n(\xi)
        \over
        1+\sigma\sqrt{2} H^n(\xi\cos\theta)
        }
        \right)
+\ {\Lambda\over\sqrt{2}} \xi\sin\theta
\]
\[
f_{H^n}(\xi)= {1\over 2r\sqrt{\pi}\xi}
\ {\rm e}^{- r\xi^2 -
\left(g_{H^n}(\xi,{\pi\over 2}) +{1\over \xi} {1\over\sigma\sqrt{2}}
\ln\left({r\over q}\right)
\right)^2 }.
\]
Notice that the function $g_H$ is bounded provided that $H$ is nonnegative
and Lipschitz continuous on $[0,\sqrt{T}]$. Recall that we have assumed
$r>q>0$.  Then the function $\xi\mapsto f_{H}(\xi)$ is bounded for
$\xi\in[0,\sqrt{T}]$ and it vanishes at $\xi=0$. Moreover, if $H$ is smooth
then $f_H$ is a flat function at $\xi=0$, i.e. $f_H(\xi)=o(\xi^n)$ as
$\xi\to 0^+$ for all $n\in N$. From the numerical point of view such a flat
function can be omitted from computations. For small values
of $\theta$ we approximate the function
$\hbox{cotg}\,\theta \, g_{H^n}(\xi,\theta)$ by its limit as $\theta\to 0^+$.
It yields the approximation of the singular term in (\ref{euro-4.2}) in the form
\[
\hbox{cotg}\,\theta g_{H}(\xi,\theta)\approx
{1\over2}{H'(\xi)\over 1+\sigma\sqrt{2}H(\xi)} +
{\Lambda\over\sqrt{2}}\xi
\qquad \hbox{for}\ \ 0<\theta \ll 1,
\]
where $H=H^n$. In the following numerical simulations we chose
$\varepsilon\approx 10^{-5}$.

\begin{figure}
\begin{center}
\includegraphics[width=0.45\textwidth]{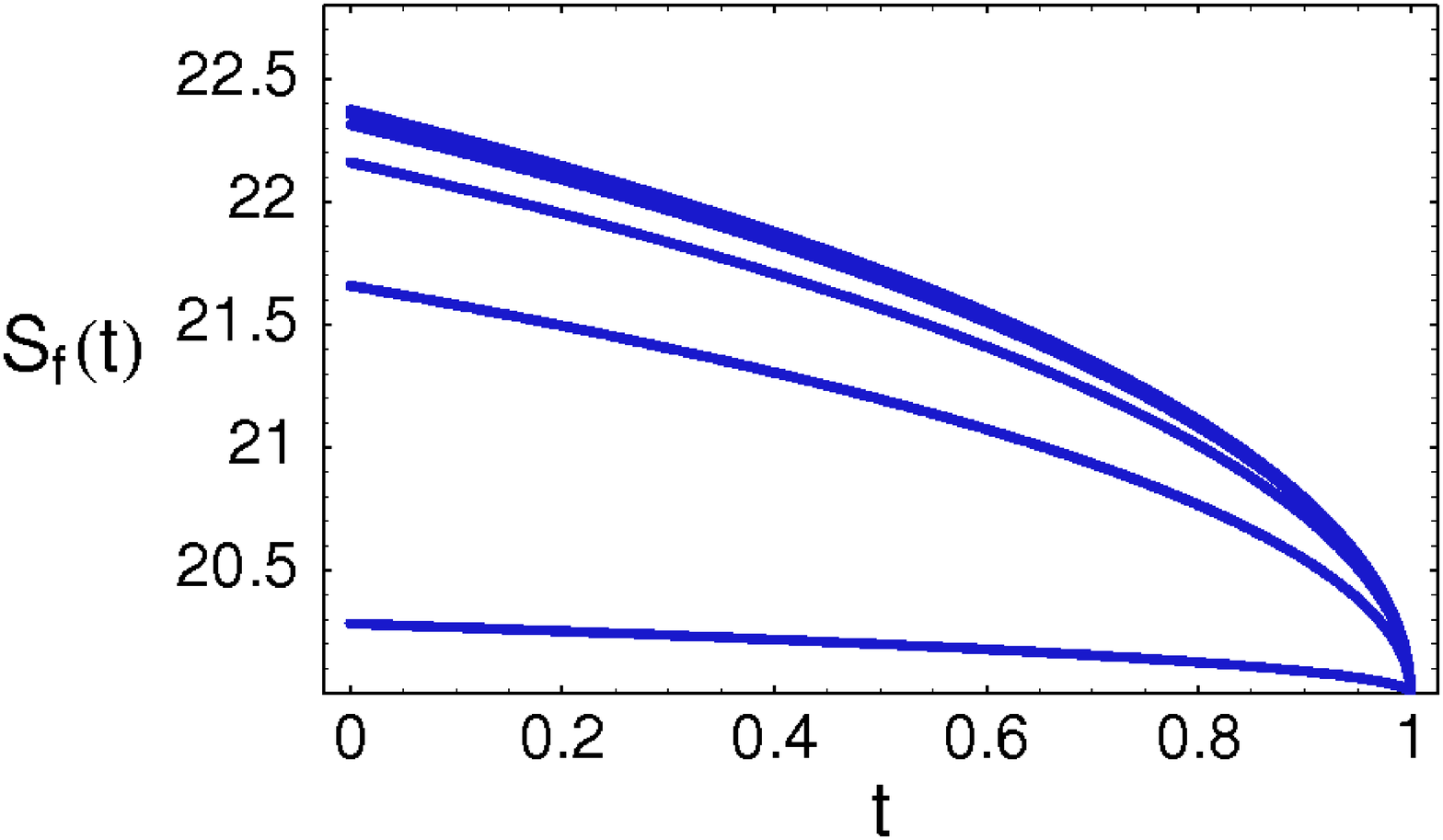}
\includegraphics[width=0.45\textwidth]{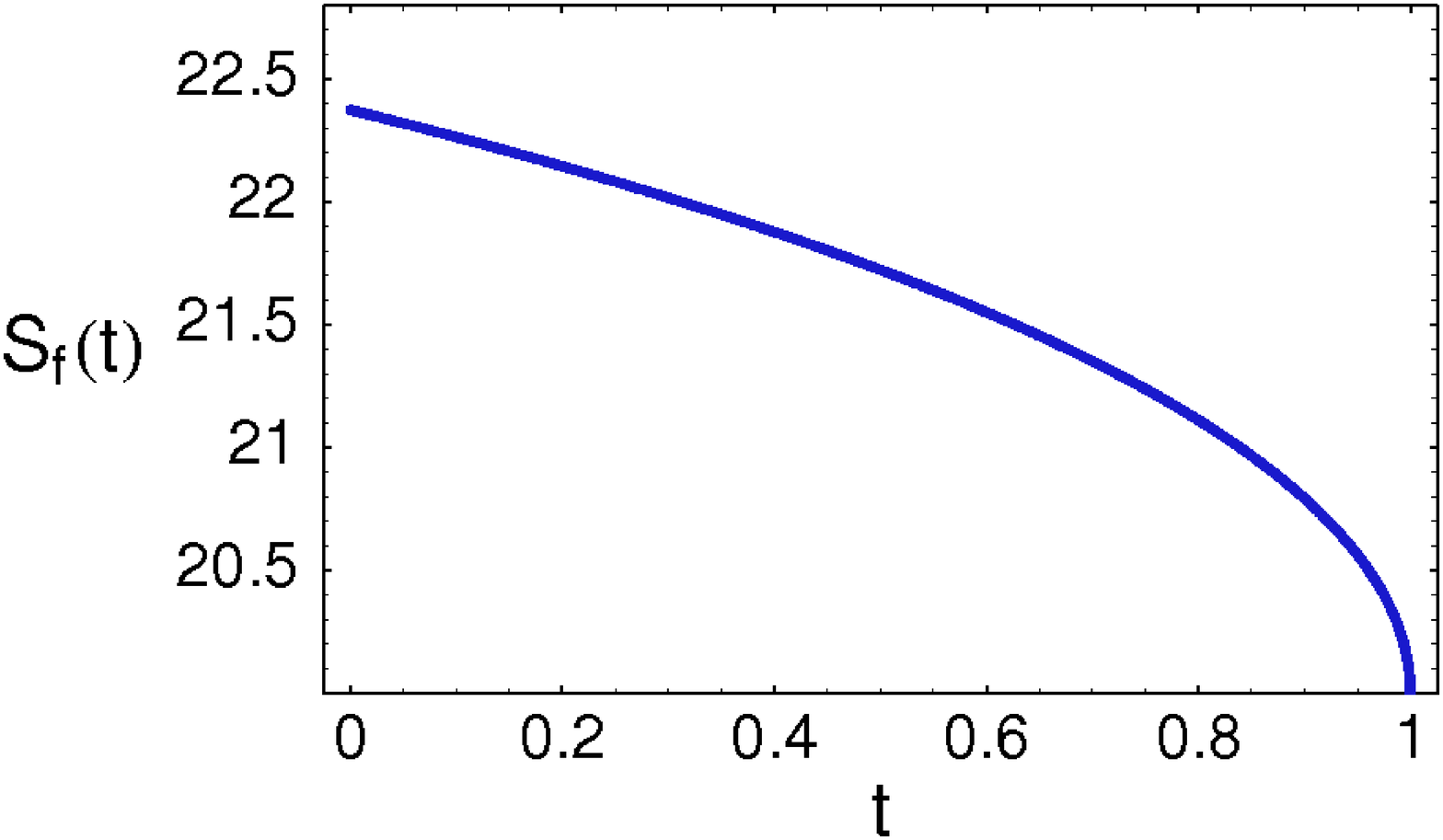}
\end{center}
\caption{\small
Five successive approximations of the free boundary  $S_f(t)$ obtained from equation (\ref{euro-3.10}). The profile of the solution $S_f(t)$.}
\label{euro-fig-12}
\end{figure}

In what follows, we present several computational examples. In Fig.~\ref{euro-fig-12} we show five iterates of
the free boundary function $S_f(t)$, where the auxiliary function $H(\xi)$
is constructed by means of successive iterations of the nonlinear integral
operator defined by the right-hand side of (\ref{euro-3.10}). This sequence converges to a fixed point of such a map, i.e. to a solution of (\ref{euro-3.10}).  Parameter values were chosen as $E=10, r=0.1, \sigma=0.2, q=0.05, T=1$. Fig.~\ref{euro-fig-12} depicts the final tenth iteration of approximation of the function $S_f(t)$. The mesh contained 100 grid points. One can observe very rapid convergence of iterates to a fixed point. In practice, no more than six iterates are sufficient to obtain the fixed point of (\ref{euro-3.10}). It is worth noting that in all our numerical simulations the convergence was monotone, i.e. the curve moves only up in the iteration process.

\begin{figure}
\begin{center}
\includegraphics[width=0.45\textwidth]{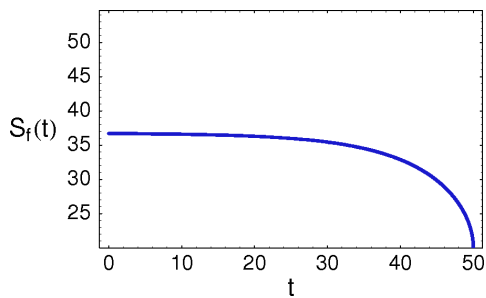}
\end{center}
\caption{\small
Long-time behavior of $S_f(t)$.}
\label{euro-fig-3}
\end{figure}

In Fig.~\ref{euro-fig-3} we show the long time behavior of the free boundary $S_f(t), t\in[0,T],$ for large values of the expiration time $T$. For the parameter values $T=50, r=0.1, q=0.05, \sigma=0.35$ and $E=10$ the theoretical value of $S_f(+\infty)$ is 36.8179 (see Dewynne {\em et al}.~\cite{DH}).

In Table~\ref{tab-1} we show a  comparison of results obtained by our  method based on the semi-explicit formula (\ref{euro-3.17}) and those obtained by known methods based on trinomial trees (both with the depth of the tree equal to 100), finite difference approximation (with 200 spatial and time grids) and analytic approximation of Barone-Adesi \& Whaley (cf. \cite{BW}, \cite[Ch. 15, p. 384]{H}), resp. It also turned out that the method based on solving the integral equation (\ref{euro-3.10}) is 5-10 times faster than other methods based on trees or finite differences. The reason is that the computation of $V(S,t)$ for a wide range of values of $S$ based on the semi-explicit pricing formula (\ref{euro-3.17}) is very fast provided that the free boundary function $\varrho$ has already been computed.

\begin{table}
\caption{\small 
Comparison of the method based on formula (\ref{euro-3.16}) with other methods for the parameter values $E=10, T=1, \sigma=0.2, r=0.1, q=0.05$. The position
$S_f(0)=\varrho(T)$ of the free boundary at $t=0$ (i.e. at $\tau=T$) was computed
as $S_f(0) = \varrho(T)= 22.3754$}
    \tabcolsep=4pt
\begin{center}
\small
    \begin{tabular}{llllll}
    \hline\hline
{Method $\backslash$ \  The asset value $S$}   & 15 & 18 & 20 & 21 &  22.3754 \\
    \hline
Our method    & 5.15 & 8.09 & 10.03 & 11.01 & 12.37
\\
Trinomial tree& 5.15 & 8.09 & 10.03 & 11.01 & 12.37
\\
Finite differences  & 5.49 & 8.48 & 10.48 & 11.48 & 12.48
\\
Analytic approximation & 5.23 & 8.10 & 10.04 & 11.02 & 12.38
\\
\hline\hline
\end{tabular}
\end{center}
\label{tab-1}
\end{table}

\begin{figure}
\begin{center}
\includegraphics[width=0.45\textwidth]{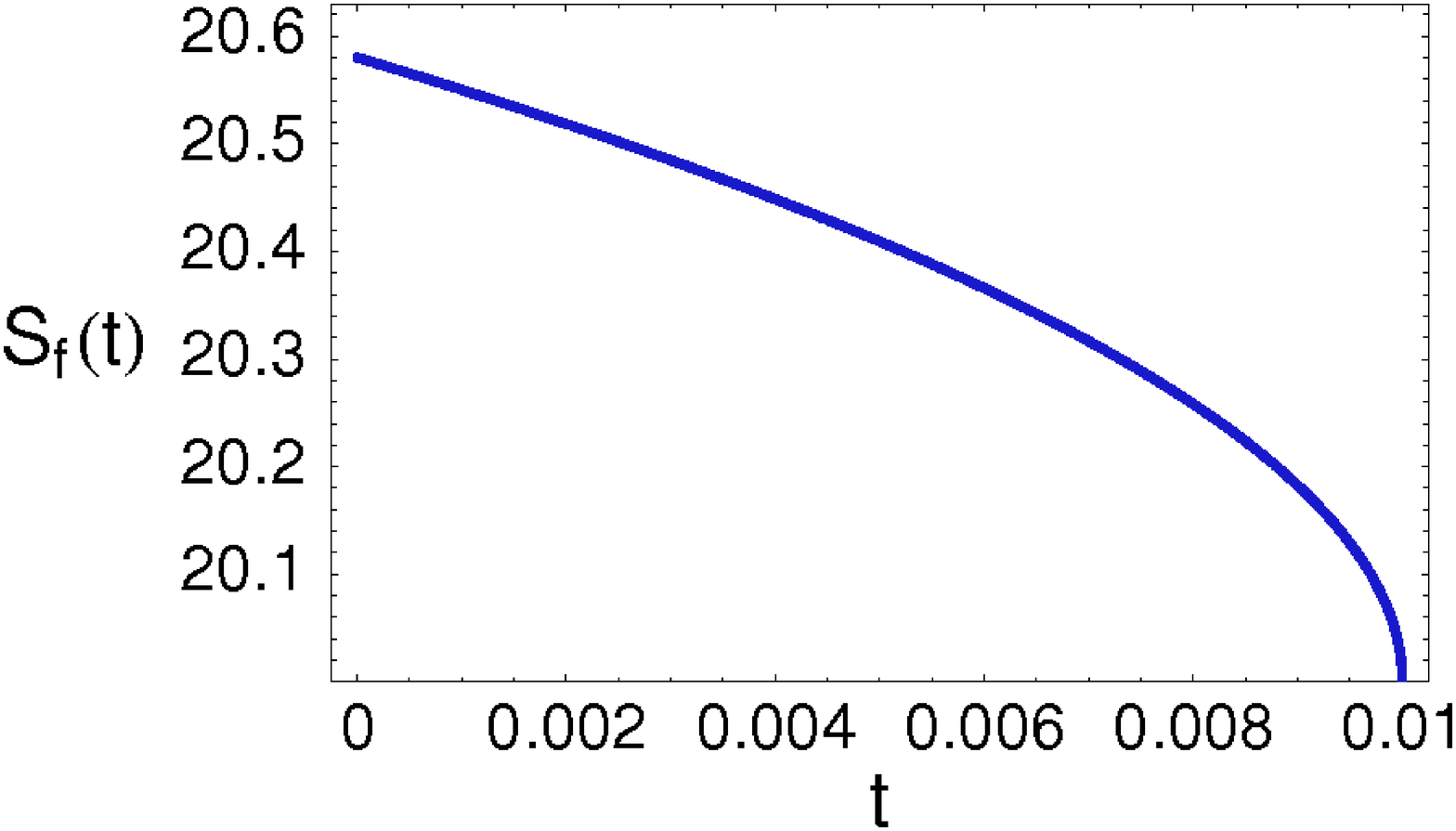}
\includegraphics[width=0.45\textwidth]{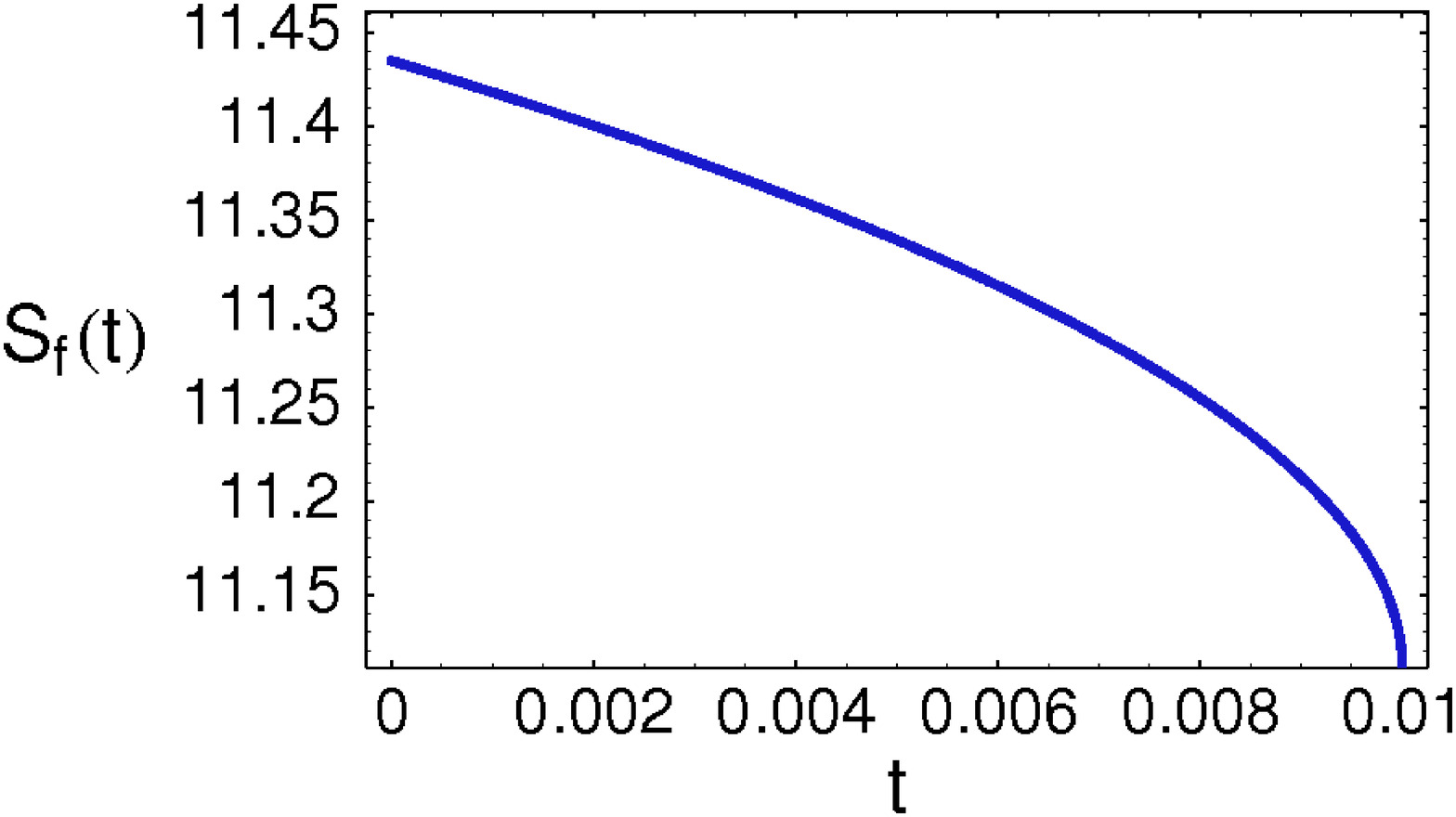}
\end{center}
\caption{\small
The early exercise boundary $S_f(t)$ for
$r=0.1, q=0.05$. The early exercise boundary $S_f(t)$ for
$r=0.1, q=0.09$.}
\label{euro-fig-45}
\end{figure}

In Fig.~\ref{euro-fig-45} the early exercise boundary $S_f(t)$ is computed for various values of the parameter $q$. In these computations we chose $E=10, T=0.01, \sigma=0.45$. Of interest is the case where $q$ is close to $r$ ($q=0.09$ and $r=0.1$).

\subsection{Early exercise boundary for an American put option}

In this section we present results of transformation method applied to valuation of the early exercise boundary for American style of a put option. Recall that the early exercise boundary problem for American put option can be formulated as follows:

\begin{eqnarray}
&&\d{V}{t} +  r S\d{V}{S} + {\sigma^2\over 2} S^2 \dd{V}{S} - r V =0\,,
\qquad 0<t<T,\ S_f(t) < S <\infty \,,
\nonumber
\\
&&V(+\infty ,t)=0,\ V(S_f(t), t)= E - S_f(t)\,, \ \d{V}{S}(S_f(t),t)=-1\,,
\\
&&V(S,T)=\max(E-S, 0)\,,
\nonumber
\label{amer-put}
\end{eqnarray}
defined on a time-dependent domain $S\in(S_f(t), \infty )$, where $t\in(0,T)$ (cf. Kwok \cite{Kw}). Again $S>0$ stands for the stock price, $E>0$ is the exercise price, $r>0$ is the risk-free rate and $\sigma>0$ is the volatility of the underlying stock process. We shall assume that the asset pay no dividends, i.e. $q=0$.  In order to perform a  fixed domain transformation of the free boundary problem (\ref{amer-put}) we introduce  the following change of variables 
\[
x=\ln\left(\frac{S}{\ro(\tau)}\right), \quad \hbox{where} \ \ \tau=T-t, \ro(\tau)=S_f(T-\tau). 
\]
Similarly as in the case of a call option we define a synthetised portfolio $\Pi$ for the put option $\Pi(x,\tau) = V - S \frac{\partial V}{\partial S}$. Then it is easy to verify that $\Pi$ is a solution to the following the parabolic equation
\begin{eqnarray}
&&\frac{\partial \Pi}{\partial \tau}  
- a(\tau )\frac{\partial \Pi}{\partial x}
- \frac{\sigma^2}{2}\frac{\partial^2 \Pi}{\partial x^2}
+ r \Pi  = 0,\quad x>0,\tau\in(0,T),    \nonumber \\
&&\Pi (0,\tau ) = E, \quad  \Pi (\infty ,\tau ) = 0, \label{bc1} \\
&&\frac{\sigma^2}{2} \frac{\partial \Pi}{\partial x}(0,\tau ) = - r E,\quad \hbox{for}\ \tau\in(0,T),\label{bc2} \\
&& \Pi (x,0) = 0\quad \hbox{for}\quad x>0,   \nonumber
\end{eqnarray}
where $a(\tau ) =\frac{\dot{\varrho}(\tau )}{\varrho (\tau )}+r - \frac{\sigma^2}{2}$ (see Stamicar \emph{et al.} \cite{SSC}). 
Now, following the same methodology of applying the Fourier transform we are able to derive an integral equation for the free boundary position. It is just the equation 
$\frac{\sigma^2}{2} \frac{\partial \Pi}{\partial x}(0,\tau ) = - r E$ in which the left hand side is expressed by the inverse Fourier transform of the solution $\Pi$ as a weakly singular integral depending on the free boundary position $\ro$  (see \cite{SSC} for details). We omit the technical details here. We just recall that 
the integral equation for the free boundary function $\ro(\tau)$ yields the following expression for $\ro(\tau)$
\[
\varrho (\tau )=Ee^{-(r-\frac{\sigma^2}{2})\tau }e^{\sigma \sqrt{2 \tau }\eta (\tau )}
\]
in terms of a new auxiliary function $\eta (\tau )$ (see Stamicar {\em et al.} \cite{SSC} for details). Further asymptotic analysis of the integral equation enables us to derive an asymptotic formula for $\eta(\tau)$ as $\tau\to 0$. Namely, 
\begin{equation}
\eta (\tau )\sim -\sqrt{-\ln \left[ \frac{2 r}{\sigma} \sqrt{2\pi \tau } e^{r \tau
}\right] } \quad \hbox{as}\ \tau\to 0^+.
\label{canad1-eta2}
\end{equation}

\begin{figure}
\begin{center}
\includegraphics[width=0.45\textwidth]{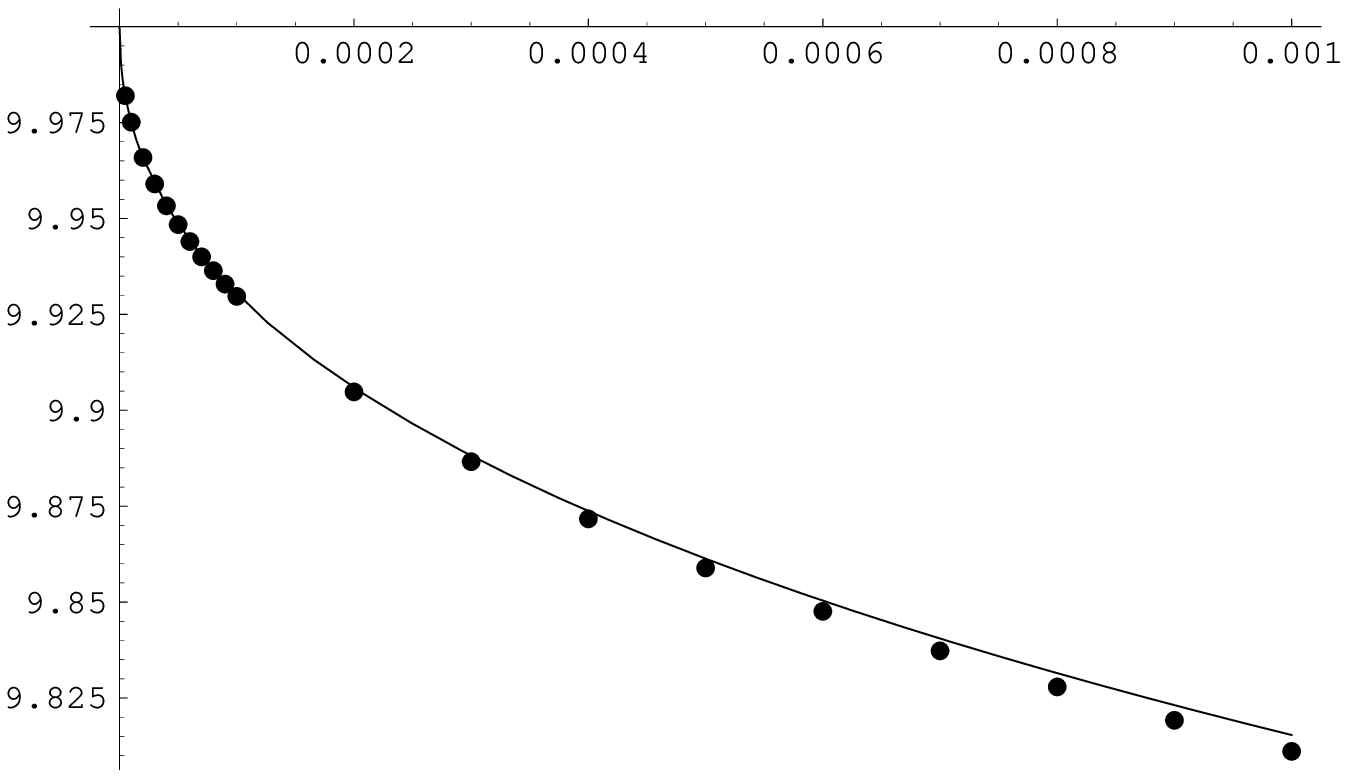}
\includegraphics[width=0.45\textwidth]{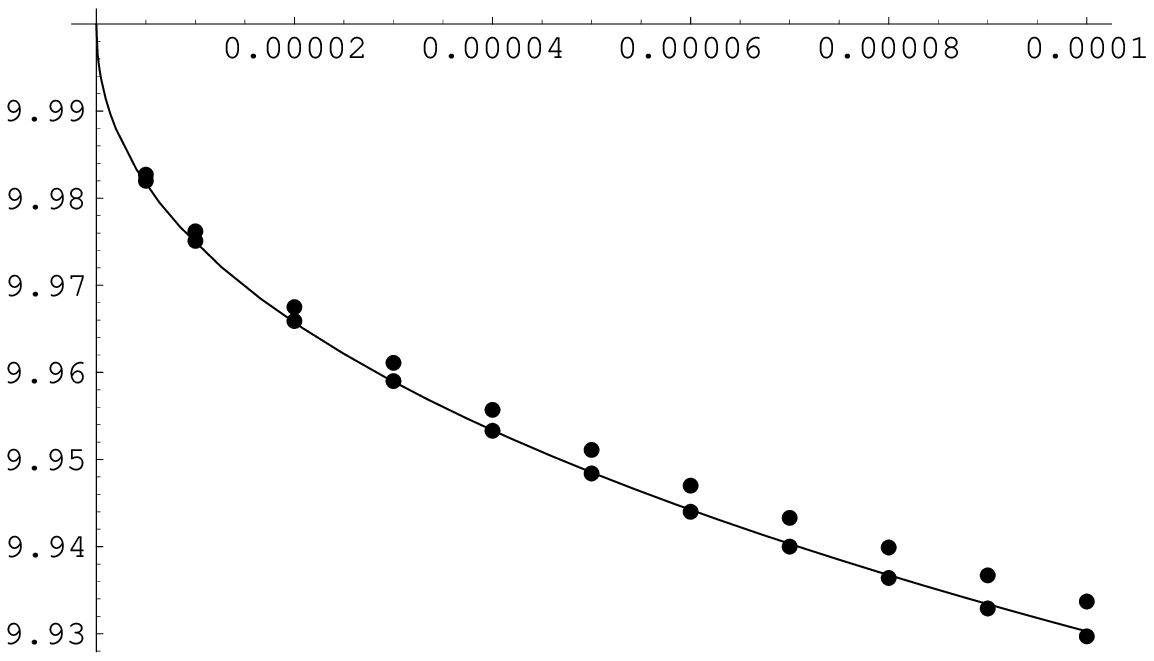}
\end{center}
\caption{\small 
Left: asymptotic approximation vs. binomial method for $\sigma=0.25,r=0.1,E=10$ and $T-t = 8.76$~hrs.  (0.001 of a year),  
MBW approximation vs. the asymptotic solution (\ref{canad1-eta2}) for
 $T-t =  0.876$~hrs (right).}
\label{fig-put}
\end{figure}

Next we examine how accurately our asymptotic approximation matches the data
from the binomial method (cf. Kwok \cite{Kw}). Near expiry at about one hour, the asymptotic approximation matches the data from the binomial method  (see Fig.~\ref{fig-put}). With $\sigma =0.25,r=0.1,E=10$ at 8.76 hours the approximation gives an overestimate but of only 0.4 cents (see \cite{SSC}). We also compared our asymptotic solution with MacMillan, Barone-Adesi, and Whaley's \cite{BW}, [4, 384--386], \cite{Mac} numerical approximation of the American put free boundary (see Fig.~\ref{fig-put}). They apply a transformation that results in a Cauchy-Euler equation that can be solved analytically. For times very close to expiry, one can see that our approximation of the free boundary matches the data from the binomial and trinomial methods more accurately.  

The numerical and analytical results obtained by the transformation method for solving free boundary problem for American put options are in agreement with those obtained recently by Zhu \cite{Zhu1,Zhu2}, Mallier \emph{et al.} \cite{Ma,MA}, Kuske and Keller \cite{KK}, Knessl \cite{K}.

\section{Transformation method for a nonlinear Black--Scholes equation}

The main goal of this section is to perform a fixed domain transformation of the free boundary problem for the nonlinear Black--Scholes 
equation (\ref{c-BS}) into a parabolic equation defined on a fixed spatial domain. 
For the sake of simplicity we will present a detailed derivation of an equation only 
for the case of an American call option. Derivation of the corresponding equation for  the American put option is similar. Throughout this section we shall assume the volatility $\sigma^2$ appearing in the Black--Scholes equation to be a function of the option price $S$, time to expiry $T-t$ and $S^2\partial^2_S V$, i.e. 
\[
\sigma = \sigma(S^2\partial^2_S V, S, T-t)\,.
\]
Following (\ref{euro-2.1}) we shall consider the following change of variables:
\[
\tau= T-t, \quad x=\ln\left(\ro(\tau)/S\right)
\ \ \hbox{where}\ \ \ro(\tau)=S_f(T-\tau).
\]
Then $\tau\in(0,T)$ and $x\in(0,\infty)$ iff $S\in(0,S_f(t))$. The boundary value $x=0$ corresponds to
the free boundary position $S=S_f(t)$ whereas $x\approx +\infty$ corresponds to the default
value $S=0$ of the underlying asset. Following Stamicar {\em et al.}  \cite{SSC} and 
\v{S}ev\v{c}ovi\v{c} \cite{Se,Se2} 
we construct the so-called synthetic portfolio function $\Pi=\Pi(x,\tau)$ 
defined as follows:
\begin{equation}
\Pi(x,\tau)= V(S,t) - S \frac{\partial V}{\partial S}(S,t)\,.
\label{c-transformacia}
\end{equation}
Again, it represents a synthetic portfolio consisting of one long positioned option and
$\Delta= \partial_S V$ underlying short stocks. Similarly as in Section 3 we have
\[
\frac{\partial\Pi}{\partial x} = S^2 \frac{\partial^2 V}{\partial S^2},  
\ \ 
\frac{\partial\Pi}{\partial\tau} + \frac{\dot\ro}{\ro} \frac{\partial\Pi}{\partial x} =
- \frac{\partial}{\partial t} \left( V - S \frac{\partial V}{\partial S}\right)
\]
where we have denoted $\dot\ro=d\ro/d\tau$. 
Assuming sufficient smoothness of a solution $V=V(S,t)$ to (\ref{c-BS}) 
we can deduce from (\ref{c-BS}) a parabolic equation for the synthetic portfolio 
function $\Pi=\Pi(x,\tau)$
\[
\frac{\partial\Pi}{\partial\tau}  + (b(\tau) - \frac12 \sigma^2 ) \frac{\partial\Pi}{\partial x}
- \frac12 \frac{\partial}{\partial x}\left( \sigma^2 \frac{\partial \Pi}{\partial x} \right) + r \Pi=0
\]
defined on a fixed domain $x\in \R,\, t\in(0,T),$ with a time-dependent coefficient
\begin{equation}
b(\tau)= {\dot\ro(\tau)\over\ro(\tau)} + r -q 
\label{c-bfun}
\end{equation}
and a diffusion coefficient given by: 
$\sigma^2=\sigma^2( \partial_x\Pi(x,\tau), \ro(\tau)e^{-x}, \tau)$
depending on $\tau, x$ and the gradient $\partial_x\Pi$ 
of a solution $\Pi$. Now the boundary conditions $V(0, t)=0, V(S_f(t), t)= S_f(t) - E$ 
and $\partial_S V(S_f(t), t)= 1$ imply
\begin{equation}
\Pi(0, \tau) = -E, \quad \Pi(+\infty, \tau) =0\,, \quad 0<\tau<T\,,
\end{equation}
and, from the terminal pay-off diagram for $V(S,T)$, we deduce
\begin{equation}
\Pi(x,0)=
\left\{
   \begin{array}{lll}
-E & \hbox{for} \ x<\ln\left({\ro(0)\over E}\right)\hfil
\\
\ \ 0 & \hbox{otherwise.}\hfil
\end{array}
\right.
\end{equation}
In order to close up the system of equations that determines the value of 
a synthetic portfolio $\Pi$ we have to construct an equation for the free
boundary position $\ro(\tau)$. Indeed, both the coefficient $b$ as well as the
initial condition $\Pi(x,0)$ depend on the function $\ro(\tau)$. 
Similarly as in the case of a constant volatility $\sigma$ (see \cite{Se,SSC})
we proceed as follows: since $S_f(t)-E=V(S_f(t),t)$ and $\partial_S V(S_f(t),t)=1$ we have
${\scriptstyle \frac{d}{dt}} S_f(t) 
= \partial_S V(S_f(t),t) {\scriptstyle \frac{d}{dt}} S_f(t) + \partial_t V(S_f(t), t)$
and so $\partial_t V(S,t)=0$ along the free boundary $S=S_f(t)$. Moreover, assuming $\partial_x\Pi$ is 
continuous  up to the boundary $x=0$ we obtain $S^2\partial^2_S V(S,t) \to  \partial_x\Pi(0,\tau)$
and $S\partial_S V(S,t)\to \ro(\tau)$ as $S\to S_f(t)^-$. Now, by taking the limit 
$S\to S_f(t)^-$ in the Black--Scholes equation (\ref{c-BS}) we obtain
$(r-q)\ro(\tau) +\onehalf \sigma^2 \partial_x \Pi(0,\tau) -r(\ro(\tau)-E) =0$.
Therefore
\[
\ro(\tau)= {rE\over q} 
+ \frac{1}{2q} \sigma^2(\partial_x\Pi(0,\tau), \ro(\tau), \tau)
\frac{\partial \Pi}{\partial x}(0,\tau)
\]
for $0<\tau\le T$. The value of $\ro(0)$  can be easily derived from the smoothness
assumption made on $\partial_x\Pi$ at the origin $x=0, \tau=0$ under the structural assumption 
\begin{equation}
0<q <  r
\label{c-rD}
\end{equation}
made on the interest and dividend yield rates $r,q$ (cf. \cite{Se,Se2}). Indeed, continuity of  $\partial_x\Pi$ at the origin $(0,0)$ implies $\lim_{\tau\to0^+}\partial_x\Pi(0,\tau) =\partial_x\Pi(0,0) =\lim_{x\to0^+}\partial_x \Pi(x,0) = 0$ because  $\Pi(x,0)=-E$ for $x$ close to $0^+$ provided $\ln(r/q)>0$. From the above equation  for $\ro(\tau)$ we deduce $\ro(0)= {rE\over q}$  by taking the limit $\tau\to 0^+$.  Putting all the above equations together we end up with a closed system of equations for $\Pi=\Pi(x,\tau)$ and $\ro=\ro(\tau)$
\begin{eqnarray}
&&\frac{\partial\Pi}{\partial \tau}  + \bigl( b(\tau) -\frac{\sigma^2}{2} \bigr)\frac{\partial \Pi}{\partial x}
- \frac12  \frac{\partial}{\partial x}\bigl( \sigma^2 \frac{\partial \Pi}{\partial x}\bigr) + r \Pi=0\,,
\nonumber \\
&&\Pi(0, \tau) = -E\,, \ \  \Pi(+\infty, \tau) =0\,,\  x>0\,, \tau\in(0,T)\,,
\nonumber\\
&&\Pi(x,0)= \left\{
\begin{array}{lll}
-E & \quad\hbox{for} \ x<\ln(r/q)\hfil
\\
\ \ \  0 & \quad\hbox{otherwise\,,}\hfil
\end{array}
\right.
\label{c-tBS}
\end{eqnarray}
where
$\sigma=\sigma(\partial_x\Pi(x,\tau), \ro(\tau)e^{-x},\tau )\,,\ 
b(\tau)= {\dot\ro(\tau)\over\ro(\tau)} + r - q$ and the free boundary position 
$\ro(\tau)=S_f(T-\tau)$ satisfies an implicit algebraic equation
\begin{equation}
\ro(\tau)={r E\over q} + {\sigma^2(\partial_x\Pi(0,\tau), \ro(\tau), \tau )\over 2q}
\frac{\partial \Pi}{\partial x}(0,\tau)\,,\quad \hbox{with }\ \ro(0)={rE\over q},
\label{c-ro}
\end{equation}
where $\tau\in(0,T)$.
Notice that, in order to guarantee parabolicity of equation (\ref{c-tBS}) we have
to assume that the function $p\mapsto \sigma^2(p, \ro(\tau)e^{-x}, \tau) p$ is strictly
increasing. More precisely, we shall assume that there exists a positive constant $\gamma>0$
such that 
\begin{equation}
\sigma^2(p, \xi, \tau)  + p\partial_p\sigma^2(p, \xi, \tau)\ge \gamma>0
\label{c-strict} 
\end{equation}
for any $\xi>0, \tau\in(0,T)$ and $p\in \R$. Notice that condition (\ref{c-strict}) is satisfied for the RAPM model in which $\sigma^2=\hat\sigma^2(1+\mu p^{\frac13} \xi^{-\frac13})$ for any $\mu\ge 0$ and $p\ge 0$. Clearly $p=S^2\partial^2_S V >0$ for the case of plain vanilla call or put options. As far as the Barles and Soner model is concerned, we have $\sigma^2=\hat\sigma^2(1+ \Psi(a^2 e^{r\tau} p) )$ and condition (\ref{c-strict}) is again satisfied because the function $\Psi$ is a positive and increasing function in the Barles and Soner model.

\medskip
\begin{remark}\label{rem-Vprice}
Following exactly the same argument as in (\ref{euro-3.14}) one can derive an explicit expression for the option price $V(S,t)$:
\begin{equation}
V(S,T-\tau)= {S\over\ro(\tau)} \left(
\ro(\tau) -E + \int_0^{\ln{\ro(\tau)\over S}} {\rm e}^x \Pi(x,\tau) \, dx
\right)\,.
\end{equation}
\end{remark}

\subsection{An iterative algorithm for approximation of the early exercise 
boundary}

The idea of the iterative numerical algorithm for solving the problem (\ref{c-tBS}), (\ref{c-ro})  
is rather simple: we use the backward Euler method of 
finite differences in order to discretize the parabolic equation (\ref{c-tBS}) in time. In each time
level we find a new approximation of a solution pair $(\Pi, \ro)$. First we determine a new position
of $\ro$ from the algebraic equation (\ref{c-ro}). 
We remind ourselves that (even in the case $\sigma$ is constant) the free boundary function $\ro(\tau)$ 
behaves like $r E/q + O(\tau^{1/2})$ for $\tau\to 0^+$ (see e.g. \cite{DH} or \cite{Se}) and so $b(\tau) = 
O(\tau^{-1/2})$. Hence the convective term $b(\tau)\partial_x\Pi$ becomes 
a dominant part of equation (\ref{c-tBS}) for small values of $\tau$. In order to overcome this difficulty we employ the operator splitting technique for successive solving of the convective and diffusion parts of equation (\ref{c-tBS}). 
Since the diffusion coefficient $\sigma^2$ depends on the derivative $\partial_x\Pi$ of a solution $\Pi$ itself we make several micro-iterates
to find a solution of a system of nonlinear algebraic equations. 

Now we present our algorithm in more details.  We restrict the spatial domain $x\in (0,\infty)$ to a finite interval of values $x\in (0,L)$ where $L>0$ is sufficiently large. For practical purposes one can take $L\approx 3$ as it corresponds to the interval  $S\in (S_f(t) e^{-L}, S_f(t))$ in the original asset price variable $S$. The value $S_f(t) e^{-L}$ is then could be a good  approximation for the default value $S=0$ if $L\approx 3$.  Let us denote by $k>0$ the time step, $k=T/m$, and, by $h>0$ the spatial step, $h=L/n$ where $m, n\in N$ stand for the number of time and space discretization steps, resp.  We denote by $\Pi_i^j$ an approximation of $\Pi( x_i, \tau_j)$, $\ro^j \approx \ro(\tau_j)$, $b^j \approx b(\tau_j)$ where $x_i=i h, \tau_j=j k$. We approximate the value of the  volatility $\sigma$ at  the node $(x_i,\tau_j)$ by finite difference as follows:
\[
\sigma_i^j = \sigma_i^j(\ro^j, \Pi^j) =\sigma((\Pi_{i+1}^j - \Pi_i^j)/h, \ro^j e^{-x_i}, \tau_j)\,.
\]
Then for the Euler backward in time finite difference approximation of equation (\ref{c-tBS}) we have
\begin{equation}
\frac{\Pi^j-\Pi^{j-1}}{k} + \left( b^j -\frac12 (\sigma^j)^2 \right) \partial_x\Pi^j 
-\frac12 \partial_x \left( (\sigma^j)^2 \partial_x\Pi^j\right) +r \Pi^j = 0
\label{c-timediscret}
\end{equation}
and the solution $\Pi^j=\Pi^j(x)$ is subject to Dirichlet boundary conditions at $x=0$  and $x=L$. We set
$\Pi^0(x)=\Pi(x,0)$.
Now we decompose the above problem into two parts - a convection part and a diffusive part by introducing 
an auxiliary intermediate step $\Pi^{j-\onehalf}$:

\medskip

({\it Convective part})
\begin{equation}
\frac{\Pi^{j-\onehalf}-\Pi^{j-1}}{k} + b^j  \partial_x\Pi^{j-\onehalf} = 0\,,
\label{c-convective}
\end{equation}

({\it Diffusive part})
\begin{equation}
\frac{\Pi^{j}-\Pi^{j-\onehalf}}{k}   - \frac{(\sigma^j)^2}{2}  \partial_x\Pi^j
-\frac12 \partial_x \left( (\sigma^j)^2 \partial_x\Pi^j\right) +r \Pi^j = 0\,.
\label{c-diffusion}
\end{equation}
The idea of the operator splitting technique consists in comparison the sum of solutions to 
convective and diffusion 
part to a solution of (\ref{c-timediscret}). Indeed, if $\partial_x\Pi^j \approx \partial_x\Pi^{j-\onehalf}$ then 
it is reasonable to assume that $\Pi^j$ computed from the system (\ref{c-convective})--(\ref{c-diffusion})
is a good approximation of the system (\ref{c-timediscret}).

The convective part can be approximated by an explicit solution to the transport equation:
\begin{equation}
\partial_\tau\tilde\Pi + b(\tau) \partial_x\tilde\Pi =0 \qquad \hbox{for } x>0,\  \tau\in(\tau_{j-1},\tau_j] 
\label{c-convective-anal}
\end{equation}
subject to the boundary condition $\tilde\Pi(0,\tau) =-E$ and initial condition   $\tilde\Pi(x,\tau_{j-1})=\Pi^{j-1}(x)$. For American style of call option the free boundary $\ro(\tau)=S_f(T-\tau)$ must be an increasing function in $\tau$ and we have assumed $0<q < r$ we have $b(\tau)=\dot\ro(\tau)/\ro(\tau) +r-q >0$ and so prescribing the in-flowing boundary condition $\tilde\Pi(0,\tau) =-E$ is consistent with the transport equation. Let us denote by $B(\tau)$ the primitive function to $b(\tau)$, i.e. $B(\tau)=\ln \ro(\tau) + (r-q)\tau$. Equation (\ref{c-convective-anal}) can be integrated to obtain its explicit solution:
\begin{equation}
\tilde\Pi(x,\tau)=\left\{ 
\begin{matrix}
\Pi^{j-1}(x-B(\tau)+B(\tau_{j-1})) \hfill & \quad \hbox{if } x-B(\tau)+B(\tau_{j-1})>0\,, \hfill\cr 
-E \hfill & \quad \hbox{otherwise.} \hfill
\end{matrix}
\right.
\label{c-convective-explicit}
\end{equation}
Thus the spatial approximation $\Pi^{j-\onehalf}_i$ can be constructed from the formula
\begin{equation}
\Pi^{j-\onehalf}_i=\left\{ 
\begin{matrix}
\Pi^{j-1}(\xi_i) \hfill & \quad \hbox{if } \xi_i=x_i- \ln\ro_j + \ln\ro_{j-1} - (r-q)k>0\,, \hfill\cr 
-E \hfill & \quad \hbox{otherwise,} \hfill
\end{matrix}
\right.
\label{c-convective-discrete}
\end{equation}
where a linear approximation between discrete values $\Pi^{j-1}_i, i=0,1, ..., n,$ 
is being used to compute the value
$\Pi^{j-1}(x_i - \ln\ro_j + \ln\ro_{j-1} - (r-q)k)$.

The diffusive part can be solved numerically by means of finite differences. Using central finite difference
for approximation of the derivative $\partial_x\Pi^j$ we obtain

\begin{eqnarray}
\frac{\Pi_i^j - \Pi_i^{j-\onehalf}}{k} + r\Pi_i^j
&-& \frac{(\sigma^j_i)^2}{2} \frac{\Pi_{i+1}^j-\Pi_{i-1}^j}{2h}
\nonumber \\
&-& \frac{1}{2 h} \left( 
(\sigma_i^j)^2 \frac{\Pi_{i+1}^j - \Pi_{i}^j}{h}
-
(\sigma_{i-1}^j)^2 \frac{\Pi_{i}^j - \Pi_{i-1}^j}{h}
\right)  =0\,.
\nonumber
\end{eqnarray}
Hence, the vector of discrete values $\Pi^j=\{\Pi_i^j, i=1,2, ..., n\}$ at the time level 
$j\in\{1,2, ..., m\}$ satisfies the tridiagonal system of equations
\begin{equation}
\alpha_i^j \Pi_{i-1}^j + \beta_i^j \Pi_{i}^j + \gamma_i^j \Pi_{i+1}^j = \Pi_i^{j-\onehalf},
\label{c-eq-tridiag}
\end{equation}
for $i=1,2, ..., n,$  where
\begin{eqnarray}
\label{c-abc}
\alpha_i^j &\equiv& \alpha_i^j(\ro^j, \Pi^j)= -\frac{k}{2h^2} (\sigma_{i-1}^j)^2
+ \frac{k}{2h} \frac{(\sigma_{i}^j)^2}{2},
\nonumber \\
\gamma_i^j &\equiv& \gamma_i^j(\ro^j, \Pi^j) = -\frac{k}{2h^2} (\sigma_{i}^j)^2
- \frac{k}{2h} \frac{(\sigma_i^j)^2}{2},
\\
\beta_i^j &\equiv& \beta_i^j(\ro^j, \Pi^j)=  1 + r k - (\alpha_i^j +\gamma_i^j)\,.
\nonumber
\end{eqnarray}
The initial and boundary conditions at $\tau=0$ and $x=0, L,$ resp., can be approximated as
follows:
\[
\Pi_i^0 = \left\{
\begin{array}{lll}
-E & \ \ \hbox{for} \ x_i <\ln\left({r/q}\right),\hfil
\\
\ \ \ 0 & \ \ \hbox{for} \ x_i \ge \ln\left({r/q}\right), \hfil
\end{array}
\right.
\]
for $i=0,1, ..., n,$ and $\Pi_0^j = -E, \quad \Pi_n^j = 0$.

Next we proceed by approximation of equation (\ref{c-ro}) which introduces a nonlinear
constraint condition between the early exercise boundary function $\ro(\tau)$ and
the trace of the solution $\Pi$ at the boundary $x=0$ ($S=S_f(t)$ in the 
original variable). Taking a finite difference approximation of $\partial_x\Pi$ at 
the origin $x=0$ we obtain
\begin{equation}
\ro^j = \frac{rE}{q} + \frac{1}{2q} \sigma^2 \left((\Pi_1^j-\Pi_0^j)/h, \ro^j, \tau_j\right)
\frac{\Pi_1^j-\Pi_0^j}{h}\,.
\label{c-eq-ro}
\end{equation}

Now, equations (\ref{c-convective-discrete}), (\ref{c-eq-tridiag}) and (\ref{c-eq-ro}) can be written 
in an abstract form as a system of nonlinear equations:
\begin{eqnarray}
\ro^j &&= {\mathcal F}(\Pi^j, \ro^j),\nonumber\\
\Pi^{j-\onehalf} &&={\mathcal T}(\Pi^j, \ro^j),\label{c-abstract}\\
{\mathcal A}(\Pi^j, \ro^j) \Pi^j &&= \Pi^{j-\onehalf},\nonumber
\end{eqnarray}
where 
${\mathcal F}(\Pi^j, \ro^j)$ is the right-hand side of the algebraic equation (\ref{c-eq-ro}),
${\mathcal T}(\Pi^j, \ro^j)$ is the transport equation solver given by 
the right-hand side of (\ref{c-convective-discrete}) and
${\mathcal A}={\mathcal A}(\Pi^j, \ro^j)$ is a tridiagonal matrix with coefficients given 
by (\ref{c-abc}). The system (\ref{c-abstract}) can be approximately solved by means of successive iterates 
procedure. We define, for $j\ge 1,$  $\Pi^{j,0} = \Pi^{j-1}, \ro^{j,0} = \ro^{j-1}$. Then the 
$(p+1)$-th approximation of $\Pi^j$ and $\ro^j$ is obtained as a solution to the system:
\begin{eqnarray}
\ro^{j,p+1} &&= {\mathcal F}(\Pi^{j,p}, \ro^{j,p}),\nonumber\\
\Pi^{j-\onehalf, p+1} &&={\mathcal T}(\Pi^{j,p}, \ro^{j,p+1}), \label{c-abstract-iter}\\
{\mathcal A}(\Pi^{j,p}, \ro^{j,p+1}) \Pi^{j,p+1} &&= \Pi^{j-\onehalf, p+1} \,.\nonumber
\end{eqnarray}
Notice that the last equation is a linear tridiagonal equation for the vector $\Pi^{j,p+1}$
whereas $\ro^{j,p+1}$ and $\Pi^{j-\onehalf, p+1}$ can be directly computed from (\ref{c-eq-ro}) and 
(\ref{c-convective-discrete}), resp.
If the sequence of approximate solutions $\{(\Pi^{j,p}, \ro^{j,p}) \}_{p=1}^\infty$ converges to 
some limiting value $(\Pi^{j,\infty}, \ro^{j,\infty})$ as $p\to\infty$ then this limit is a solution to a nonlinear 
system of  equations (\ref{c-abstract}) at the time level $j$ and we can  proceed by computing 
the approximate solution the next time level $j+1$.

\subsection{Numerical approximations of the early exercise boundary}

In this section we focus on numerical experiments based on the iterative  
scheme described 
in the previous section. The main purpose is to compute the free boundary profile 
$S_f(t)=\ro(T-t)$ for different (non)linear Black--Scholes models and for various model
parameters. A solution $(\Pi,\ro)$ has been computed by our iterative algorithm for the following basic model parameters:
$E=10, T=1$ (one year), $r=0.1$ (10\% p.a) , $q=0.05$ (5\% p.a.) and  $\hat\sigma=0.2$. We used $n = 750$ spatial points and $m = 225000$ time discretization steps. Such a time step $k=T/m$ corresponds to 140 seconds between consecutive time levels when expressed in real time scale. In average we needed $p\le6$ micro-iterates (\ref{c-abstract-iter}) in order to solve the nonlinear 
system (\ref{c-abstract}) with the precision $10^{-7}$.

\subsubsection{Case of a constant volatility -- comparison study}

\begin{figure}

\begin{center}
\includegraphics[width=0.45\textwidth]{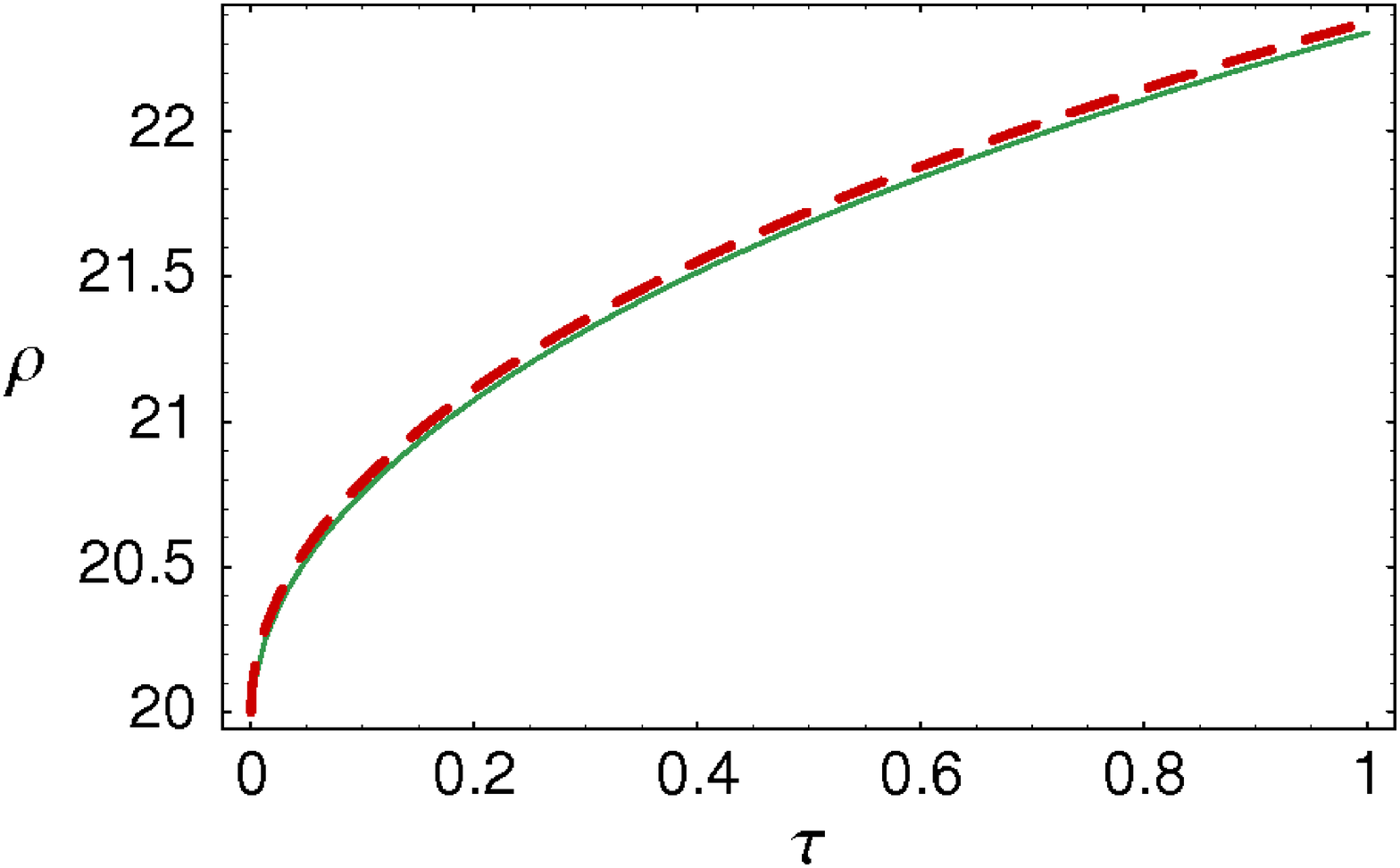}
\includegraphics[width=0.45\textwidth]{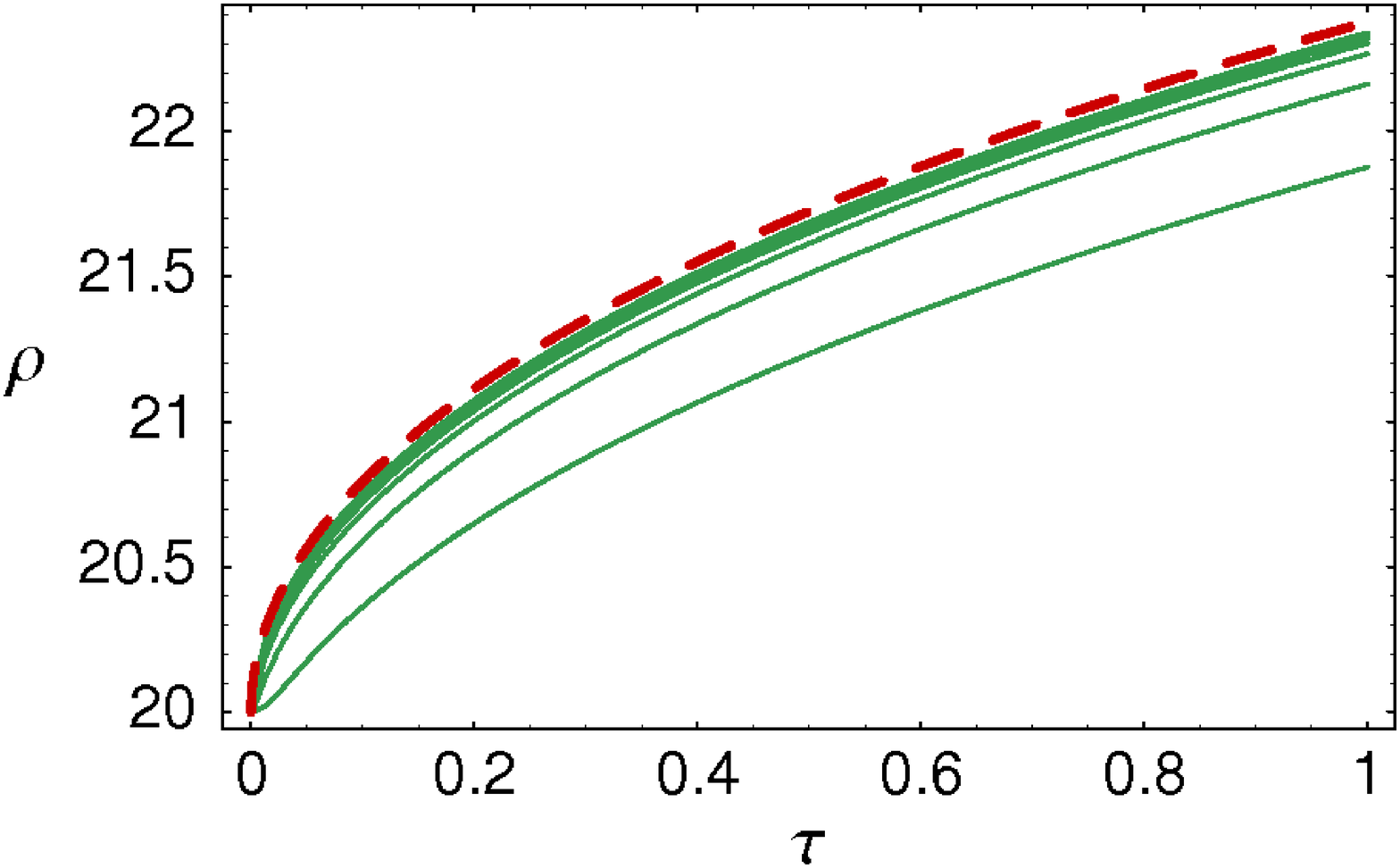}

\centerline{\small a) \hglue 0.4\textwidth b)}

\vglue 0.3truecm
\includegraphics[width=0.45\textwidth]{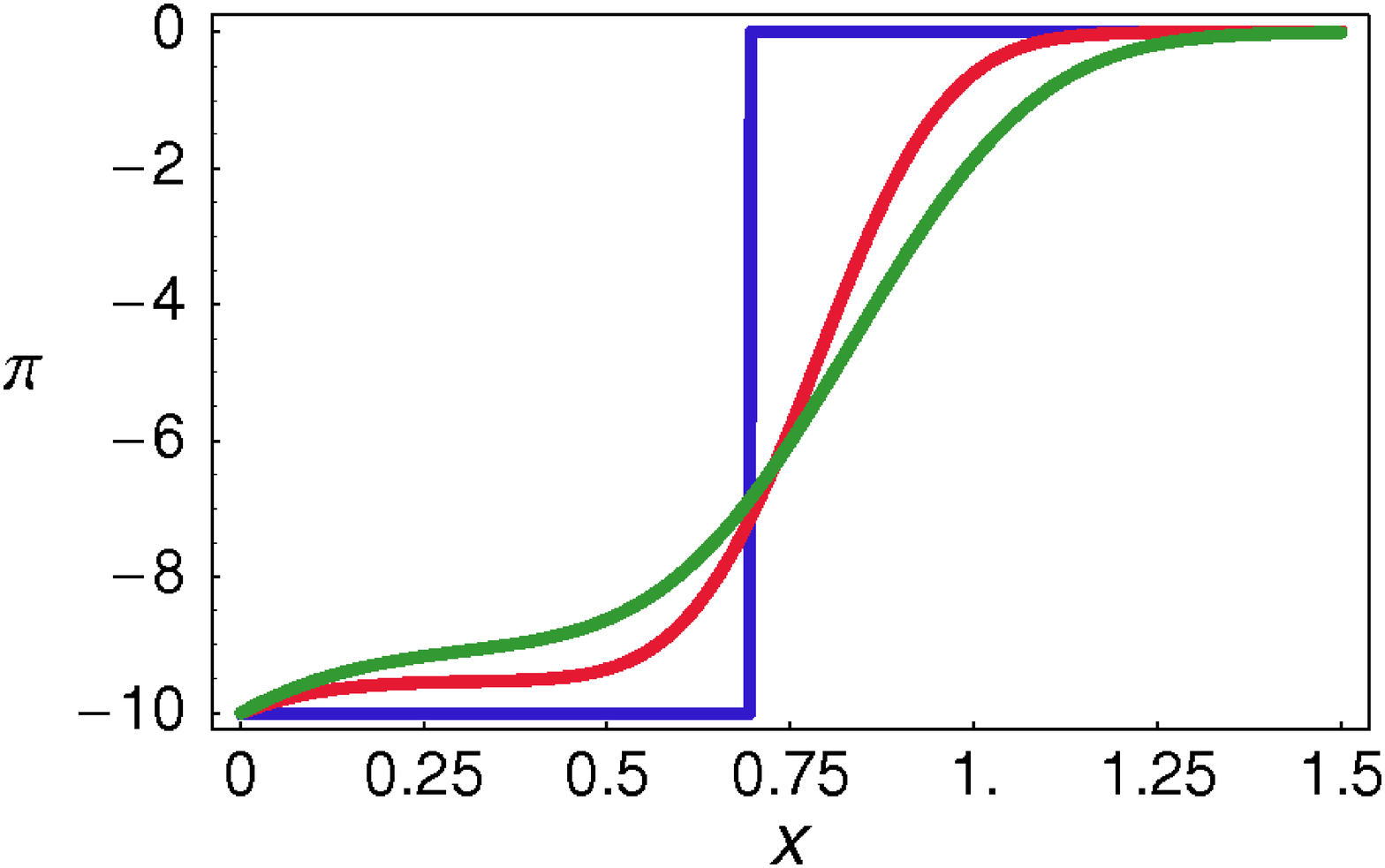}

\centerline{\small c)}

\includegraphics[width=0.45\textwidth]{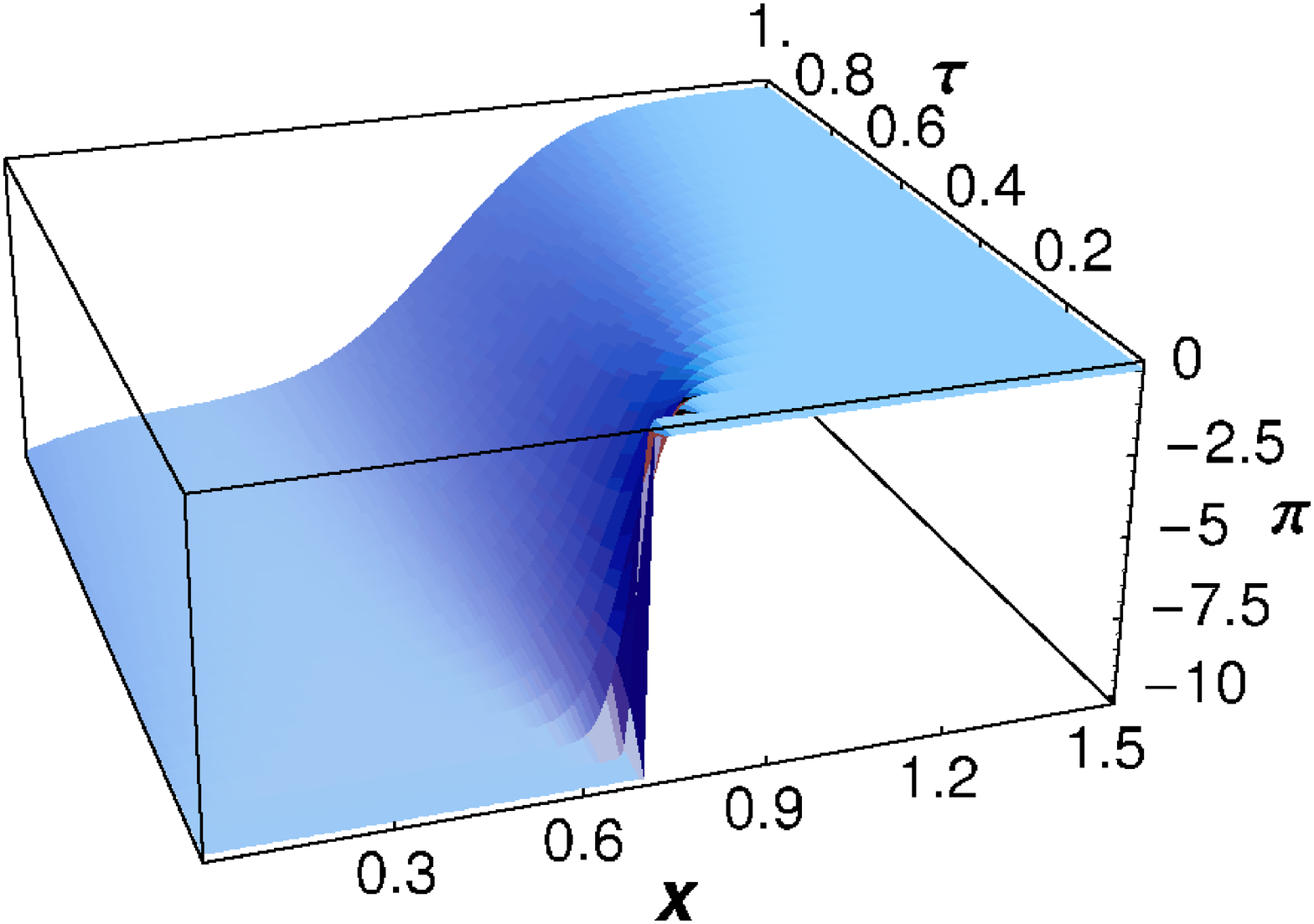}
\includegraphics[width=0.35\textwidth]{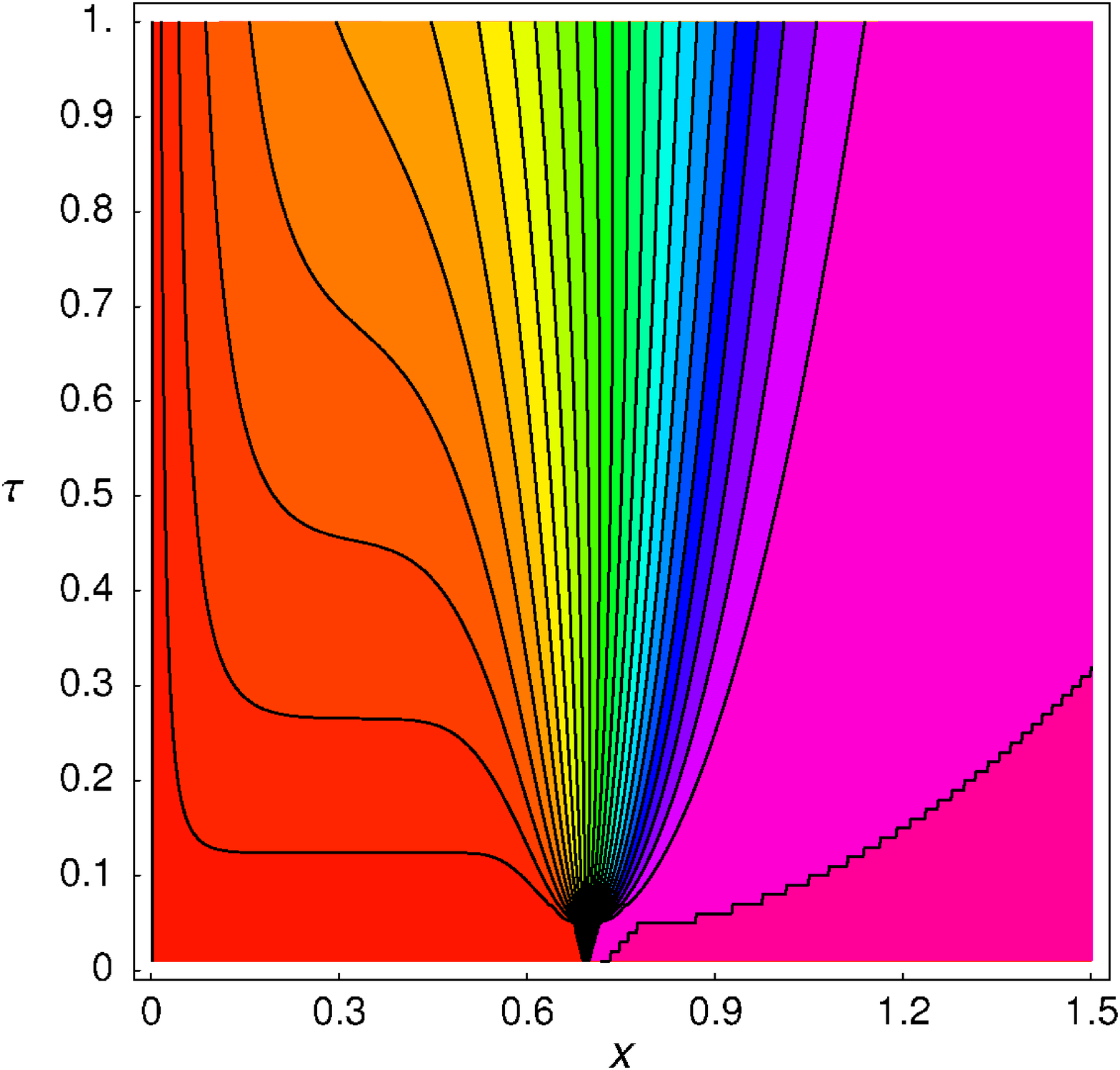}

\centerline{\small d) \hglue 0.4\textwidth e)}
\end{center}

\caption{\small
a) A comparison of the free boundary function $\ro(\tau)$ computed by the iterative algorithm 
(green solid curve) to the integral equation based  approximation (dashed red curve); 
b) free boundary positions computed for various mesh sizes;
c) a solution profile $\Pi(x,\tau)$ for $\tau=0$ (blue line), $\tau=T/2$ (red curve),
$\tau=T$ (green curve); d) 3D plot and e) contour plot of the function $\Pi(x,\tau)$.
}
\vspace{2mm}
\label{c-fig-noRAPM}
\end{figure}

In our first numerical experiment we make attempt to compare our iterative approximation scheme for  solving the free boundary problem for an American call option to known schemes in the case when the volatility $\sigma>0$ is constant. We compare our solution to the one computed by means of a solution to a nonlinear integral equation for $\ro(\tau)$  (see also  \cite{Se,SSC}). This comparison can be also considered as a benchmark or test example for which we know a solution that can be computed by a another justified algorithm.  In Fig.~\ref{c-fig-noRAPM}, part a),   we show the function $\ro$ computed by our iterative algorithm for  $E=10, T=1, r=0.1,  q=0.05, \sigma=0.2$.  At the expiry $T=1$ the value of $\ro(T)$ was computed as: $\ro(T)=22.321$. The corresponding value $\ro(T)$ computed from the integral equation (\ref{euro-3.8}) (cf. \cite{Se}) was  $\ro(T)=22.375$. The relative error is less than 0.25\%. In the part b) we present 7 approximations of the free boundary function $\ro(\tau)$ computed for different mesh sizes $h$ (see Tab.~\ref{c-tab1} for details).  The sequence of approximate free boundaries $\ro_h, h=h_1, h_2, ... ,$ converges monotonically from below to the free boundary function $\ro$ as $h\downarrow 0$. The next part c) of Fig.~\ref{c-fig-noRAPM} depicts various solution profiles of a function $\Pi(x,\tau)$. In order to achieve a reasonable approximation to equation (\ref{c-eq-ro}) we need very accurate approximation of $\Pi(x,\tau)$ for $x$ close to the origin $0$. The parts d) and e) of Fig.~\ref{c-fig-noRAPM} depict the contour and 3D  plots of the function  $\Pi(x,\tau)$.

In Tab.~\ref{c-tab1} we present the numerical error analysis for the distance $\Vert\ro_h -\ro\Vert_p$ measured in two different norms ($L^\infty$ and $L^2$)  of a computed free boundary position $\ro_h$ corresponding to the mesh size $h$ and the solution $\ro$ computed from the integral equation described in (\ref{euro-3.8}) (cf. \cite{Se}). The time step $k$ has been adjusted to the spatial mesh size $h$ in order to satisfy CFL condition $\hat\sigma^2k/h^2  \approx 1/2$. We also computed the experimental order of convergence $\hbox{eoc}(L^p)$ for $p=2,\infty$. Recall that the experimental order of convergence can be defined as the ratio:
\[
\hbox{eoc}(L^p) = \frac{\ln(\Vert\ro_{h_i} -\ro\Vert_p) - \ln(\Vert\ro_{h_{i-1}} -\ro\Vert_p)}{\ln h_i - \ln h_{i-1}}\,.
\] 
It can be interpreted as an exponent $\alpha=\hbox{eoc}(L^p)$ for which we have $\Vert\ro_{h} -\ro\Vert_p = O(h^\alpha)$. It turns out from Tab.~\ref{c-tab1} that the conjecture on the order of convergence  $\Vert\ro_{h} -\ro\Vert_\infty = O(h)$ whereas $\Vert\ro_{h} -\ro\Vert_2 = O(h^{3/2})$ as $h\to 0^+$ could be reasonable. 

\begin{table}

\caption{\small Experimental order of convergence of the iterative algorithm for approximating the free
boundary position.}
\smallskip
\begin{center}
\small
\begin{tabular}
{l|l|l|l|l}
\hline
\hline
$h$& err($L^\infty$) & eoc($L^\infty$)& err($L^2$)& eoc($L^2$) \\
\hline
\hline
0.03& 0.5& -& 0.808& - \\
0.012& 0.215& 0.92& 0.227& 1.39 \\
0.006& 0.111& 0.96& 0.0836& 1.44 \\
0.004& 0.0747& 0.97& 0.0462& 1.46 \\
0.003& 0.0563& 0.98& 0.0303& 1.47 \\
0.0024& 0.0452& 0.98& 0.0218& 1.48 \\
0.002& 0.0378& 0.98& 0.0166& 1.48 \\
\hline
\end{tabular}
\end{center}
\label{c-tab1}
\end{table}

\subsubsection{Risk Adjusted Pricing Methodology model}

\begin{figure}

\begin{center}
\includegraphics[width=0.45\textwidth]{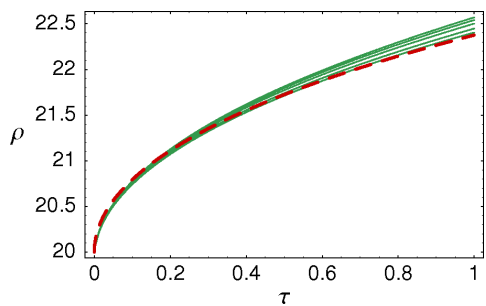}
\end{center}

\caption{\small
A comparison of the free boundary function $\ro^R(\tau)$ computed for the Risk Adjusted Pricing
Methodology model. Dashed red curve represents a solution corresponding to 
$R=0$, whereas the green curves represent a solution $\ro^R(\tau)$ 
for different values of the risk premium coefficients 
$R=5, 15, 40, 70, 100$.
}
\vspace{2mm}
\label{c-fig-rapm}
\end{figure}

In the next example we computed the position of the free boundary $\ro(\tau)$ in the case of the Risk Adjusted Pricing Methodology model - a nonlinear Black--Scholes type model derived  by Janda\v{c}ka and \v{S}ev\v{c}ovi\v{c} in \cite{JS}. In this model the volatility $\sigma$ is a nonlinear function of the asset price $S$ and the second derivative $\partial_S^2 V$ of the  option price, and it is given by  formula (\ref{c-RAPM}). In Fig.~\ref{c-fig-rapm} we present results of numerical approximation of the free boundary position $\ro^R(\tau)=S_f^R(T-\tau)$ in the case when  the coefficient of transaction costs $C=0.01$ is fixed and the risk premium measure $R$  varies from  $R=5, 15, 40, 70,$ up to $R=100$. We compare  the position of the free boundary $\ro^R(\tau)$ to the case when there are no transaction costs and no risk from volatile portfolio, i.e. we compare it with the free boundary position $\ro^0(\tau)$ for the linear Black--Scholes equation (see Fig.~\ref{c-fig-rapm}).  An increase in the  risk premium coefficient $R$ resulted in an increase of the free boundary position  as it can be expected.

\begin{table}
\caption{\small 
Distance  $\Vert \ro^R-\ro^0\Vert_p$ ($p=2,\infty$) of the free boundary position 
$\ro^R$ from the reference free boundary position $\ro^0$ and experimental orders $\alpha_\infty$
and $\alpha_2$ of convergence.}
\smallskip
\begin{center}
\small
\begin{tabular}
{r|l|l|l|l}
\hline
\hline
$R$& $\Vert \ro^R-\ro^0\Vert_\infty$ & $\alpha_\infty$ & $\Vert \ro^R-\ro^0\Vert_2$& $\alpha_2$ \\
\hline
\hline
1 & 0.0601& -& 0.0241& -\\
2 & 0.0754& 0.33& 0.0303& 0.328\\
5 & 0.102& 0.33& 0.0408& 0.326\\
10 & 0.128& 0.33& 0.0511& 0.324\\
15 & 0.145& 0.32& 0.0582& 0.323\\
20 & 0.16& 0.32& 0.0639& 0.322\\
30 & 0.182& 0.32& 0.0727& 0.321\\
40 & 0.2& 0.32& 0.0798& 0.32\\
50 & 0.214& 0.32& 0.0856& 0.319\\
60 & 0.227& 0.32& 0.0907& 0.318\\
70 & 0.239& 0.32& 0.0953& 0.317\\
80 & 0.249& 0.32& 0.0994& 0.317\\
90 & 0.259& 0.32& 0.103& 0.316\\
100 & 0.268& 0.32& 0.107& 0.316\\
\hline
\end{tabular}
\end{center}
\label{c-tab2}
\end{table}

\begin{figure}

\begin{center}
\includegraphics[width=0.45\textwidth]{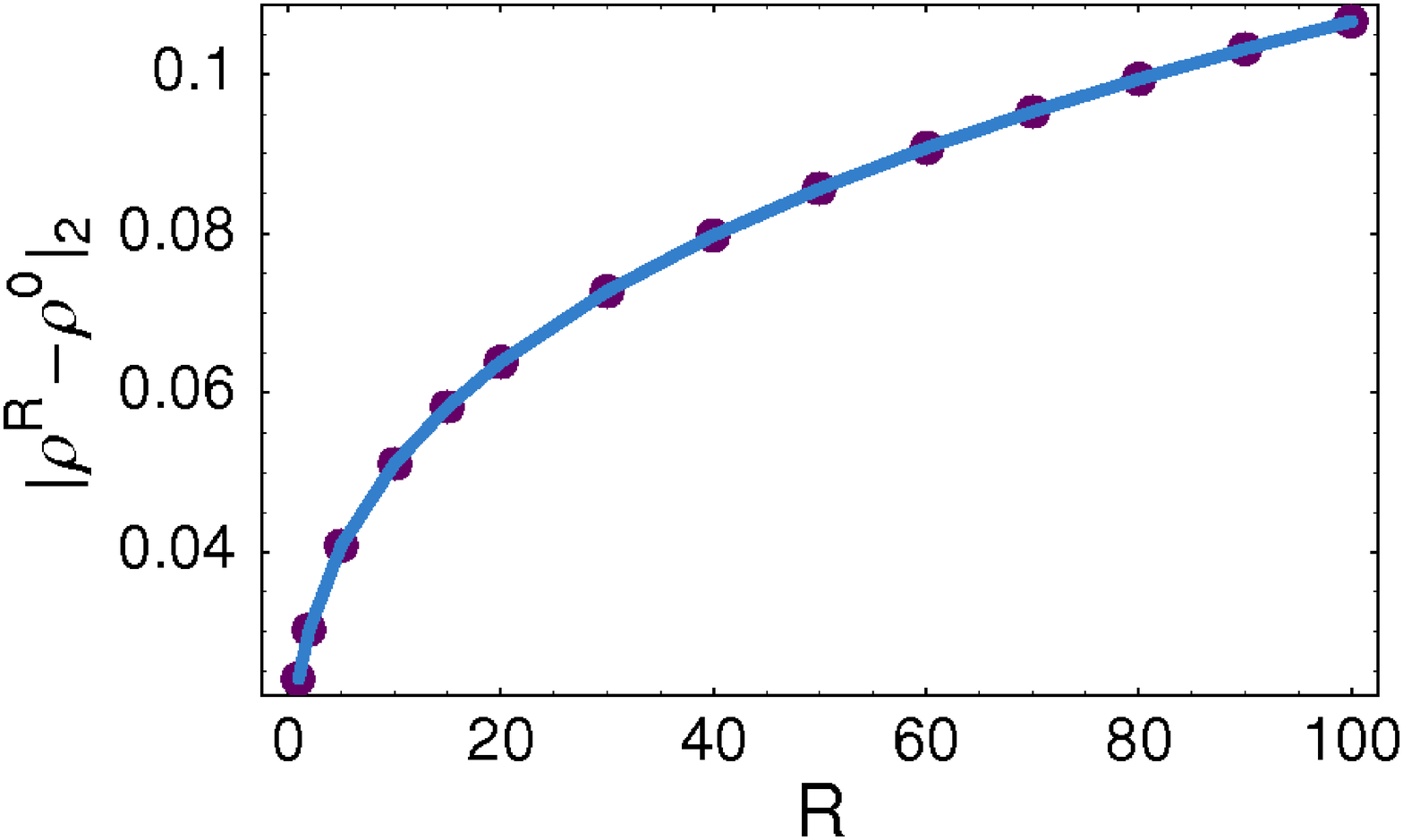}
\includegraphics[width=0.45\textwidth]{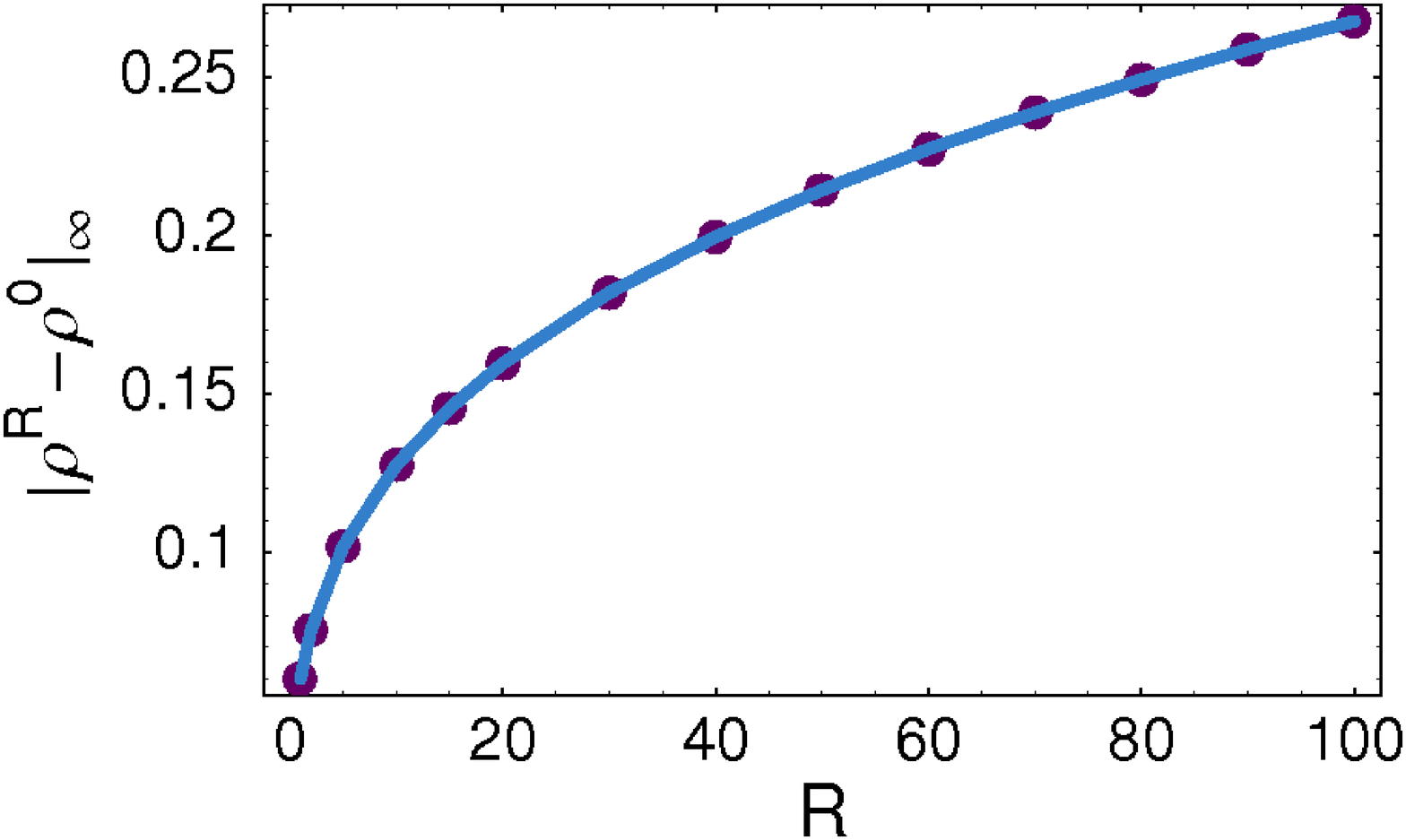}
\end{center}

\caption{\small
Dependence of the norms $\Vert \ro^R -\ro^0\Vert_p$ ($p=\infty, 2$) of the deviation of the free boundary 
$\ro=\ro^R(\tau)$ for the RAPM model on the risk premium coefficient $R$. 
}
\vspace{2mm}
\label{c-fig-rapm-err}
\end{figure}

In Tab.~\ref{c-tab2} and Fig.~\ref{c-fig-rapm}  we summarize results of comparison of the free boundary position $\ro^R$ for various values of the risk premium coefficient to the reference position $\ro=\ro^0$ computed from the Black--Scholes model with a constant volatility $\sigma=\hat\sigma$, i.e. $R=0$. The experimental order $\alpha_p$ of the distance function  $\Vert \ro^R-\ro^0\Vert_p = O(R^{\alpha_p})$ has been computed for $p=2,\infty$,  as follows:
\[
\alpha_p = \frac{\ln(\Vert\ro^{R_i} -\ro^0\Vert_p) - \ln(\Vert\ro^{R_{i-1}} -\ro^0\Vert_p)}{\ln R_i - \ln R_{i-1}}\,.
\]
According to the values presented in Tab.~\ref{c-tab2} it turns out that a reasonable conjecture on the order of convergence is that $\Vert \ro^R-\ro^0\Vert_p = O(R^{1/3})$  for both norms $p=2$ and $p=\infty$. Since the transaction cost coefficient $C$ and risk premium measure $R$ enter the expression for the RAPM volatility (\ref{c-RAPM}) only in the product $C^2 R$ we can conjecture that $\Vert \ro^{R,C}-\ro^{0,0}\Vert_p = O(C^{2/3} R^{1/3})$ as either $C\to 0^+$ or $R\to 0^+$.

\subsubsection{Barles and Soner model}

\begin{figure}

\begin{center}
\includegraphics[width=0.45\textwidth]{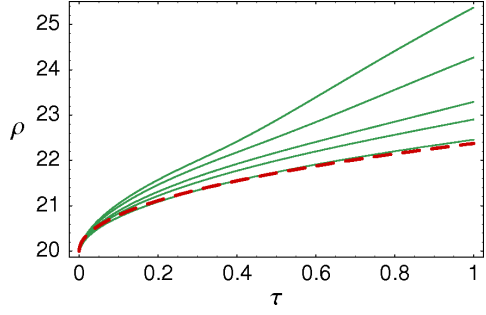}
\end{center}

\caption{\small
A comparison of the free boundary function $\ro(\tau)$ computed for the Barles and Soner
model. Dashed red curve represents a solution corresponding to 
$R=0$, whereas the green curves represents a solution $\ro(\tau)$ 
for different values of the risk aversion coefficient 
$a=0.01, 0.07, 0.13, 0.25, 0.35$.
}
\vspace{2mm}
\label{c-fig-barles}
\end{figure}

\begin{figure}

\begin{center}
\includegraphics[width=0.45\textwidth]{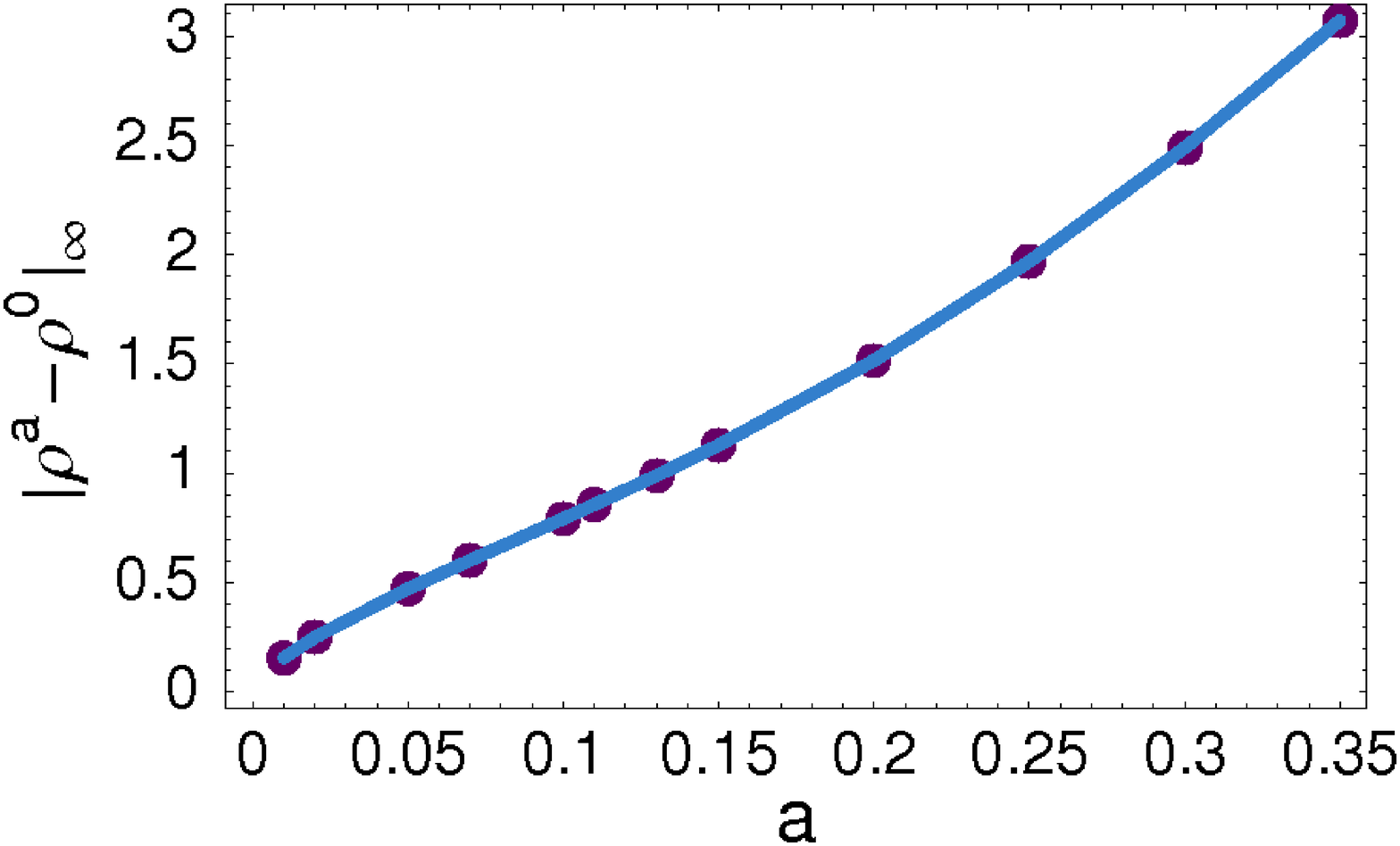}
\includegraphics[width=0.45\textwidth]{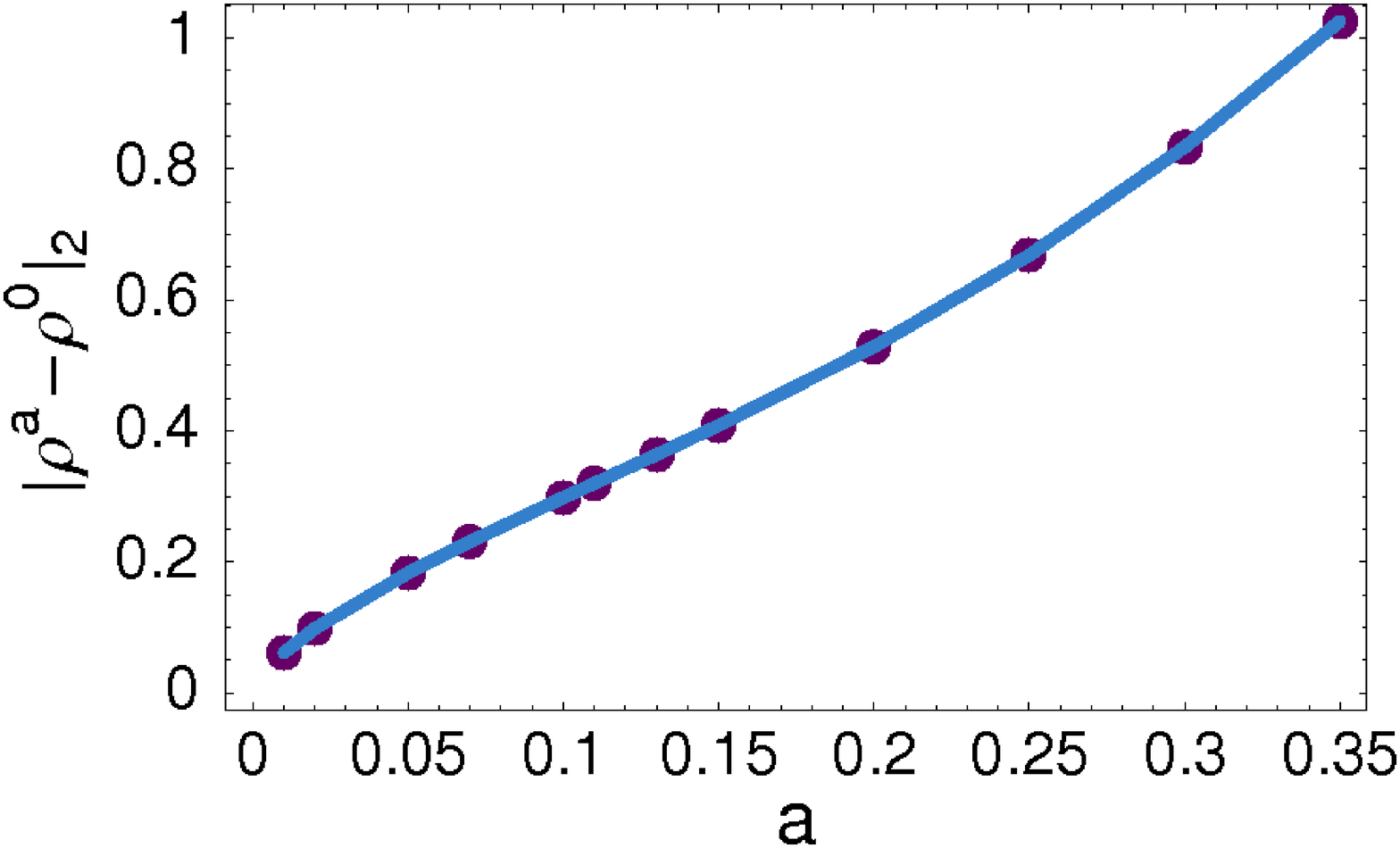}
\end{center}

\caption{\small
Dependence of the norms $\Vert \ro^a -\ro^0\Vert_p$ ($p=\infty, 2$) of the deviation of the free boundary 
$\ro=\ro^a(\tau)$ for the Barles-Soner model on the risk aversion parameter $a$. 
}
\vspace{2mm}
\label{c-fig-barles-err}
\end{figure}

\begin{table}
\caption{\small 
Distance  $\Vert \ro^a-\ro^0\Vert_p$ ($p=2,\infty$) of the free boundary position 
$\ro^a$ from the reference free boundary position $\ro^0$ and experimental orders $\alpha_\infty$
and $\alpha_2$ of convergence.}
\smallskip
\begin{center}
\small
\begin{tabular}
{l|l|l|l|l}
\hline
\hline
$a$& $\Vert \ro^a-\ro^0\Vert_\infty$ & $\alpha_\infty$ & $\Vert \ro^a-\ro^0\Vert_2$& $\alpha_2$ \\
\hline
\hline
0.01& 0.156& -& 0.0615& -\\
0.02& 0.25& 0.68& 0.0985& 0.68\\
0.05& 0.472& 0.69& 0.184& 0.679\\
0.07& 0.602& 0.72& 0.232& 0.69\\
0.1& 0.793& 0.77& 0.298& 0.712\\
0.11& 0.857& 0.82& 0.32& 0.74\\
0.13& 0.99& 0.86& 0.364& 0.766\\
0.15& 1.13& 0.92& 0.409& 0.807\\
0.2& 1.52& 1.& 0.529& 0.897\\
0.25& 1.97& 1.2& 0.669& 1.05\\
0.3& 2.49& 1.3& 0.833& 1.21\\
0.35& 3.07& 1.4& 1.03& 1.35\\
\hline
\end{tabular}
\end{center}
\label{c-tab3}
\end{table}

Our next example is devoted to the nonlinear Black--Scholes model due to Barles and Soner  (see \cite{BaSo}). In this model the volatility is given by equation (\ref{c-barles}).  Numerical results are depicted in  Fig.~\ref{c-fig-barles}. Choosing a larger value of the risk aversion  coefficient $a>0$ resulted in increase of the free boundary position $\ro^a(\tau)$. The position of the early exercise boundary $\ro^a(\tau)$ has  considerably increased in comparison to the linear Black--Scholes equation with constant volatility $\sigma=\hat\sigma$. In contrast to the case of constant volatility as well as the RAPM model, there is, at least a numerical evidence (see Fig.\ref{c-fig-barles} and $\ro^a$ for the largest value  $a=0.35$) that  the free boundary profile $\ro^a(\tau)$ need not be necessarily convex. Recall that that convexity of the free boundary profile has been proved analytically by Ekstr\"om  {\em et al.} and Chen  {\em et al.} in a recent papers \cite{CC,E,ET} in the case of a American put option and constant  volatility $\sigma=\hat\sigma$.

Similarly as in the previous model we also investigated  the dependence of the free boundary 
position $\ro=\ro^a(\tau)$ on the risk aversion parameter $a>0$. 
In Tab.~\ref{c-tab3} and Fig.~\ref{c-fig-barles-err} we present results of comparison of the free boundary position $\ro^a$ for various values
of the risk aversion coefficient $a$ to the reference
position $\ro=\ro^0$. Inspecting values $\alpha_p$ of the order of distance $\Vert \ro^a-\ro^0\Vert_p$ 
it can be conjectured that $\Vert \ro^a-\ro^0\Vert_p = O(a^{2/3})$ as $a\to 0$
for both norms $p=2$ and $p=\infty$.

\section{Transformation methods for Asian call options}

Path dependent options are options whose pay-off diagram depends on the path history of the underlying asset. Among path dependent options Asian options plays an important role as they are quite common in currency and commodity markets like e.g. oil industry (cf. \cite{HJ,DK}). Asian options may depend on the averaged path history in several ways. We shall restrict our attention to the so-called floating strike Asian call options. The floating strike price is assumed to be an arithmetic average of the underlying asset prices over the entire time interval $[0,T]$ where $T$ is the expiration time. 

Let us define the arithmetic average $A=A_t$ of the underlying asset $S=S_t$ by
\[
A_t = \frac{1}{t} \int_0^t S_\tau\, d\tau\,.
\]
For the case of the so-called Asian floating strike call option the pay-off diagram at expiry $T$  reads as follows: $V(S,A,T) = \max( S-A, 0)$. It means the price $V$ of an option contract will depend not only on the underlying asset price $S$, time $t,$ but also on the underlying asset path average $A$, i.e. $V=V(S,A,t)$.

\subsection{Governing equations for Asian options}

As it is usual in the option pricing theory, we shall describe the asset price dynamics by  a geometric Brownian with drift $\varrho$, dividend yield $q\ge 0$ and volatility $\sigma$, i.e. $d S = (\varrho -q)S dt + \sigma S dW$ where $W$ is the standard Wiener process. If we apply It\^o's formula to the function $V=V(S,A,t)$  we obtain
\[
d V = \left(
   \frac{\partial V}{\partial t}
+\frac{\sigma^2}{2}S^2 \frac{\partial^2 V}{\partial S^2}
\right)
+
\frac{\partial V}{\partial S} dS
+
\frac{\partial V}{\partial A} dA\,.
\]
In the case of arithmetic averaging we have $dA = t^{-1} (S - A) dt$. Hence the differential $dA$ is of the order of $dt$ and this is why the above expression for $dV$ indeed represents its lowest order approximation when taking into account stochastic character of the dynamics of the asset price $S$. Therefore, following standard arguments from the Black--Scholes theory one can derive the governing equation for pricing Asian option with arithmetic averaging in the form:
\begin{equation}
\frac{\partial V}{\partial t}
+\frac{\sigma^2}{2}S^2 \frac{\partial^2 V}{\partial S^2}
+S(r-q)\frac{\partial V}{\partial S}
+\frac{S-A}{t}\frac{\partial V}{\partial A} - r V = 0, 
\label{asian-eq}
\end{equation}
where $0<t<T, \ S, A>0$ (see e.g. \cite{DK}). For Asian call option the above equation is subject to the terminal pay-off condition
\begin{equation}
V(S,A,T) = \max(S- A,0), \quad S,A>0\,.
\label{asian-payoff}
\end{equation}
It is well known (see e.g. \cite{Kw,DK}) that for Asian options with floating strike we can achieve dimension reduction by introducing the following similarity variable:
\[
x=\frac{S}{A}, \qquad W(x,\tau) = \frac{1}{A} V(S,A,t)
\]
where $\tau=T-t$. It is straightforward to verify that $V(S,A,t) =  W(S/A, T-t) A $ is a solution of (\ref{asian-eq}) iff $W=W(x,\tau)$ is a solution to the following parabolic PDE:
\begin{equation}
\frac{\partial W}{\partial \tau}
-\frac{\sigma^2}{2}x^2 \frac{\partial^2 W}{\partial x^2}
-(r-q)x\frac{\partial W}{\partial x}
-\frac{x-1}{T-\tau}\left( W-x\frac{\partial W}{\partial x}\right)+rW=0, 
\label{asian-redeq}
\end{equation}
where $x>0$ and $0<\tau<T$. The initial condition for $W$ immediately follows from the terminal pay-off diagram for the call option,
\begin{equation}
W(x,0) = \max(x-1,0).
\label{asian-init}
\end{equation}

\subsection{American style of Asian options}

Following Dai and Kwok \cite{DK}, American style of Asian options is characterized by the exercise region 
\[
{\mathcal E} = \{ (S,A,t) \in [0,\infty)\times[0,\infty)\times[0,T),\ 
V(S,A,t) = V(S,A,T)\}.
\]
In the case of a call option this region can be described by an early exercise boundary function $S_f=S_f(A,t)$ such that 
\[
{\mathcal E} = \{ (S,A,t) \in [0,\infty)\times[0,\infty)\times[0,T),\ 
S\ge S_f(A,t)\}.
\]
For American style of an Asian call option we have to impose a homogeneous Dirichlet boundary condition $V(0,A,t)=0$ at $S=0$. According to \cite{DK} the $C^1$ continuity condition at the point  $(S_f(A,t),A, t)$ of a contact of a solution $V$ with its pay-off diagram implies the following boundary condition at the free boundary position $S_f(A,t)$: 
\begin{equation}
\frac{\partial  V}{\partial S}(S_f(A,t),A,t)=1,\quad 
V(S_f(A, t),A,t)=S_f(A, t) - A
\label{asian-bc}
\end{equation}
for any $A>0$ and $0<t<T$. It is important to emphasize that the free boundary function $S_f$ can be also reduced to a function of one variable by introducing a new state function $x_f(t)$ as follows:
\[
S_f(A,t) = A x_f(t).
\]
The function  $x_f=x_f(t)$ is a free boundary function for the transformed state variable $x=S/A$. For American style of Asian call options the spatial domain for the reduced equation (\ref{asian-redeq}) is given by
\[
0<x<\ro(\tau),\ \ \tau\in(0,T),\quad \hbox{where}\ \ro(\tau)=x_f(T-\tau)\,.
\]
Taking into account boundary conditions (\ref{asian-bc}) for the option price $V$ we end up with corresponding boundary conditions for the function $W$:
\begin{equation}
W(0,\tau) =0, \qquad W(x,\tau)=x-1,\ \ \frac{\partial W}{\partial x} (x,\tau) =1 \ \ \hbox{at}\ \ x=\ro(\tau)
\label{asian-redbc}
\end{equation}
for any $0<\tau<T$ and the initial condition 
\begin{equation}
W(x,0) =\max(x-1,0)
\label{asian-redic}
\end{equation}
for any $x>0$. 

\subsection{Fixed domain transformation for American style of Asian call options}

Similarly as in Section 3, in order to apply the fixed domain transformation for the  free boundary problem (\ref{asian-redeq}), (\ref{asian-redbc}), (\ref{asian-redic}) we introduce a new variable $\xi$ and an auxiliary function $\Pi=\Pi(\xi,\tau)$  (again representing a  synthetic portfolio) defined as follows:
\begin{equation}
\xi = \ln\left(\frac{\ro(\tau)}{x}\right),\qquad 
\Pi(\xi,\tau)=W(x,\tau)-x\frac{\partial W}{\partial x}\,.
\label{asian-portf}
\end{equation}
Clearly, $x\in (0,\ro(\tau))$ iff $\xi\in (0,\infty)$ for $\tau\in(0,T)$. The value $\xi=\infty$ of the transformed variable corresponds to the value $x=0$ ($S=0$) expressed in the original variable. On the other hand, the value $\xi=0$ corresponds to the free boundary position $x=\ro(\tau)$ ($S=A S_f(A,t)$). 

A straightforward calculation similar to that of Section 4 enables us to to justify that the function $\Pi=\Pi(\xi,\tau)$ is a solution to the following parabolic PDE
\begin{equation}
\frac{\partial \Pi}{\partial \tau}
+a(\xi,\tau)\frac{\partial \Pi}{\partial \xi} 
-\frac{\sigma^2}{2}\frac{\partial^2 \Pi}{\partial \xi^2}
+\left(r+\frac{1}{T-\tau}\right)\Pi=0, 
\label{asian-fixedeq} 
\end{equation}
where the term $a(\xi,\tau)$ is given by
\begin{equation}
a(\xi,\tau)=
\frac{\dot{\rho}(\tau)}{\rho(\tau)}
+ r - q 
-\frac{\sigma^2}{2}-\frac{\rho e^{-\xi}-1}{T-\tau}.
\end{equation}
Notice the spatial dependence of the coefficient $a=a(\xi,\tau)$ in comparison to the case of transformed equations for a linear or nonlinear Black--Scholes equations (see (\ref{euro-2.9}), (\ref{c-bfun})). 
The initial condition for the solution $\Pi$ follows from (\ref{asian-redic})
\[
\Pi(\xi,0)    =  \left\{ \begin{array}{ll}
        -1 & \mbox{$\xi<\ln \rho(0)$}, \\
        0 & \mbox{$\xi>\ln \rho(0)$}.
        \end{array} \right.
\]
Since $\partial_x W(x,\tau) = 1$ for $x=\ro(\tau)$ and $W(0,\tau) = 0$ we  conclude  the Dirichlet boundary conditions for the function $\Pi$
\[
\Pi(0,\tau) = - 1, \qquad \Pi(\infty,\tau)=0.
\]
It remains to determine an algebraic constraint between the free boundary function $\ro(\tau)$ and the solution $\Pi$. Similarly as in the case of a linear or nonlinear Black--Scholes equation we obtain, by differentiation the condition $W(\rho(\tau),\tau)=\rho(\tau)-1$ with respect to $\tau,$ the following identity:
\[
 \frac{d}{d\tau}\rho(\tau)=\frac{\partial W}{\partial x}(\rho(\tau),\tau)\frac{d}{d\tau}\rho(\tau)+\frac{\partial W}{\partial \tau}(\rho(\tau),\tau).
\]
Since $\frac{\partial W}{\partial x}(\rho(\tau),\tau)=1$ we have 
$\frac{\partial W}{\partial \tau}(x ,\tau)=0$ at $x=\ro(\tau)$. 
Assuming continuity of the function  $\Pi$ and its derivative $\Pi_\xi$ up to the boundary $\xi=0$ we have
\[
x^2\frac{\partial^2 W}{\partial x^2}(x,\tau)
\to
\frac{\partial \Pi}{\partial \xi}(0,\tau),
\quad 
x\frac{\partial W}{\partial x}(x,\tau)\to\rho(\tau)
\]
as $x\to \rho(\tau)$. Passing to the limit $x\to \rho(\tau)$ in equation (\ref{asian-redeq}) we end up with the equation
\[
-(r-q)\rho(\tau)
-\frac{\sigma^2}{2}\frac{\partial \Pi}{\partial        \xi}(0,\tau)
+\frac{\rho(\tau)-1}{T-\tau}+r[\rho(\tau)-1]=0.
\]
It yields an algebraic nonlocal expression for the free boundary position $\ro(\tau)$
\begin{equation}
\rho(\tau)
=\frac{r + \frac{1}{T-\tau} + \frac{\sigma^2}{2}\frac{\partial \Pi}{\partial        \xi}(0,\tau)}{q+\frac{1}{T-\tau}}.
\end{equation}
Next we determine the starting point of the free boundary function $\ro(0)$. It means that we have to find the terminal value of the original state variable $x_f(T)$. 
We shall assume a structural assumption 
\begin{equation}
r>q\ge0
\label{asian-rq}
\end{equation}
on the interest and dividend rates $r,q$.
If $\ro(0)>1$ then $\ln\ro(0) >0$ and this is why the initial function $\Pi(\xi,0)$ is equal to $-1$ in some right neighborhood of $\xi=0$. Thus $\partial_\xi \Pi(\xi,0) =0$ for $0<\xi \ll 1$. Again, assuming continuity of $\partial_\xi\Pi(\xi,\tau)$ at $(\xi,\tau)=(0,0)$ we obtain
\[
\lim_{\tau\rightarrow 0^+}
\frac{\partial \Pi}{\partial \xi}(0,\tau)
=\lim_{\tau\rightarrow 0^+,x\rightarrow 0^+}
\frac{\partial \Pi}{\partial \xi}(x,\tau)
=\lim_{x\rightarrow 0^+}\frac{\partial \Pi}{\partial \xi}(x,0)=0.
\]
As a consequence we can conclude
\[
\rho(0)=\frac{r+\frac{1}{T}}{q+\frac{1}{T}}=\frac{1+rT}{1+qT}>1.
\]
This initial condition for $\ro(0)$ is exactly the same as the one derived recently by Dai and Kwok in \cite{DK}. They proved, for a general choice of $r,q\ge 0$ that the initial condition for the function $\ro$ is given by
\[
\ro(0)=\max\left(\frac{1+r T}{1+q T},1\right).
\]
In summary, we have transformed the free boundary problem for pricing American style of Asian call option with floating strike price into the following nonlocal parabolic PDE with an algebraic constraint
\begin{eqnarray}
\label{asian-system}
&&\frac{\partial \Pi}{\partial
        \tau}+a(\xi,\tau)\frac{\partial \Pi}{\partial \xi}
        -\frac{\sigma^2}{2}\frac{\partial^2 \Pi}{\partial
        \xi^2}+\left(r+\frac{1}{T-\tau}\right)\Pi=0,\quad 0< \tau <T,\ 
        \xi>0,\nonumber \\
&& \rho(\tau) = \frac{1 + r (T-\tau) + \frac{\sigma^2}{2}(T-\tau)\frac{\partial \Pi}{\partial        \xi}(0,\tau)}{1 + q(T-\tau)}, \ \ 0<\tau<T, \nonumber \\
&&\hskip-1truecm\hbox{subject to initial and boundary conditions}\nonumber \\
&& \Pi(0,\tau) =-1, \qquad \Pi(\infty,\tau)=0, \nonumber \\
&& \Pi(\xi,0)   =  \left\{ \begin{array}{ll}
        -1 & \mbox{$\xi<\ln((1+r T)/(1+ q T))$}, \\
        0 & \mbox{$\xi>\ln((1+r T)/(1+ q T))$},
        \end{array} \right. \\
&&\ro(0) = (1+r T)/(1+ q T),  \nonumber \\
&&\hskip-1truecm\hbox{where} \nonumber \\
&& a(\xi,\tau) = \frac{\dot{\rho}(\tau)}{\rho(\tau)}+ r-q -\frac{\sigma^2}{2}-\frac{\rho e^{-\xi}-1}{T-\tau}.
\nonumber
\end{eqnarray}

\subsection{An approximation scheme for pricing American style of Asian options}

Similarly as in the case of a nonlinear Black--Scholes equation (see Section 4) we restrict the spatial domain $\xi\in (0,\infty)$ to a finite interval of values $\xi\in (0,L)$ where $L>0$ is sufficiently large. Let $k>0$ denote by the time step, $k=T/m$, and, by $h>0$ the spatial step, $h=L/n$ where $m, n\in N$ again stand for the number of time and space discretization steps, resp. 
We denote by $\Pi_i^j$ an approximation of $\Pi( \xi_i, \tau_j)$, $\ro^j \approx \ro(\tau_j)$ where $\xi_i = i h$ and $\tau_j = j k$. 
Then for the Euler backward in time finite difference approximation of equation (\ref{c-tBS}) we have
\begin{equation}
\frac{\Pi^j-\Pi^{j-1}}{k} +  b^j \frac{\partial \Pi^j}{\partial \xi} 
- \left( \frac{\sigma^2}{2} + \frac{\rho^j e^{-\xi}-1}{T- \tau_j} \right) \frac{\partial \Pi^j}{\partial \xi}
-\frac{\sigma^2}{2} \frac{\partial^2 \Pi^j}{\partial^2 \xi} 
+\left(r +\frac{1}{T- \tau_j}\right) \Pi^j = 0
\label{asian-timediscret}
\end{equation}
where $b^j$ is an approximation of the value $b(\tau_j)$ where the function $b(\tau)$ is defined as in (\ref{c-bfun}), i.e. $b(\tau)= {\dot\ro(\tau)\over\ro(\tau)} + r - q $. The solution $\Pi^j=\Pi^j(x)$ is subject to Dirichlet boundary conditions at $\xi=0$  and $\xi=L$. We set $\Pi^0(\xi)=\Pi(\xi,0)$ (see (\ref{asian-system}). Again we split  the above problem into  a convection part and a diffusive part by introducing  an auxiliary intermediate step $\Pi^{j-\onehalf}$:

({\it Convective part})
\begin{equation}
\frac{\Pi^{j-\onehalf}-\Pi^{j-1}}{k} + b^j  \partial_x\Pi^{j-\onehalf} = 0\,,
\label{asian-convective}
\end{equation}

({\it Diffusive part})
\begin{equation}
\frac{\Pi^{j}-\Pi^{j-\onehalf}}{k}  
- \left( \frac{\sigma^2}{2} + \frac{\rho^j e^{-\xi}-1}{T- \tau_j} \right) \frac{\partial \Pi^j}{\partial \xi}
-\frac{\sigma^2}{2} \frac{\partial^2 \Pi^j}{\partial^2 \xi} 
+\left(r +\frac{1}{T- \tau_j}\right) \Pi^j = 0.
\label{asian-diffusion}
\end{equation}
The convective part can be approximated by an explicit solution to the transport equation $\partial_\tau\tilde\Pi + b(\tau) \partial_\xi\tilde\Pi =0$ for $\xi>0$ and $\tau\in(\tau_{j-1},\tau_j]$ subject to the boundary condition $\tilde\Pi(0,\tau) =-1$ and the initial condition $\tilde\Pi(\xi,\tau_{j-1})=\Pi^{j-1}(\xi)$. In contrast to classical plain vanilla call or put options the free boundary function $\ro(\tau)$ need not be monotonically increasing (see e.g. \cite{DK} or \cite{HJ}). Therefore depending on whether the term $b(\tau)= {\dot\ro(\tau)\over\ro(\tau)} + r - q $ is positive or negative the boundary condition $\tilde\Pi(0,\tau) =-1$ at $\xi=0$ is either in-flowing ($b>0$)  or out-flowing ($b<0$). It means that the boundary condition $\Pi(0,\tau) = -1$ can be prescribed only if $b(\tau)\ge 0$. Let us denote by $B(\tau)$ the primitive function to $b(\tau)$, i.e. $B(\tau)=\ln \ro(\tau) + (r-q)\tau$.
Solving the equation  $\partial_\tau\tilde\Pi + b(\tau) \partial_\xi\tilde\Pi =0$ we obtain:
$\tilde\Pi(\xi,\tau)=\Pi^{j-1}(\xi-B(\tau)+B(\tau_{j-1}))$ if $\xi-B(\tau)+B(\tau_{j-1})>0$ and  $\tilde\Pi(\xi,\tau)=-1$ otherwise. Hence the full time-space approximation of the  half-step solution $\Pi^{j-\onehalf}_i$ can be obtained from the formula
\begin{equation}
\Pi^{j-\onehalf}_i=\left\{ 
\begin{matrix}
\Pi^{j-1}(\eta_i) \hfill & \quad \hbox{if } \eta_i=\xi_i- \ln\ro_j + \ln\ro_{j-1} - (r-q)k>0\,, \hfill\cr 
-1 \hfill & \quad \hbox{otherwise.} \hfill
\end{matrix}
\right.
\label{asian-convective-discrete}
\end{equation}
In order to compute the value $\Pi^{j-1}(\eta_i)$ we make use of a linear approximation between discrete values $\Pi^{j-1}_i, i=0,1, ..., n$.

Using central finite differences for approximation of the derivative $\partial_x\Pi^j$ we can approximate the diffusive part of a solution (\ref{asian-diffusion}) as follows:
\begin{eqnarray}
\frac{\Pi_i^j - \Pi_i^{j-\onehalf}}{k} &+& \left(r + \frac{1}{T- \tau_j}\right) \Pi_i^j
 \\
&-& \left( \frac{\sigma^2}{2}  + \frac{\rho^j e^{-\xi_i}-1}{T- \tau_j} \right)
\frac{\Pi_{i+1}^j-\Pi_{i-1}^j}{2h}
- 
\frac{\sigma^2}{2} 
\frac{ \Pi_{i+1}^j - 2 \Pi_{i}^j + \Pi_{i-1}^j}{h^2}
 =0\,.\nonumber
\end{eqnarray}
Therefore the vector of discrete values $\Pi^j=\{\Pi_i^j, i=1,2, ..., n\}$ at the time level  $j\in\{1,2, ..., m\}$ is a solution to a tridiagonal system of equations
\begin{equation}
\alpha_i^j \Pi_{i-1}^j + \beta_i^j \Pi_{i}^j + \gamma_i^j \Pi_{i+1}^j = \Pi_i^{j-\onehalf}
\label{asian-eq-tridiag}
\end{equation}
for $i=1,2, ..., n,$  where
\begin{eqnarray}
\label{asian-abc}
\alpha_i^j &\equiv& \alpha_i^j(\ro^j)
= 
- \frac{k}{2h^2} \sigma^2
+ \frac{k}{2h} \left( \frac{\sigma^2}{2} + \frac{\rho^j e^{-\xi_i}-1}{T- \tau_j}\right),
\nonumber \\
\gamma_i^j &\equiv& \gamma_i^j(\ro^j) = 
- \frac{k}{2h^2} \sigma^2
- \frac{k}{2h} \left( \frac{\sigma^2}{2} + \frac{\rho^j e^{-\xi_i}-1}{T- \tau_j}\right),
\\
\beta_i^j &\equiv& \beta_i^j(\ro^j)=  1 
+ \left(r + \frac{1}{T-\tau_j}\right) k - (\alpha_i^j +\gamma_i^j)\,.
\nonumber
\end{eqnarray}
The initial and boundary conditions at $\tau=0$ and $x=0, L,$ can be approximated as
follows:
\[
\Pi_i^0 = \left\{
\begin{array}{lll}
-1 & \ \ \hbox{for} \ \xi_i <\ln\left((1+r T)/(1+ q T)\right),\hfil
\\
\ \ \ 0 & \ \ \hbox{for} \ \xi_i \ge \ln\left({(1+r T)/(1+ q T)}\right), \hfil
\end{array}
\right.
\]
for $i=0,1, ..., n,$ and $\Pi_0^j = -E, \quad \Pi_n^j = 0$.
The equation for the free boundary position $\ro$ can be approximated by means of a finite difference approximation of $\partial_x\Pi$ at 
the origin $\xi=0$ as follows:
\begin{equation}
\ro^j 
= \frac{ 1  +  r(T-\tau_j)  + (T-\tau_j)\frac{\sigma^2}{2} \frac{\Pi_1^j-\Pi_0^j}{h}}{ 1+ q (T-\tau_j)}.
\label{asian-eq-ro}
\end{equation}
We formally rewrite discrete equations (\ref{asian-convective-discrete}), (\ref{asian-eq-tridiag}) and (\ref{asian-eq-ro}) in the operator form:
\begin{eqnarray}
\ro^j &&= {\mathcal F}(\Pi^j),\nonumber\\
\Pi^{j-\onehalf} &&={\mathcal T}(\Pi^j, \ro^j),\label{asian-abstract}\\
{\mathcal A}(\ro^j) \Pi^j &&= \Pi^{j-\onehalf},\nonumber
\end{eqnarray}
where 
${\mathcal F}(\Pi^j)$ is the right-hand side of the algebraic equation (\ref{asian-eq-ro}),
${\mathcal T}(\Pi^j, \ro^j)$ is the transport equation solver given by 
the right-hand side of (\ref{asian-convective-discrete}) and
${\mathcal A}={\mathcal A}(\ro^j)$ is a tridiagonal matrix with coefficients given 
by (\ref{asian-abc}). The system (\ref{asian-abstract}) can be approximately solved by means of successive iterates 
procedure. We define, for $j\ge 1,$  $\Pi^{j,0} = \Pi^{j-1}, \ro^{j,0} = \ro^{j-1}$. Then the 
$(p+1)$-th approximation of $\Pi^j$ and $\ro^j$ is obtained as a solution to the system:
\begin{eqnarray}
\ro^{j,p+1} &&= {\mathcal F}(\Pi^{j,p}),\nonumber\\
\Pi^{j-\onehalf, p+1} &&={\mathcal T}(\Pi^{j,p})\ro^{j,p+1}),\label{asian-abstract-iter}\\
{\mathcal A}( \ro^{j,p+1}) \Pi^{j,p+1} &&= \Pi^{j-\onehalf, p+1}\,.\nonumber
\end{eqnarray}
Now, if the sequence of approximate discretized solutions $\{(\Pi^{j,p}, \ro^{j,p}) \}_{p=1}^\infty$ converges to 
some limiting value $(\Pi^{j,\infty}, \ro^{j,\infty})$ as $p\to\infty$ then this limit is a solution to a nonlinear  system of  equations (\ref{asian-abstract}) at the time level $j$ and we can  proceed by computing  the approximate solution in the next time level $j+1$.

\subsection{Computational examples of the free boundary approximation}

We end this section with several computational examples documenting the capability of the new method for valuing early exercise boundary for American style of Asian call options with arithmetically averaged  floating strike.

\begin{figure}
\begin{center}
\includegraphics[width=0.45\textwidth]{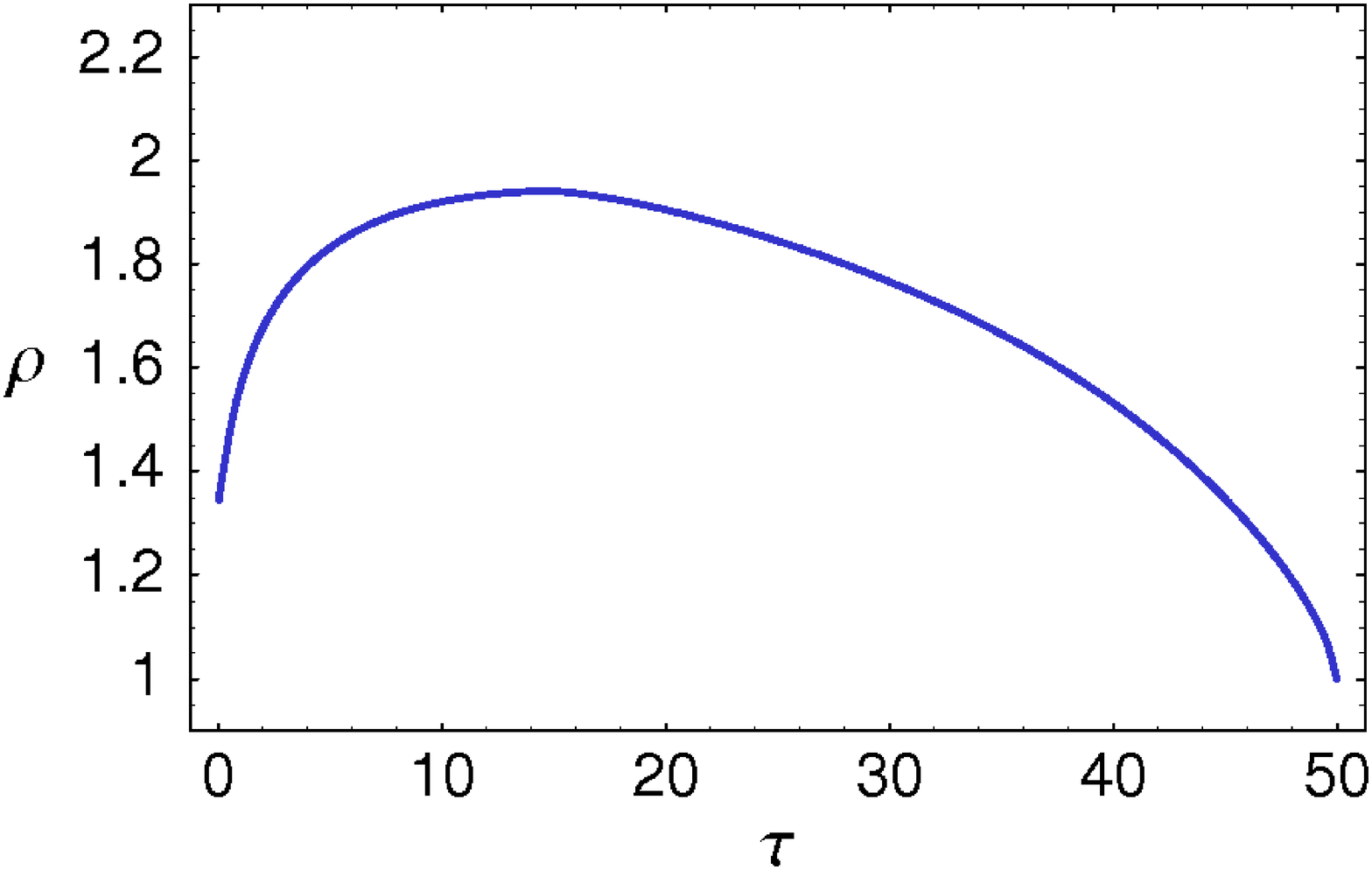}
\smallskip

\includegraphics[width=0.45\textwidth]{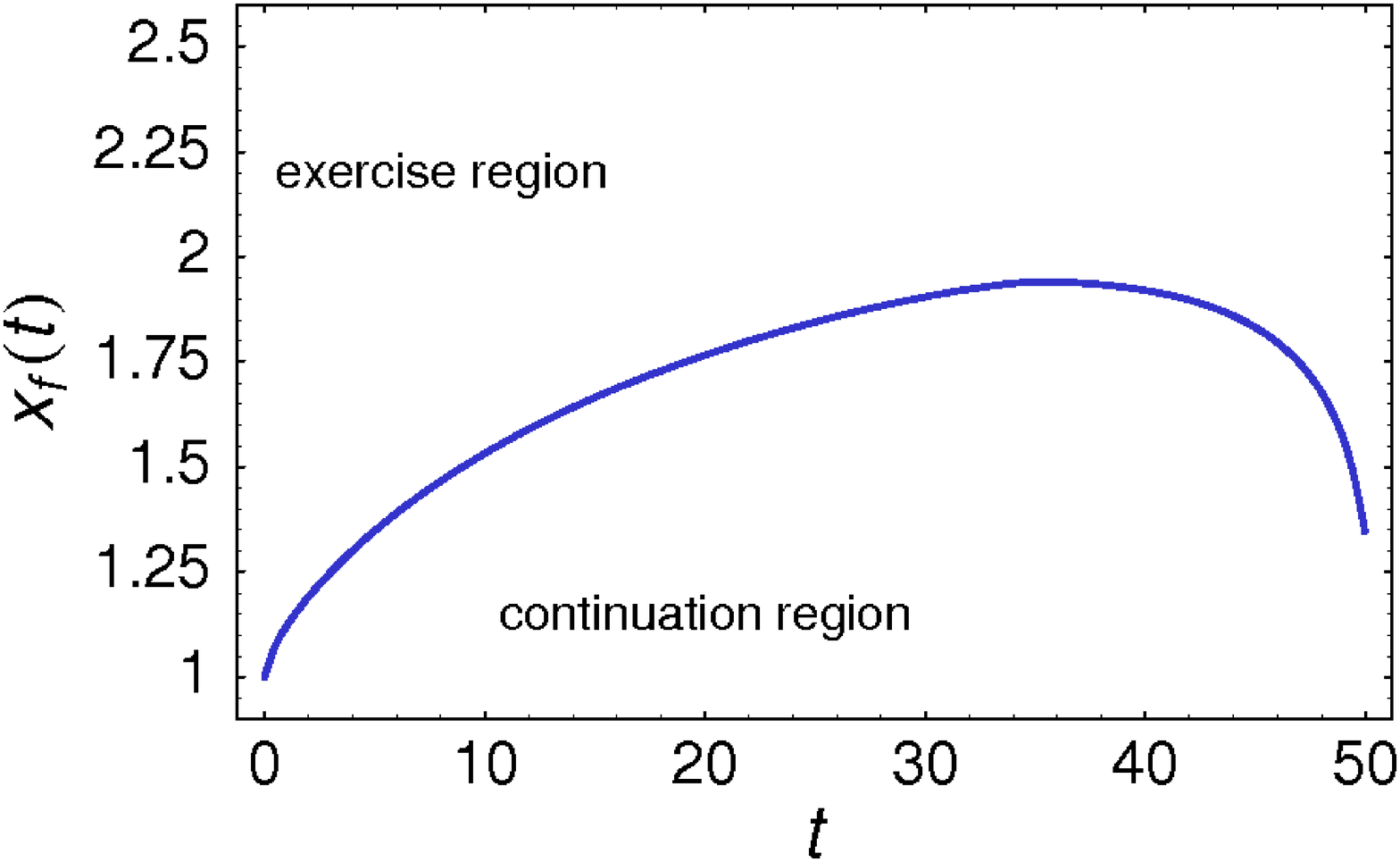}
\includegraphics[width=0.45\textwidth]{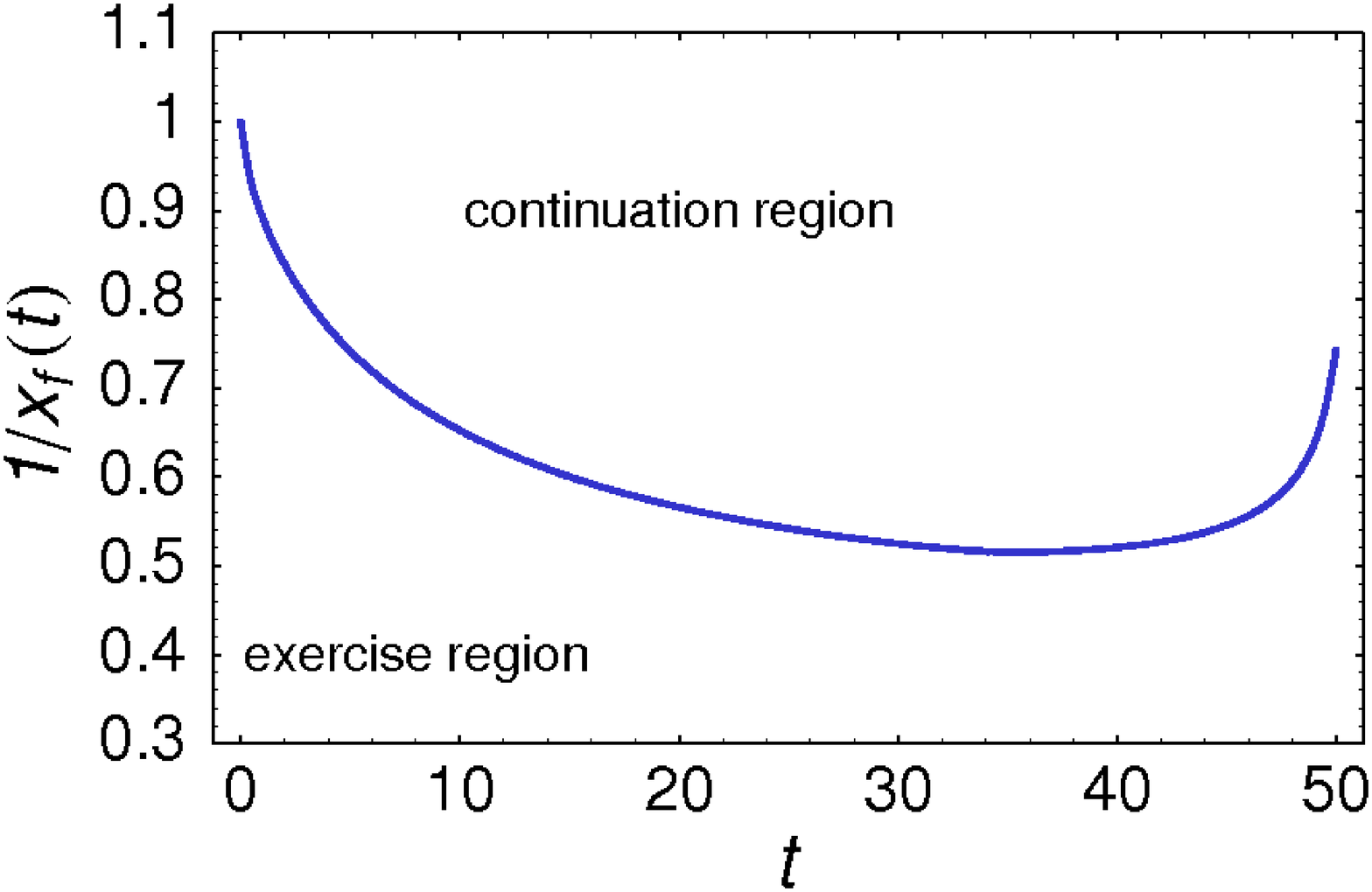}
\end{center}
\caption{\small 
The function $\varrho(\tau)$ (above) 
and the free boundary position $x_f(t)=\varrho(T-t)$ (bottom-right)
and the plot of the function $1/x_f(t)=1/\varrho(T-t)$ (bottom-left).}
\label{asian-fig-1}
\end{figure}

In Fig.~\ref{asian-fig-1} we show the behavior of the early exercise boundary function $\ro(\tau)$ and the function $x_f(t)=\ro(T-t)$. In these numerical experiments we chose $r = 0.06, q= 0.04, \sigma =0.2$ and very long expiration time $T=50$ years. These parameters correspond to the example presented in the preprint version of the paper \cite{DK} by Dai and Kwok. As far as other numerical parameters are concerned, we chose the mesh of $n=100$ spatial grid points and we considered the  number of time steps $m = 2 \times 10^6$ in other to achieve very fine time stepping corresponding to 13 minutes between consecutive time steps when expressed in the original time scale of the problem.
In order to make a graphical comparison of the state variable $x=S/A$ to its reciprocal value $A/S$ considered in \cite{DK} we also plot the function $1/x_f(t)$. 

\begin{figure}
\begin{center}
\includegraphics[width=0.45\textwidth]{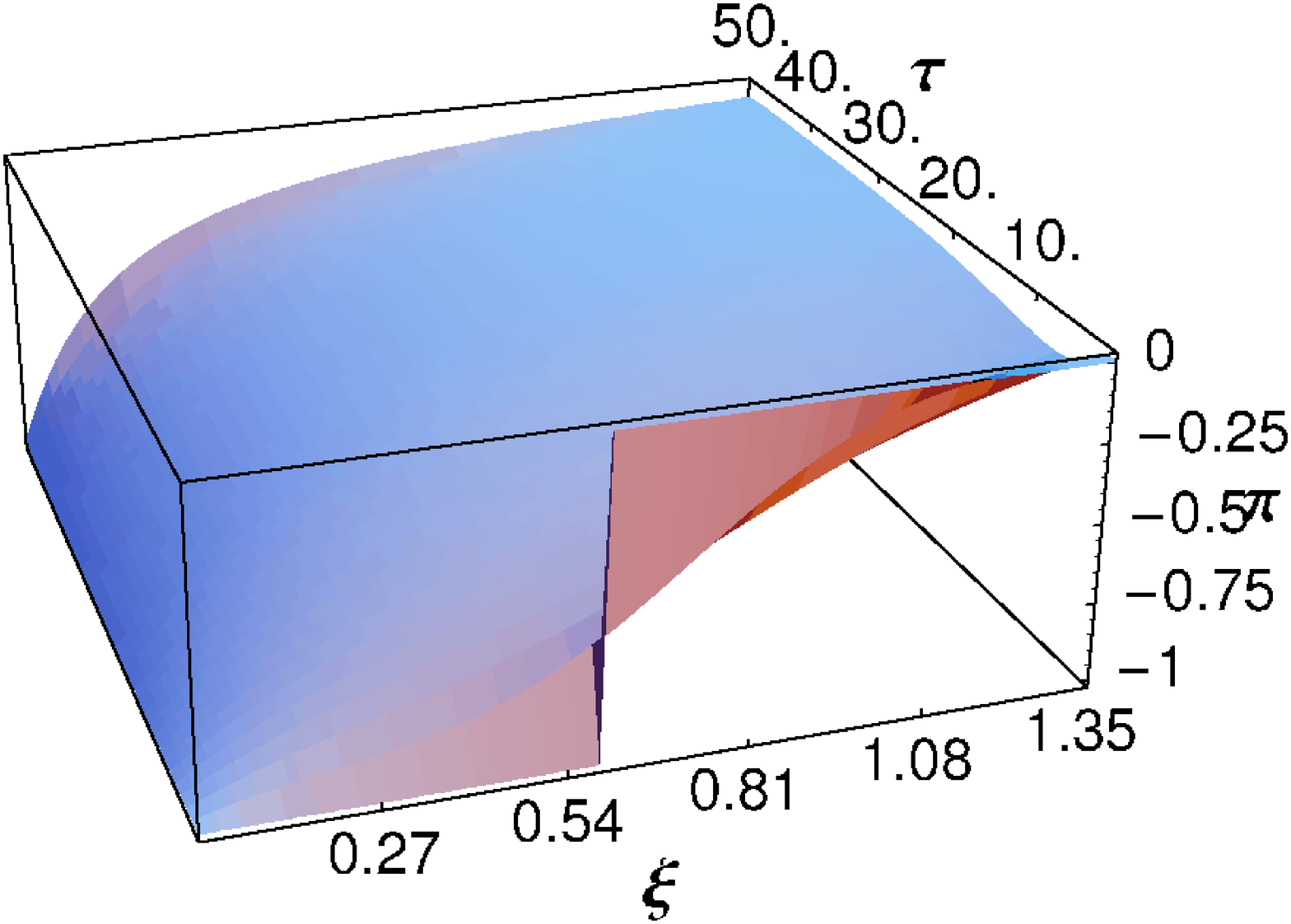}
\includegraphics[width=0.35\textwidth]{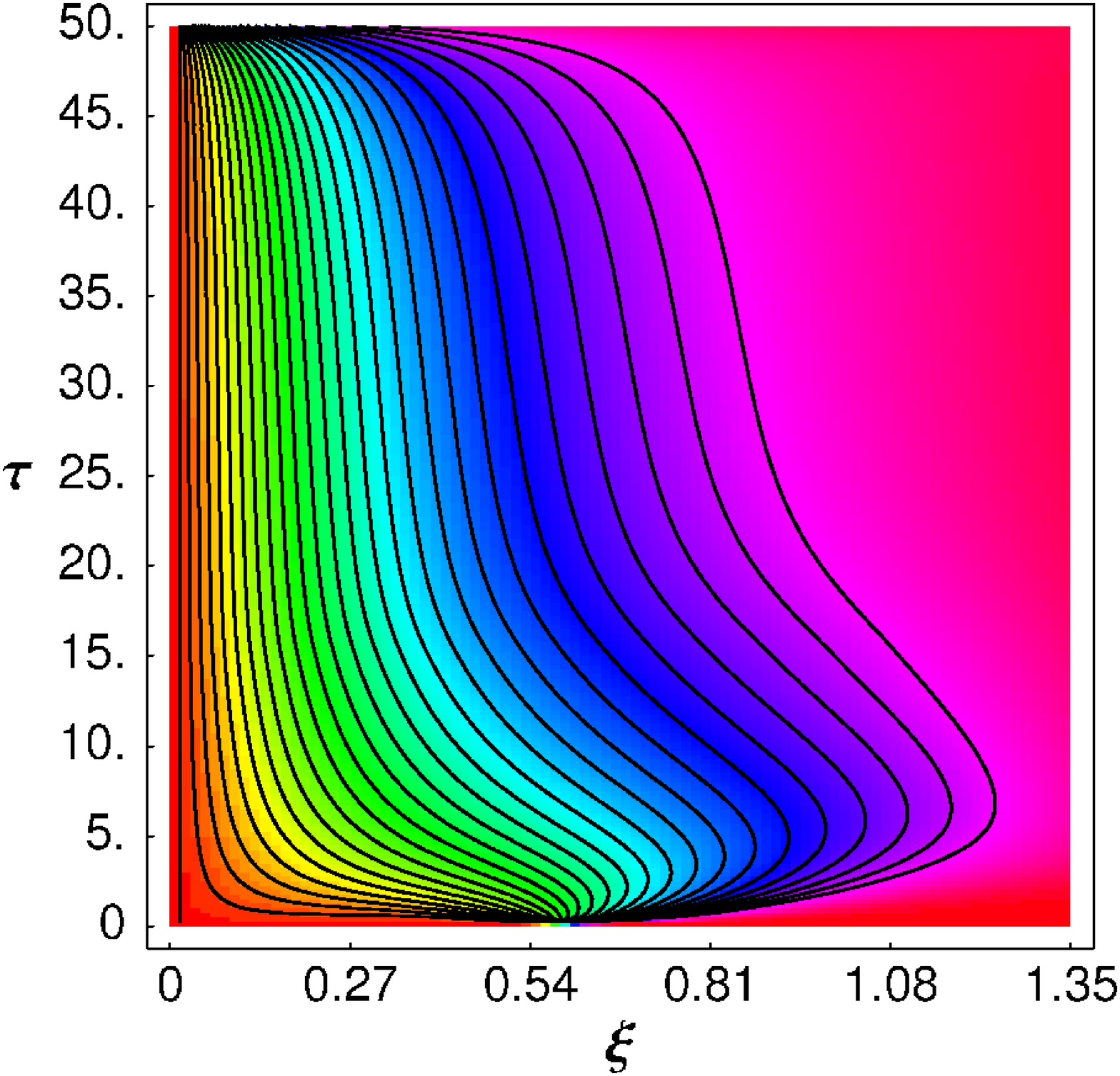}
\end{center}
\caption{\small 
A 3D plot (left) and contour plot (right) of the function $\Pi(\xi,\tau)$.}
\label{asian-fig-2}
\end{figure}

\begin{figure}
\begin{center}
\includegraphics[width=0.45\textwidth]{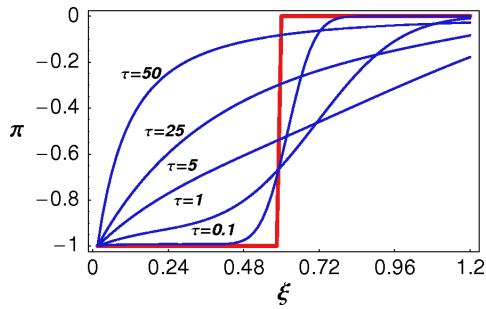}
\end{center}
\caption{\small 
Profiles of the function $\Pi(\xi,\tau)$ for various times $\tau\in[0,T]$.}
\label{asian-fig-3}
\end{figure}

\begin{figure}
\begin{center}
\includegraphics[width=0.45\textwidth]{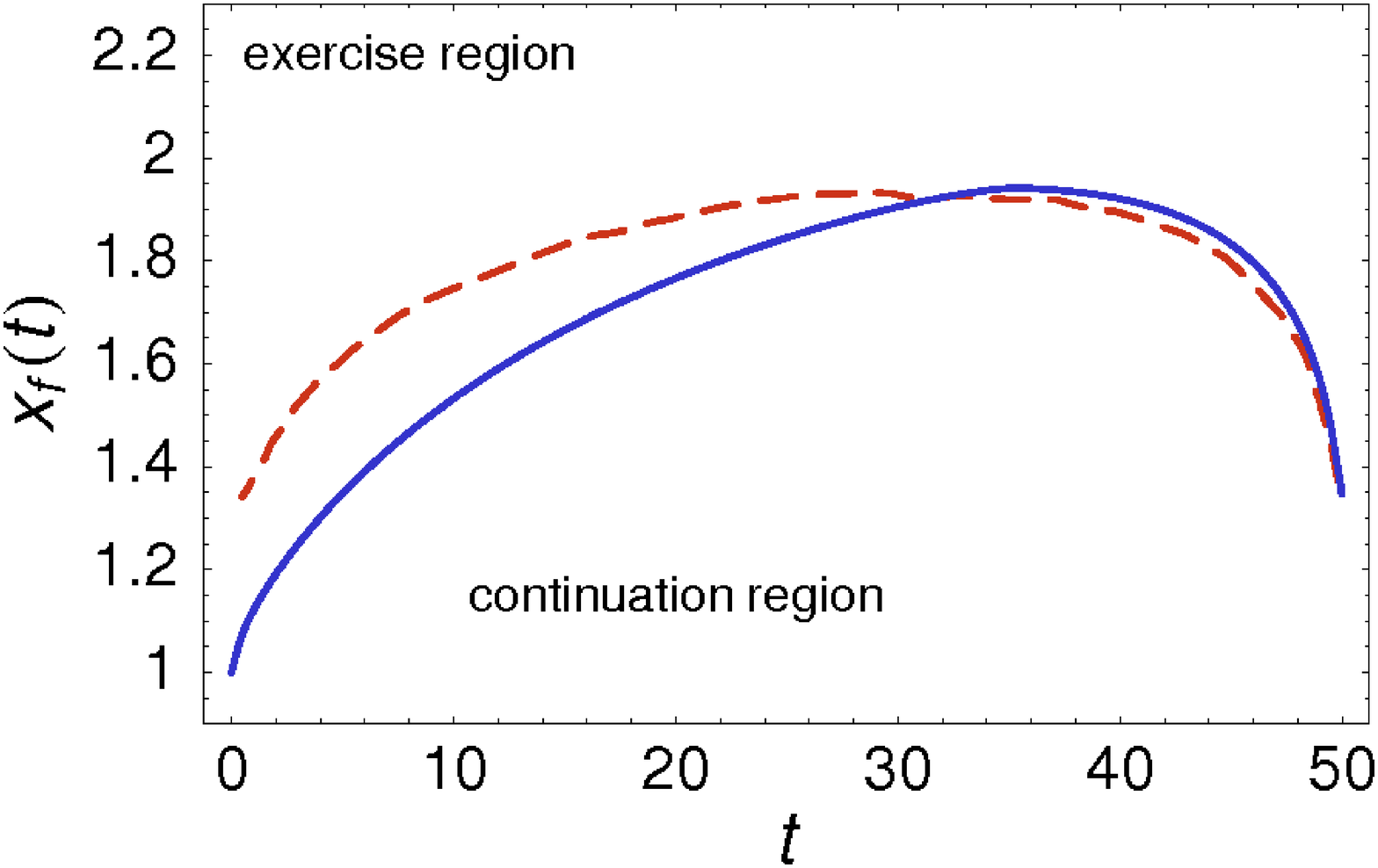}
\includegraphics[width=0.45\textwidth]{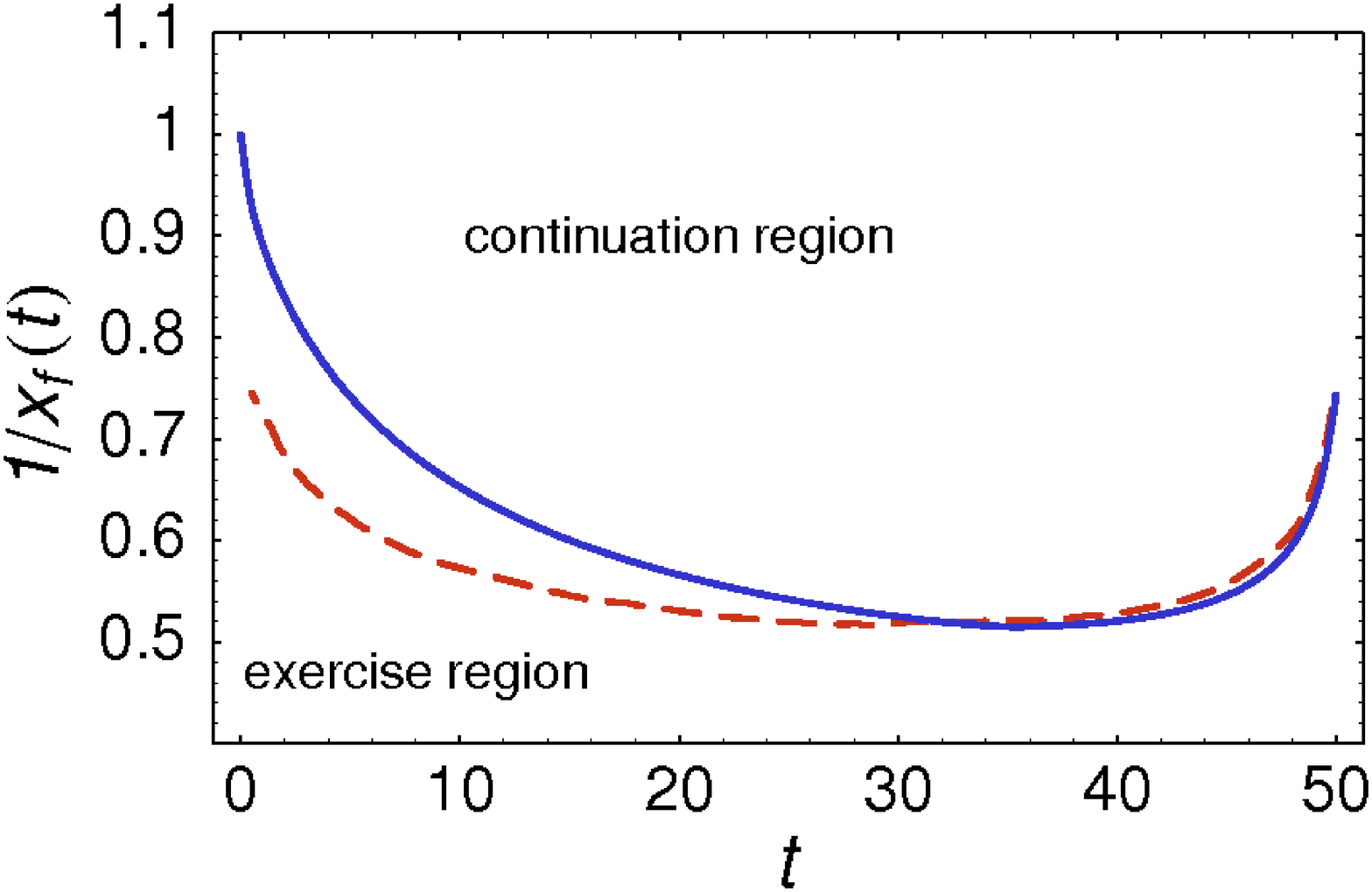}
\end{center}
\caption{\small 
A comparison of the free boundary position 
$x_f(t)=\varrho(T-t)$ (left) and $1/x_f(t)=1/\varrho(T-t)$ (right) 
obtained by our method (solid curve) and that of the projected successive over relaxation algorithm from
\cite{DK} (dashed curve).}
\label{asian-fig-4}
\end{figure}

\begin{figure}
\begin{center}
\includegraphics[width=0.45\textwidth]{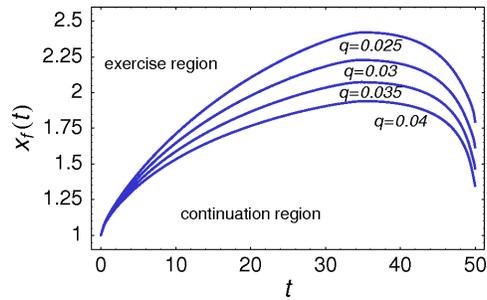}
\end{center}
\caption{\small 
A comparison of the free boundary position $x_f(t)=\varrho(T-t)$
for various dividend yield rates $q=0.04, 0.035, 0.3, 0.25$.}
\label{asian-fig-5}

\end{figure}

In Fig.~\ref{asian-fig-2} we can see the behavior of the transformed function $\Pi$ in both 3D as well as contour plot perspectives. Fig.~\ref{asian-fig-3} depicts the initial condition $\Pi(\xi,0)$ and five time steps of the function $\xi\mapsto \Pi(\xi,\tau_j)$ for $\tau_j= 0.1,1,5,25,50$. A comparison of the free boundary position 
$x_f(t)=\varrho(T-t)$ as well as of its reciprocal value $1/x_f(t)=1/\varrho(T-t)$  
obtained by our method (solid curve) and that of the projected successive over relaxation algorithm from \cite{DK} (dashed curve) is shown in Fig.~\ref{asian-fig-4}. It is clear that our method and that of \cite{DK} give almost the same result in the one third of the time interval $[0,T]$ close to the expiration time $T=50$. On the other hand, the long time behavior when $\tau\to T$ is quite different. Notice that $\lim_{\tau\to T}\ro(\tau)$ can be analytically computed and is equal to $1$ (see \cite{HJ} or \cite{Se3}). Hence the long time behavior of $\ro$ seems to be better approximated by our method. A comparison of early exercise profiles with respect to varying dividend rate $q$ is shown in Fig.~\ref{asian-fig-5}.

Finally, we present numerical experiments for shorter expiration times $T=0.5833$ (seven months) and $T=1$ (one year) with zero dividend rate $q=0$ and $r=0.05, \sigma=0.2$.

\begin{figure}
\begin{center}
\includegraphics[width=0.45\textwidth]{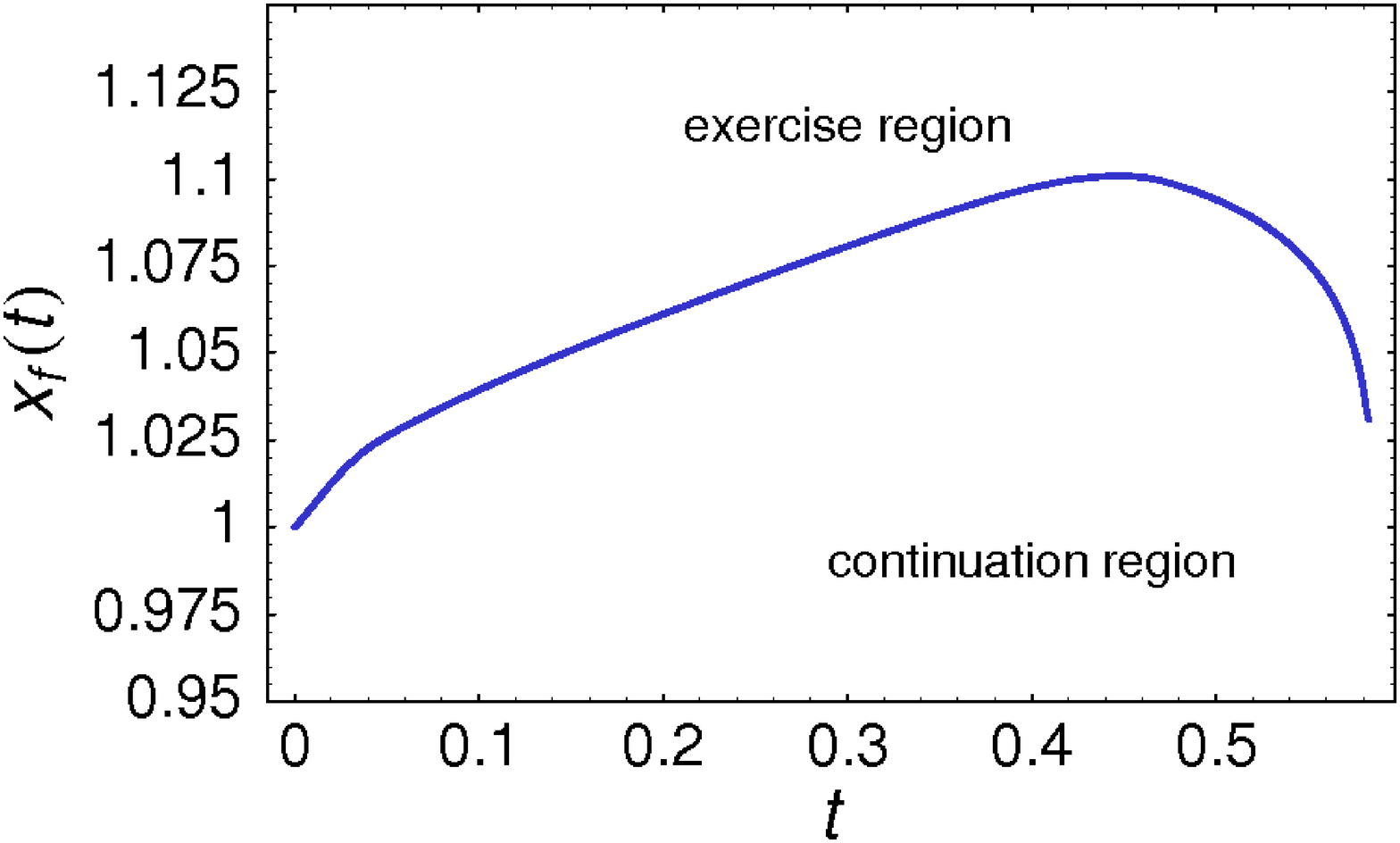}
\includegraphics[width=0.45\textwidth]{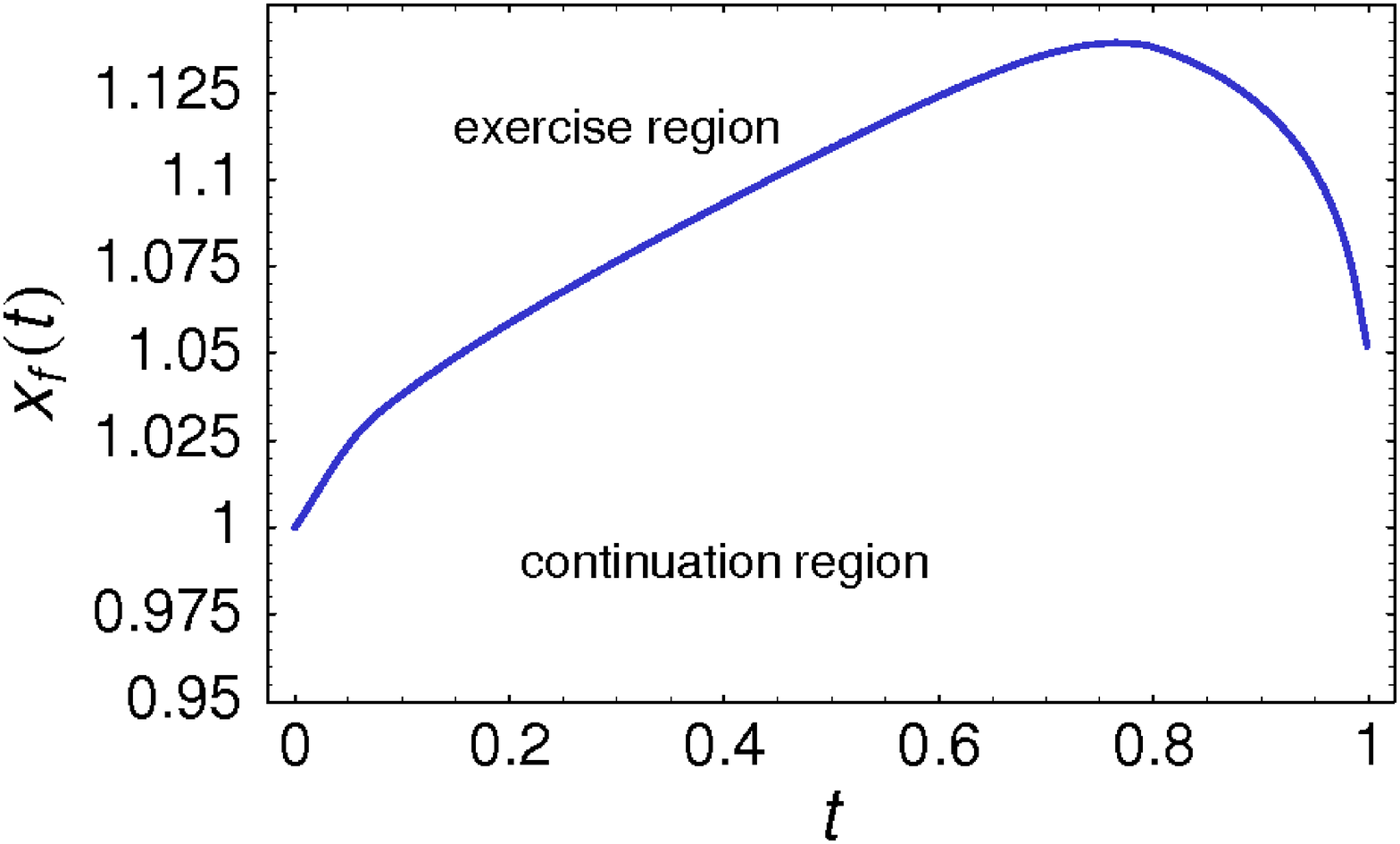}
\end{center}

\caption{\small 
The free boundary position $x_f(t)=\varrho(T-t)$ for Asian call on assets paying no dividends ($q=0$) with expiration time $T=0.7$ (left) and $T=1$  (right).}
\label{asian-fig-6}
\end{figure}

\section*{Conclusion}

In this survey paper we presented recent developments in fixed domain transformation methods applied to evaluation of the early exercise boundary for American style of options. We discussed an iterative numerical scheme for approximating of the
early exercise boundary for a class of Black--Scholes equations with a volatility which may depended on the asset price as well as the second derivative of the option price. The method consisted of transformation the free boundary problem for the early exercise boundary  position into a solution of a nonlinear parabolic equation and a nonlinear algebraic constraint equation. The transformed problem has been solved by means of operator splitting iterative technique. We also presented results of numerical approximation of the free boundary for several nonlinear Black--Scholes equation including, in particular, Barles and Soner model and the Risk adjusted pricing methodology model. The method of fixed domain transformation has been also applied for evaluation of early exercise boundary for American style of Asian option with arithmetically averaged strike price.

\section*{Acknowledgments}
The author thanks Matthias Ehrhardt for fruitful discussions and his encouragement to write this survey chapter. I also appreciate help of my student B.~Kuchar\v{c}\'{\i}k with preparation of the last section. The work was supported by  grants VEGA 1/3767/06 and DAAD-MSSR-11/2006.


\end{document}